\documentclass[iop]{emulateapj}
\usepackage{natbib}
\shorttitle{Type II SNe Environment Metallicities}
\shortauthors{Stoll et al.}

\begin{document}

\title{Probing the low-redshift star formation rate as a function of 
metallicity through the local environments of type~II supernovae}

\author{R.~Stoll\altaffilmark{1}, J.L.~Prieto\altaffilmark{2,3}, 
        K.Z.~Stanek\altaffilmark{1,4}, R.W.~Pogge\altaffilmark{1,4}
        }

\altaffiltext{1}{Department of Astronomy, The Ohio State University 
}
%                 140 W. 18th Ave., Columbus OH 43210}
%\altaffiltext{2}{Carnegie Observatories, 813 Santa Barbara Street, 
%                 Pasadena, CA 91101}
\altaffiltext{2}{Department of Astrophysical Sciences, Princeton University
} 
%                 4 Ivy Lane, Peyton Hall, Room 3, Princeton, NJ 08544}
\altaffiltext{3}{Hubble and Carnegie-Princeton Fellow}
\altaffiltext{4}{Center for Cosmology and AstroParticle Physics, 
                 The Ohio State University}
%                 The Ohio State University, 191 West Woodruff Avenue, 
%                 Columbus OH 43210}

\begin{abstract}

Type~II SNe can be used as a star formation tracer to probe the 
metallicity distribution of global low-redshift star formation.  
We present oxygen and iron abundance distributions of type~II 
supernova progenitor regions that avoid many previous sources of bias.  
Because iron abundance, rather than oxygen abundance, is of key importance 
for the late stage evolution of the massive stars that are the progenitors 
of core-collapse supernovae, and because iron enrichment lags oxygen 
enrichment, we find a general conversion from oxygen abundance to iron 
abundance.  
The distributions we present here are the best yet 
observational standard of comparison for evaluating 
how different classes of supernovae depend on progenitor metallicity.  
We spectroscopically measure the gas-phase oxygen abundance near a 
representative subsample of the hosts of type~II supernovae from the 
first-year Palomar Transient Factory (PTF) supernova search, using a 
combination of SDSS spectra near the supernova location (9 hosts) and new 
longslit spectroscopy (25 hosts).  
The median metallicity of these 34 hosts at or near the supernova location 
is 12+log(O/H) = 8.65, with a median error of 0.09.
The median host galaxy 
stellar mass from fits to SDSS photometry is $10^{9.9}$~M$_\odot$.  They do 
not show a systematic offset in metallicity or mass from a redshift-matched 
sample of the MPA/JHU value-added catalog.
In contrast to previous supernova host metallicity studies, this sample is 
%more homogeneous 
drawn from a single survey.  It is also drawn from an areal rather than a 
targeted survey, so 
supernovae in the lowest-mass galaxies are not systematically excluded.  
Indeed, the PTF 
supernova search has a slight bias towards following up transients in low 
mass galaxies.  The progenitor region metallicity distribution 
we find is statistically indistinguishable from the metallicity distribution 
of type II supernova hosts found by targeted surveys and by 
%inhomogeneous 
samples from multiple surveys with different selection functions.  
%Though iron abundance is more central to the evolution of massive stars than 
%oxygen abundance, it cannot be measured directly in extragalactic 
%HII regions.  
Using the relationship between iron and oxygen abundances 
found for Milky Way disk, bulge, and halo stars, we translate our 
distribution of type~II SN environments as a function of oxygen abundance 
into an estimate of the iron abundance, since iron varies more steeply than 
oxygen.  We find that though this sample spans only 0.65 dex in oxygen 
abundance, the gap between the iron and oxygen abundance is 50\% wider at the 
low-metallicity end of our sample than at the high-metallicity end.
\end{abstract}

% Acceptable keywords found here:
% http://authors.aas
\keywords{(stars:) supernovae: general; galaxies: abundances; galaxies: dwarf}

\section{Introduction}\label{sec:intro}

The question of how certain rare classes of supernovae (SNe) depend on 
progenitor metallicity has been limited by the lack of an unbiased standard of 
comparison.  
It is incorrect use the metallicity distribution of galaxies 
%as a standard of comparison because 
for this because
core-collapse supernovae (CCSNe) trace star formation rather than 
stellar mass.  Existing supernova host galaxy metallicity 
distributions are selected to compare matched samples and use supernova 
surveys that do not search in the faintest hosts, and are not likely 
to be suitable as a standard of comparison for rare events that appear to be 
more common in low-mass galaxies.  
We address this by measuring a metallicity distribution of 
type~II supernova progenitor regions from a single source survey that is areal 
rather than targeted.
%, while not itself completely 
%unbiased, avoids many previous sources of bias.  

The observational properties of 
%core-collapse supernovae 
CCSNe span a broad 
range of spectral types, luminosities, apparent kinetic energies, and other 
properties. 
Interpreting this diversity remains a fundamental theoretical and 
observational challenge, particularly as to how differences in the 
stellar progenitors of the SNe 
are related to the explosions.
For example, the relative fractions of hydrogen-rich type~II SNe and 
hydrogen-poor type~Ib/c SNe can be predicted as a function 
of metallicity based on models for mass-loss from the progenitor stars 
\citep{eldridge08,georgy09,georgy12}.  
Standard mass loss models for massive stars are based on line-driven winds
\citep[e.g.][]{kudritzki00}; the efficiency of these winds depends on 
metallicity because metals, particularly iron, dominate the line opacities 
driving the winds \citep{vink05}. 
Furthermore, the explosion energy of normal 
type~II SNe may depend on the progenitor metallicity \citep{kasen09}, and
certain types of SNe may occur only for low-metallicity progenitor stars
%such as pair-instability supernovae 
\citep{ober83,heger02,langer07}.

% apparently 3rd paragraph of page 1
% point 1
A serendipitous observation of a SN progenitor star prior to explosion is  
one useful way to characterize the progenitor and determine how the 
SN properties depend progenitor properties.  Unfortunately, observing or 
strongly constraining the properties of the progenitor star is only possible 
for SNe in nearby galaxies, leading to very small samples 
\citep[e.g.][and references therein]{smartt09review}.  This limits 
the utility of this technique for constraining the 
properties of subclasses of events that are rare and therefore observed 
mostly at large distances, such as extremely optically luminous CCSNe.

% apparently 4th paragraph of page 1
% point 2
% DEALT WITH.
Observational techniques for addressing this question without pre-explosion 
data involve estimating the progenitor properties from the environments that 
remain behind.  These techniques range from estimating 
progenitor age and mass from the degree of correlation with H$\alpha$ 
\citep{anderson08} and NUV \citep{anderson12} emission, to detailed 
characterization of resolved stellar 
populations near supernova explosion sites in nearby galaxies 
\citep{badenes09,murphy11} to extrapolation of stellar population properties 
from galaxy photometry and spectroscopic observations 
\citep[e.g.][]{kelly08,arcavi10,kelly12}.

Spectroscopic observations of \ion{H}{2} regions 
%from the progenitor region of 
near
a supernova can be used to estimate the metallicity using 
strong-line oxygen abundance indicators.  There are precision and 
accuracy limitations to strong-line abundance estimates, but more rigorous 
abundance measurements using faint auroral lines are too costly (or 
completely unfeasible) for statistical surveys of SN host metallicities.

% point 3, point 4.  DEALT WITH.
% Apparently 5th paragraph of page 1
% and 1st paragraph of page 2
The frequency of several classes of 
%core-collapse SN 
CCSN types has been found 
to vary with host metallicity.   
\citet{prantzos03} found that the ratio 
between type~Ib/c and type~II SNe increases with increasing host 
luminosity, and because of the galaxy luminosity/metallicity relationship, 
they suggested this was probably a metallicity effect. 
This was confirmed and expanded upon by \citet{prieto08z} in a subsequent 
study, which found that 
hosts of type~Ib/c SNe are higher metallicity than hosts of type~II and 
type~Ia SNe based on spectroscopically measured metallicity.  
%as did \citet{anderson10}.  
\citet{stanek06} found that type~Ic SNe associated with nearby 
long gamma-ray bursts (GRBs) were in faint, metal-poor galaxies, proposing 
a progenitor metallicity cut-off above which GRBs do not occur, 
and \citet{modjaz08} found that 
type~Ic SNe that were associated with GRBs were in more metal-poor regions 
than those that were not.
\citet{anderson10} found at marginal significance that type~Ib 
and type~Ic SNe hosts have slightly higher metallicity than hosts of 
type~II and type~Ia SNe.  
The metallicity local to type~Ic SNe without broad 
lines was found by \citet{modjaz11} to be on average higher than near type~Ib 
SNe, regardless of which strong-line metallicity diagnostic was used, 
consistent with the results of \citet{kelly12}, though \citet{anderson10} 
and \citet{leloudas11} found the difference to not be statistically 
significant.  
\citet{kelly12} found that although hosts of type~Ib and type~Ic 
SNe are higher in metallicity than hosts of type~II SNe, hosts of broad-lined 
type~Ic (Ic-BL) SNe are lower in metallicity.
%, and that type~Ic-BL and 
%type~IIb SNe are found in bluer ($u'-z'$) sites in their host galaxies 
%than type~II, type~Ic, type~Ib, or type~IIn SNe, presumably due to high 
%SFR and low extinction.

CCSNe which are abnormally optically luminous appear to occur more 
often in low-mass, low-luminosity galaxies \citep{neill11}.  
Spectroscopic abundance measurements of hosts of five such luminous CCSNe 
indicate this is likely due to metallicity, as shown by \citet{stoll11} 
\citep[including data from][]{young10,kozlowski10}.  These abnormally 
luminous CCSNe are predominantly SNe Ic or IIn.  Subsequent 
spectroscopic measurements of the host of the luminous SN~2008am 
\citep{chatzopoulos11} and the luminous SN~2010ay 
\citep{2010aydisc,2010ayZ,sanders12},
and photometric limits on the hosts of the luminous SNe PS1-10awh and 
PS1-10ky \citep{chomiuk11} reinforce this conclusion.

% Apparently 3rd paragraph of page 2
% point 5
% Dealt with, but could be expanded upon
Many studies \citep[e.g.][]{neill09,sanders12,stoll11,campisi11,vergani11} 
compare host galaxies of SNe or GRBs to the overall galaxy population, which 
is a good way to put the host metallicity results in context.  It is not, 
however, a secure way to evaluate whether metallicity is a key parameter 
governing whether a supernova has a given spectral type or luminosity.  
Supernovae trace star formation rather than overall stellar mass, and 
star formation is not evenly distributed among galaxies of a given mass.  
Recent results \citep{laralopez10,mannucci10} show that the scatter in the 
galaxy mass-metallicity relationship may be reduced by considering star 
formation as a third parameter, and that star formation rates (SFRs) are 
higher in lower metallicity galaxies at a given mass 
\citep[but see also][which finds that the form of any such relation between 
mass, metallicity, and star formation rate depends on the strong-line 
metallicity diagnostic used]{yates12}.

% ******** Perhaps cite neill09 here as well?
\citet{levesque10b} and \citet{han10} suggested that instead of a metallicity 
cutoff for GRBs there is a separate luminosity-metallicity 
relationship for GRB host galaxies offset to lower metallicity than the 
normal galaxy mass-metallicity relationship of \citet{tremonti04}.  (These 
results are consistent with a simple metallicity cutoff 
%\citep[e.g.][]{stanek06} 
with the exception 
of a single high-mass GRB host galaxy that appears to be high metallicity.)  
No such offset has been observed for type Ia SNe \citep{neill09}.
This question of an offset in the mass-metallicity relationship has confused 
many subsequent discussions of a metallicity cutoff.

% Apparently 5th paragraph of page 2
% point 6
% It was intentionally vague to avoid unnecessary wordiness...
% dealt with, but could be done better
\citet{mannucci11} notes that for galaxies at a given mass, lower metallicity 
galaxies have higher average star formation rates \citep{mannucci10} and thus 
core-collapse events, which trace star formation, should have a 
mass-metallicity distribution shifted to slightly lower metallicity at a 
given mass.  This effect is qualitatively similar enough to the offset 
proposed by \citet{levesque10b} and \citet{han10} that \citet{mannucci11} 
conclude that GRB hosts do not differ substantially from the typical galaxy 
population and therefore there is no metallicity dependence to GRB hosts.  
\citet{kocevski11} quantifies the expected metallicity shift, and finds that 
the star-formation-weighted relationship between galaxy mass and metallicity 
is shifted towards lower metallicity at a given mass, but that this alone 
cannot explain the observed GRB host distribution, and a metallicity 
dependence is still required.
Many subsequent studies have claimed to disprove any dependence on metallicity 
% metallicity connection 
for GRBs or luminous SNe by showing that the host galaxy metallicity is 
consistent with its mass and star formation rate according to the 
relationship between the three \citep{laralopez10,mannucci10}.  While this 
is evidence against a distinct mass-metallicity relationship for the hosts 
of these SNe, it is not evidence against a \emph{metallicity dependence}.

% Apparently 6th paragraph of page 2
% point 7
This shift can also be 
quantified semi-analytically by convolving the galaxy mass function with the 
relationship between galaxy mass, metallicity, and star formation rate to 
find the overall distribution of star formation as a function of 
metallicity.  
\citet{niino11} does so by comparing the observed metallicity distribution 
of GRB hosts to the metallicity distribution of star formation, calculated 
two different ways; first, calculated observationally from the stellar mass 
function, the galaxy M-SFR relation, and the galaxy mass-metallicity 
relation, and second, calculated using the relationship between galaxy mass, 
metallicity, and 
%star formation rate 
SFR defined by \citet{mannucci10}, and 
assigning metallicities based on 
%star formation rate 
SFR as well as mass.
He finds the difference between these two estimates is less than 
0.5~dex in oxygen abundance on the KK04 scale, 
Regardless which method is used, he finds that the GRB host metallicity 
distribution is incompatible with the metallicity distribution of star 
formation unless the GRB fraction depends on metallicity.
(This is not the primary result of that study, which examines the fact that 
galaxies do not have one single metallicity, but show an internal spread.  
Assuming a hypothetical transient phenomenon which has a strict cutoff 
metallicity above which it cannot occur, \citet{niino11} shows 
that such a spread in internal galaxy metallicities would serve to widen the 
observed host metallicity distribution of that transient phenomenon.)

Only by comparing the metallicity distribution of a SN variety to the overall 
metallicity distribution of star formation can one rigorously test for a 
metallicity dependence.  Our metallicity distribution of type~II SN hosts can 
serve as a standard of comparison for evaluating how rare classes of SNe 
depend on progenitor metallicity.

% Apparently 8th paragraph of page 2
% point 8
%A problem for essentially all of these previous studies relating metallicity 
%and SN properties is that they draw on heterogeneous SN samples.  
Essentially all of these previous studies relating metallicity and SN 
properties draw on SN samples from multiple surveys.
A perfectly homogeneous and unbiased SN sample does not currently exist.  
At present, the Palomar Transient Factory \citep[][PTF]{rau09}  appears to 
supply the closest approximation to this, in 
the sense that the SN selection does not exclude the smallest, most vigorously 
star-forming galaxies.  This would not be true, for example, of the Lick 
Observatory Supernova Search (LOSS) survey \citep{li11}, which explicitly 
targets larger galaxies.  In contrast, the PTF biases their followup slightly 
towards supernovae in low-mass galaxies (I.~Arcavi, private communication).
%*********  
In this paper we will focus on the 52 type~II SNe found in the 
first year of PTF operations.  Because some subtypes of CCSNe are known to 
have different distributions in host metallicity, we focus on type~II SNe.  
We have measured metallicities for the environments of a 
representative subsample of 34 of these type~II SNe, 
and present the resulting metallicity distribution.

This distribution probes the low-redshift star formation rate as a function of 
metallicity in an independent way from current methods relying on 
galaxy population statistics.  This is an important distribution to 
characterize because in order to determine whether a massive star outcome 
has a metallicity dependence, we need to examine its frequency relative to 
the metallicity distribution of star formation, not to the metallicity 
distribution of existing stellar mass (the galaxy metallicity distribution).  
% backpedal a little more here, since this is not what we are doing.
% 
Because the PTF followup is biased slightly towards transients in lower mass 
host galaxies, we expect the progenitor region metallicity distribution we 
measure may be slightly more metal-poor than the true overall progenitor 
distribution of type II SNe.  This makes our distribution a rigorous 
standard of comparison for 
a population suspected to be metal-poor; if that population is significantly 
more metal-poor than the distribution we present here, it is certainly more 
metal-poor than the overall distribution of type II SNe.

% Apparently this is 3rd paragraph, page 3
% point 9.  
% Yes, but not here.  
% Fine.  Sacrificed clarity for precision.
We place the metallicity distribution of type~II SNe environments in context 
by comparing it with previous supernova host studies 
\citep{prieto08z,anderson10,kelly12}, with the SDSS DR7 MPA/JHU 
value-added catalog \citep{kauffmann03,tremonti04,brinchmann04,salim07}, 
and with estimates from galaxy population statistics \citep{stanek06}.
Noting that iron is more fundamental than oxygen to the evolutionary outcomes 
of massive stars because iron opacity drives stellar winds, we then translate 
our oxygen abundance distribution of type~II SNe environments into an iron 
abundance distribution by using the observed relationship between oxygen and 
iron abundance in Milky Way bulge, disk, and halo stellar abundances.

\section{Spectroscopic observations and analysis}

% Apparently 4th paragraph, page 3
% point 10
% dealt with
We drew our targets from the first-year of the Palomar Transient 
Factory (PTF) survey \citep{arcavi10}, an areal rather than a targeted 
survey, so supernovae in the lowest-mass galaxies are not excluded by 
selection.  The sample selection for the survey is not yet published, but 
transients in dwarf hosts are prioritized for spectroscopic followup 
(I.~Arcavi, private communication), so a slight bias towards metal-poor 
environments is expected.  In contrast, a bias towards metal-rich 
environments would be expected 
for galaxy-targeted surveys, which miss low-mass galaxies entirely.
There are 52 type~II SNe in the full first-year PTF CCSN sample, 
of which we have 
measured spectroscopic metallicity determinations for a subsample of 34. 
This subsample is representative of the overall type~II sample, as we 
show in \textsection\ref{sec:repr}.

\begin{figure}
\begin{center}
\begin{tabular}{c}
\includegraphics[width=7.62cm]{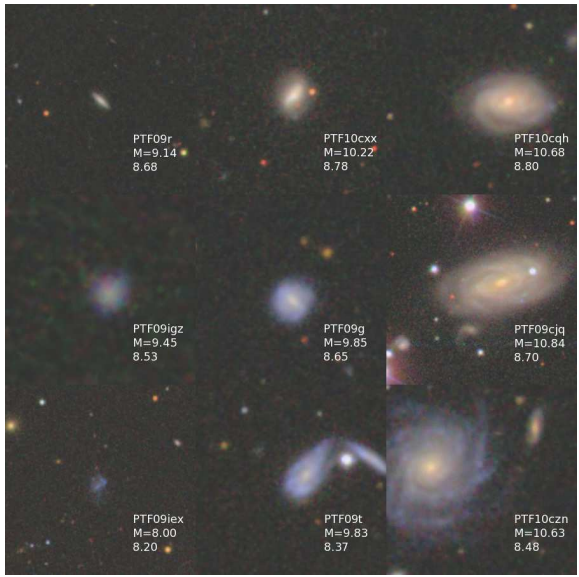}
\end{tabular}
\end{center}
\caption
{\label{fig:finding}
Nine type~II SN hosts, spanning the 
metallicity and mass range, arranged so mass increases toward the right and 
oxygen abundance increases toward the top.  Each panel is scaled to a physical 
size of 50 kpc ($H_0,\Omega_m,\Omega_\Lambda=70,0.3,0.7$) and is centered 
on the position of the supernova.  All images are from SDSS and were taken 
before the supernovae.
}
\end{figure}

We obtained host galaxy spectra of these SNe using 
the Ohio State Multi-Object Spectrograph 
\citep[OSMOS,][]{martini11,stoll10} on the 2.4-m Hiltner telescope, the Wide 
Field Reimaging CCD Camera (WFCCD) on the 2.5-m du Pont telescope, 
and the dual imaging 
spectrograph (DIS) on the 3.5-m Astrophysical Research Consortium telescope.
We also use twelve archival spectra from SDSS DR7 
\citep{dr7,uomoto99,york00,gunn06}.  
The properties of the spectroscopic observations are summarized in 
Table~\ref{table:obs}.  
We processed the data using standard techniques in IRAF\footnote{IRAF is 
distributed by the National Optical Astronomy Observatory, which is operated 
by the Association of Universities for Research in Astronomy (AURA) under 
cooperative agreement with the National Science Foundation.}, individually 
extracting each spectrum.  Relative flux calibration was done with 
observations of spectrophotometric standard stars taken each night.  
The PP04N2 metallicity diagnostic is extremely insensitive to reddening 
because it depends on the flux ratio of two very close lines, so intrinsic 
extinction corrections were not applied.
Images of nine representative type~II hosts spanning the observed range of 
metallicity and galaxy mass are shown in Figure~\ref{fig:finding}.
We also measured spectroscopic metallicities for the hosts of 
three type~Ib, two type~IIb, three type~Ic, and one type~Ic-BL from the 
first-year PTF sample.  
These numbers are too small for rigorous statistical comparison, so we 
exclude these from our analysis of the type~II sample, and discuss them 
further in \textsection~\ref{sec:sel}.

%----------
%----------

The new observations were made either at the supernova position or at a 
similar galactocentric radius to minimize any biases from metallicity 
gradients in the host galaxies.   
We include the angular distance from the host galaxy center to the supernova 
site in Table~\ref{table:AllZ} for easy comparison to the seeing and to the 
spectroscopic aperture, listed in Table~\ref{table:obs}.  We include the 
projected physical distance from the galaxy center to facilitate future 
comparison to studies that use galactocentric spectra.  
The SDSS spectra are the only spectra taken at the galaxy center
rather than at the galactocentric radius of the supernova.  Note that for
these SDSS spectra, most SN locations are within the fiber diameter, and only
one is more than 2 fiber diameters away.  We obtained new spectra for any
hosts with existing SDSS spectra more distant than this.  Any resulting 
effect from galactic metallicity gradients on the metallicity distribution 
should be minimal. 
 Line fluxes 
of H$\alpha$~$\lambda 6563$ and [\ion{N}{2}]$\lambda 6584$ are given in 
Table~\ref{table:AllZ}.

For host galaxies with multiple SDSS spectra or with spectra from multiple 
sources, we fit a metallicity gradient where possible and provide the 
best fit metallicity at the galactocentric radius of the supernova progenitor; 
these are labeled as `grad'.  Line fluxes, observed galactocentric radii, and 
derived metallicity for these measurements are given in 
Table~\ref{table:gradlineflux}.

We primarily consider host metallicities determined with the N2 diagnostic 
of \citet{pp04}, which we directly measure for each of our targets.  
This diagnostic depends solely on 
[\ion{N}{2}]$\lambda 6584/$H$\alpha \lambda 6563$, and is extremely 
insensitive to reddening, though it has a larger intrinsic scatter 
than other strong-line diagnostics based on the physical conditions of the 
\ion{H}{2} regions.  There are a number of techniques that are used to 
estimate the oxygen abundances of \ion{H}{2} regions in star-forming 
galaxies, and a substantial literature discussing their various merits and 
drawbacks \citep[e.g.][and references therein]{kewley08}.  A full 
recapitulation of this is outside the scope of this paper, but we discuss the 
consequences of 
these uncertainties for our study in Section~\ref{sec:Zdiagnostics}.

\begin{figure}
\begin{center}
\begin{tabular}{c}
\includegraphics[width=8cm]{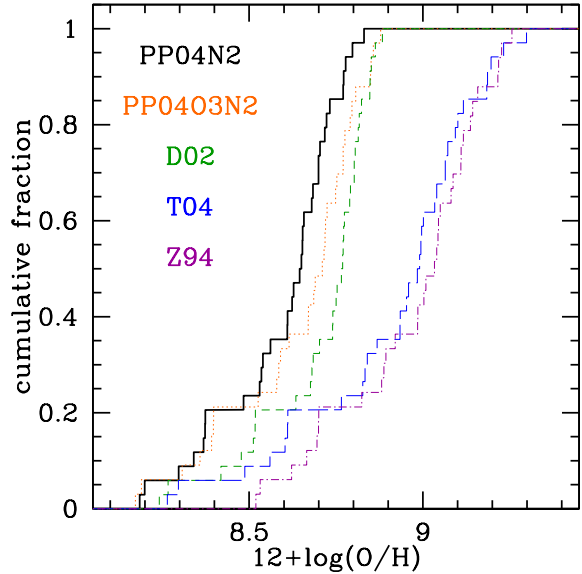}
\end{tabular}
\end{center}
\caption
{\label{fig:ptfZIIcdf}
Cumulative distribution of local gas-phase oxygen abundances for 
%our sample of 
type~II SNe.
As type~II SNe trace young stellar populations, this traces the 
metallicity distribution of 
star formation at low redshift.
%global SFR in the low-z universe as a function of metallicity.  
The solid black distribution uses the N2 diagnostic of \citet[PP04]{pp04} we adopt 
as our standard.  Subsequent figures only use these measurements.
The other distributions show how the result would change for other 
strong-line metallicity diagnostics, based on the empirical conversions of 
\citet{kewley08}.  The dotted orange curve uses the PP04 O3N2 diagnostic, 
the short-dashed green curve uses \citet{d02}, the long-dashed blue curve 
uses \citet{tremonti04}, the dot-dashed purple curve uses \citet{z94}.  
We do not show conversions to the scales of \citet{kk04}, \citet{m91}, and 
\citet{kd02}, which require external branch information.
%In black (solid) is the distribution measured with the N2 diagnostic of 
%\citet{pp04}.  We also plot several other strong-line metallicity diagnostics 
%using the empirical conversions of \citet{kewley08}.  In orange (dotted) 
%is the O3N2 diagnostic of \citet{pp04}.  In green (short dashed) is the 
%diagnostic of \citet{d02}.  In blue (long-dashed) is the diagnostic of 
%\citet{tremonti04}.  In purple (dot-dashed) is the diagnostic of \citet{z94}.
%We do not plot conversions to the scales of \citet{kk04}, \citet{m91}, and 
%\citet{kd02}, which require external branch information.  
%In subsequent figures, we plot only the PP04N2 diagnostic, shown here in black.
%The selection effects of this method of measuring the 
%metallicity distribution of star formation are mostly independent of 
%existing methods.  
}
\end{figure}

% ? 8th paragraph, page 3?
% point 11
In Table~\ref{table:AllZ} we list the metallicities of the progenitor regions 
of 34 type~II SNe, 
three type~Ib, two type~IIb, three type~Ic, and one type~Ic-BL 
from the PTF first-year core-collapse sample.  
All subsequent analysis is of the type~II hosts, for which we have 
good statistics.  
We do not include type~IIb SNe in the type~II sample because their  
spectral similarity to type~Ib SNe at all but early times can lead to some 
typing issues, and because previous studies \citep[e.g.][]{modjaz11,kelly12} 
considered them in the stripped-envelope subclass.
We show the metallicity distribution of the type~II hosts 
for several different strong-line metallicity diagnostics using the 
empirical conversions of \citet{kewley08}
in Figure~\ref{fig:ptfZIIcdf}.
In subsequent figures, we choose as our scale convention only the N2 
diagnostic of \citet{pp04}.
%which we directly measure, 
%a choice we discuss in Section~\ref{sec:Zdiagnostics}.

%----------
%----------

%******************************************************

\section{Characterizing the spectroscopic subsample}

To investigate whether we have acquired spectra of a representative subsample 
of the hosts of first-year PTF CCSNe \citep{arcavi10}, we compared the 
stellar mass and star formation rates of the hosts with and without 
metallicity estimates.  
We used SED models of the SDSS photometry of the hosts within the DR8 
footprint to estimate the masses, SFR, and characteristic stellar ages. 
Of the 52 type~II SNe in the full sample, 47 have SDSS photometry.  
We also analyzed the properties of the 19 non-type~II PTF CCSN hosts that 
fell in the DR8 footprint, but we will 
restrict our comparisons with our spectroscopic 
sample of type~II SNe to these 47 type~II SNe
to avoid any of the currently known selection effects with metallicity 
linked with supernova type (see \textsection\ref{sec:intro}).

\subsection{Extracting fluxes from SDSS imaging}\label{sec:phot}

% Apparently 3rd paragraph, page 4
% point 12
We began with 
%used 
%made use of 
%obtained 
$ugriz$ images of the 66 first-year PTF CCSN fields in the SDSS Data 
Release 8 \citep{dr8}.  
These images are fully calibrated in the SDSS natural system, 
which is close to the AB system, and sky-subtracted.   We combined the most 
sensitive SDSS bands ($gri$) for each supernova field in order to make a 
deeper stacked image that can be used to find all the galaxies and define 
their photometric apertures.  
We used these deeper stacked images as the reference image for source 
detection using SExtractor \citep{sextractor} and checked by eye the 
positions around each supernova in order to select the most likely host 
galaxy.  We were able to assign likely host 
galaxies for 64/66 SNe. The two events without host galaxy 
detections, PTF09be and PTF09gyp, have sources that are $\gtrsim 13$~kpc 
(projected) from the positions of the SNe. 
We note that the host of PTF09gyp has a reported magnitude of $r=21.75$~mag in 
\citet{arcavi10} from pre-explosion PTF survey co-adds 
(I.~Arcavi, private communication), but we cannot confirm 
this detection with DR8, although a galaxy detected approximately 15 
arcseconds away has a 
similar magnitude.  After selecting the host galaxies in the stacked images, 
we used {\it imedit} in IRAF\footnote{IRAF 
is distributed by the National Optical Astronomy Observatory, which is 
operated by the Association of Universities for Research in Astronomy (AURA)
under cooperative agreement with the National Science Foundation.} 
to mask nearby stars, which could contaminate the flux measurements, 
filling the masked regions with the local background.
Finally, we 
ran SExtractor on the individual $ugriz$ images using the apertures defined 
from the deeper stacked images to obtain total (AUTO) galaxy fluxes. 
We applied the small 
($\lesssim 0.04$~mag) corrections derived in \citet{kessler09} to transform 
the SDSS fluxes to the AB system. The resulting coordinates and fluxes of the 
host galaxies are presented in Table~\ref{table:SDSS}, and absolute magnitudes 
k-corrected and corrected for galactic extinction 
are presented in Table~\ref{table:absmags}.
We include $3\sigma$ upper limits for the hosts of PTF09be and PTF09gyp, which 
were calculated assuming a circular aperture of radius $r=5$~kpc at the 
distance of the SN.

\subsection{Galaxy properties}\label{sec:FAST}

We used the code for Fitting and Assessment of Synthetic Templates 
\citep[FAST v0.9b,][]{kriek09} to fit these host galaxy spectral energy 
distributions to estimate the stellar mass, star formation rates (SFRs), and 
characteristic ages. We chose 
the \citet{bruzual03} libraries, a Salpeter IMF, and Solar ($Z=0.02$) 
metallicity to do the fits. We also assumed an exponentially declining SFR 
model with $\tau=1$~Gyr for the star forming component of the model.  
The results are presented in Table~\ref{table:FAST}.

\begin{figure}
\begin{center}
\begin{tabular}{c}
\includegraphics[width=8cm]{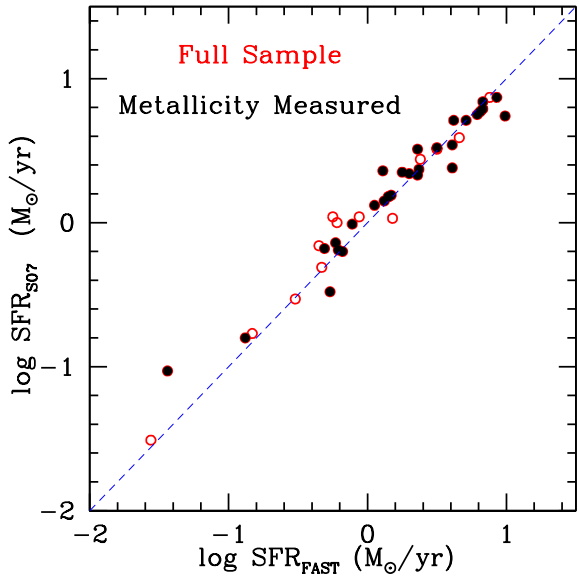}
\end{tabular}
\end{center}
\caption
{\label{fig:sfrDirComp}
The two methods of estimating SFR are 
consistent.  SFR estimates based on $u$-band photometry 
%\citep[][left]{salim07}.
\citep{salim07}
are plotted against SFR estimates based on FAST for the full sample of 
PTF type~II SN hosts (open red points) and the metallicity 
subsample (solid black points).  The blue dashed line represents a 1:1 
correspondence.  
}
\end{figure}

% apparently 5th paragraph, page 4
% revised to address referee's point 14
In order to check the SFRs calculated with FAST, we also estimated SFR based 
on the results presented by \citet{salim07}.  This method is also based on 
toy models for the SFR as a function of time, but combines bursts, constant 
star formation, and exponentially declining star formation, which make it 
more realistic than FAST.  We used the results of \citet{salim07} for 
$\sim 50000$ SDSS galaxies with GALEX photometry to derive a relation between 
the absolute $u$-band magnitudes, corrected for intrinsic attenuation, and 
their derived SFRs.  We obtain a linear relation fit between $M_u$ and SFR 
( in M$_\odot$/yr) of 
the form $\rm log(SFR) = -0.36\times M_u - 6.73$ (for a Salpeter IMF), valid 
for $\rm 2 > log(SFR) > -2$ and $\rm log(SFR/M_{*}) > -10.5$, with an rms 
scatter of 0.23~dex. We applied this relation to obtain SFRs for the 
supernova host galaxies, using the Low-Resolution Template code of 
\citet{assef08} to derive $K$-corrected $u$-band absolute magnitudes 
corrected for Galactic extinction. After further correcting these magnitudes 
by intrinsic attenuation using the values obtained with FAST and the Calzetti 
reddening law, we applied the linear relation derived from the \citet{salim07} 
data to estimate SFRs. These values are presented in Table~\ref{table:FAST}.
The agreement with the SFRs derived by FAST is fairly good in general,
with a Kolmogorov-Smirnov (K-S) test probability of 81\% of the results of 
the two different methods being drawn from the same underlying 
distribution.  The two SFR estimation methods are directly compared in 
Figure~\ref{fig:sfrDirComp}.  The SFR calculated from the $u$-band luminosity 
with aperture corrections is more consistent with the method the MPA/JHU 
value-added catalog uses to determine star formation rates than the FAST 
template fitting.

%----------
%----------

%\section{Results and Discussion}
\section{Results and Discussion}
%\section{Discussion}

This is the largest sample yet of supernova host galaxy spectra metallicity 
measurements from a single survey.  Nevertheless, limited observing time made 
following up the entire PTF type~II sample impractical, so we first show 
in \textsection\ref{sec:repr} that our spectroscopic host sample is 
representative of the full sample.  We then place the SN hosts in context in 
\textsection\ref{sec:context} by comparing their properties to those of 
galaxies in the MPA/JHU value-added catalog.  We find that while they are 
well-matched in galaxy mass and metallicity, the type~II hosts appear to be 
biased toward higher star formation rates than the galaxies in the catalog.  
We show in \textsection\ref{sec:comp} that the metallicity distribution of 
these type~II hosts is remarkably similar to that found by previous studies 
of hosts of type~II SNe, despite coming from multiple surveys with different 
selection functions.  Their 
metallicity distribution is also consistent with a distribution of star 
formation calculated from galaxy population statistics 
(\textsection\ref{sec:theodist}).  
We discuss in \textsection\ref{sec:sel} how our study avoids some selection 
effects due to supernova type and host galaxy type that might influence the 
metallicity distribution.  A key future use of the metallicity distribution 
of type~II SNe we find here will be to evaluate possible metallicity 
dependence of other subclasses of CCSNe.  We discuss the advantages and 
disadvantages of the metallicity diagnostic we choose for this 
study in \textsection\ref{sec:Zdiagnostics}.  Finally, we fit a 
relationship between oxygen and iron abundances in 
\textsection\ref{sec:iron}, and convert our observed oxygen abundance 
distribution into an assumed iron abundance distribution, as iron is more 
important to the evolution of massive stars than oxygen.

\subsection{How representative is the metallicity sample?}\label{sec:repr}

\begin{figure*}
\begin{center}
\begin{tabular}{c}
\plottwo{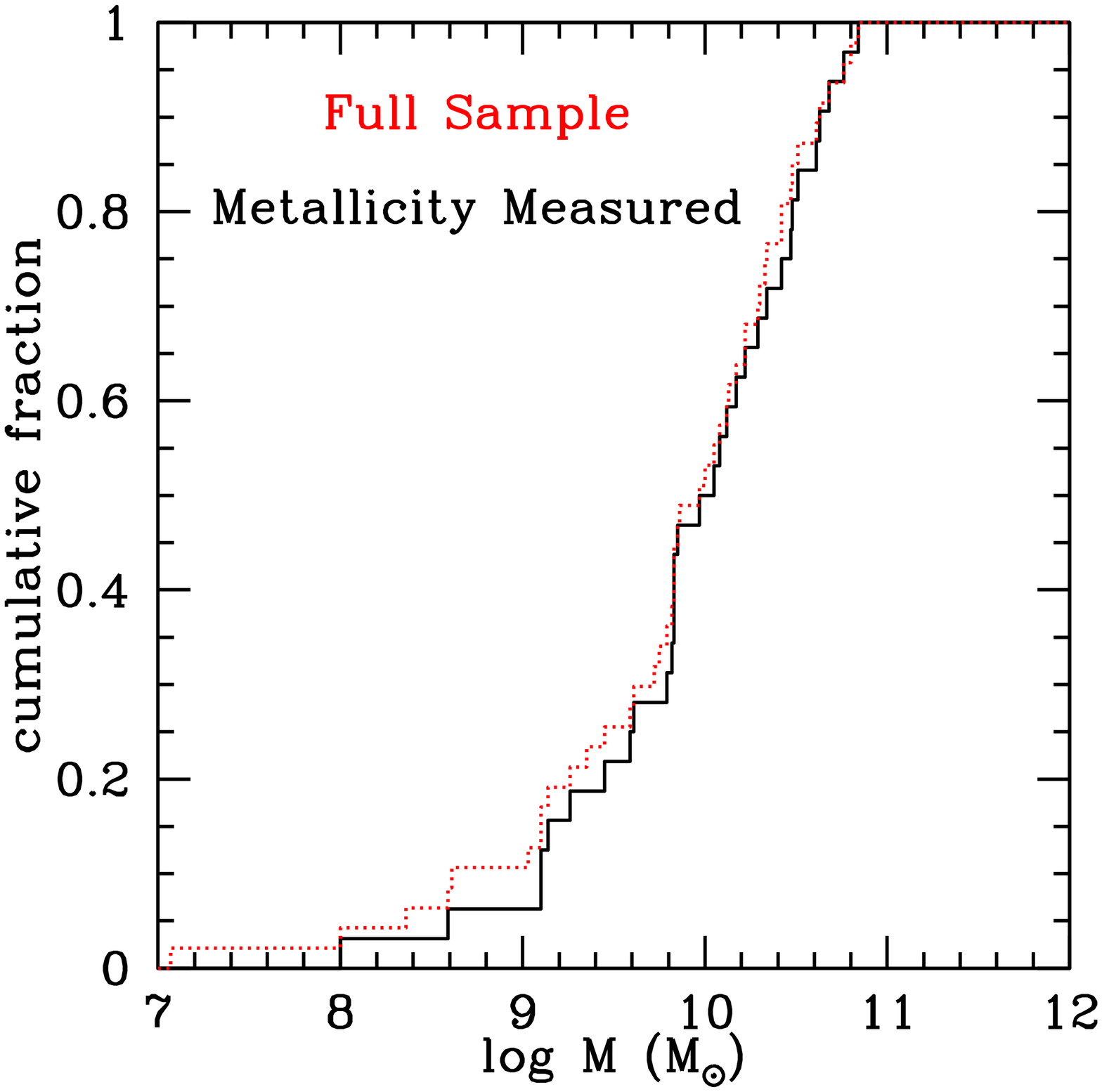}{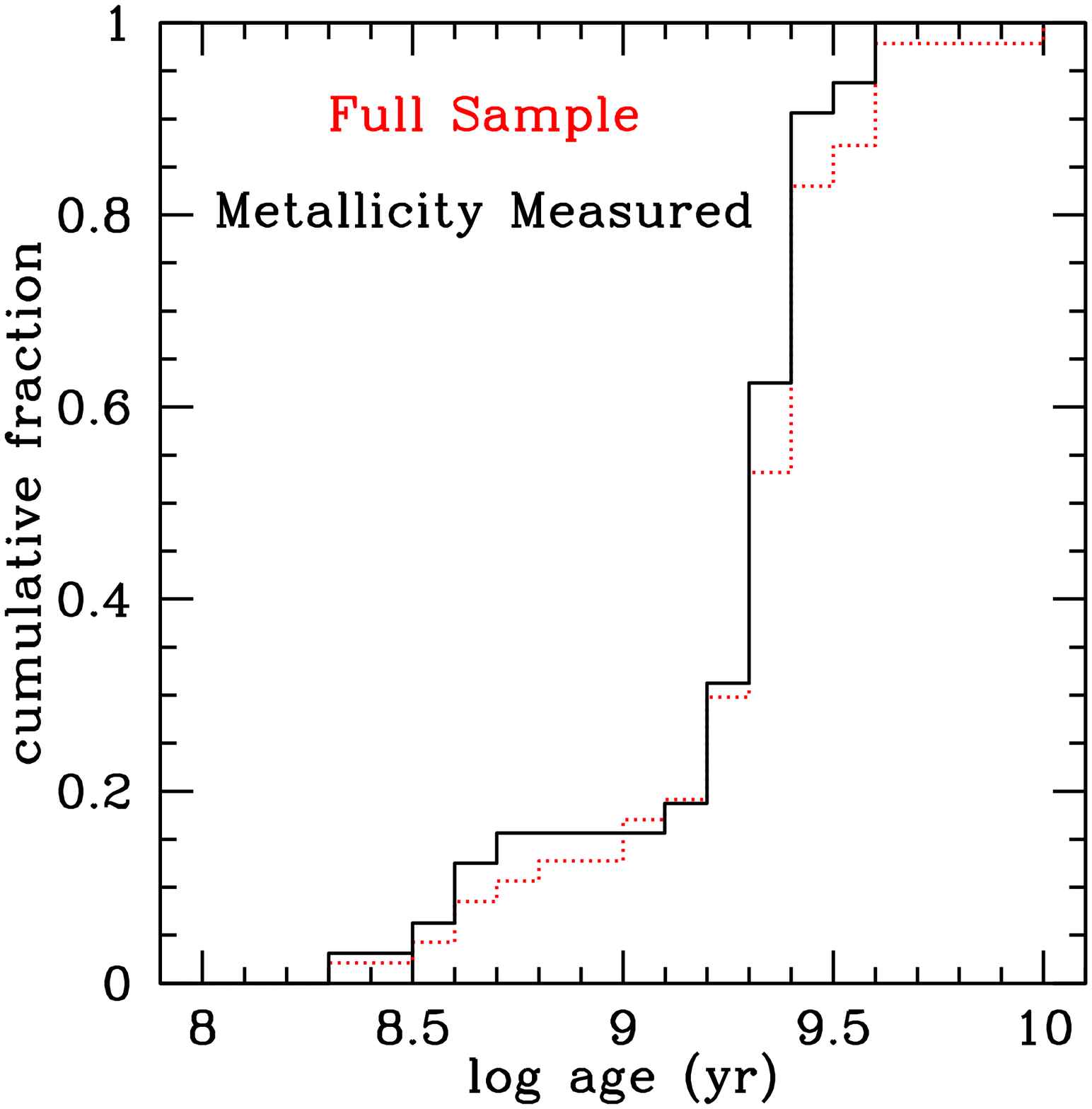}
\end{tabular}
\end{center}
\caption
{\label{fig:representative}
The distributions of the full PTF type~II sample (dotted red) and those for 
which we obtained metallicities (solid black).  Section~\ref{sec:FAST} 
describes how the host properties were estimated.
The subsample for which we have spectroscopic metallicity measurements 
is quite representative of the full sample, with K-S test probabilities of 
99.9\% (mass) and 57.8\% (age) that they are drawn from the same distribution.
%The distribution of galaxy mass on the left and characteristic stellar age 
%on the right for the full sample of first-year PTF type~II SN hosts in red 
%(dotted), and the subsample for which we have determined metallicities in 
%black (solid).  The host 
%properties were fit using FAST on SDSS photometry, as described in 
%Section~\ref{sec:FAST}. 
%
%Performing
%K-S tests yields a 99.9\% probability that the galaxy masses are drawn
%from the same distribution and a 57.8\% probability that the characteristic
%ages are drawn from the same distribution.
%The subsample for which we have measured metallicities 
% appears to be
%is 
%quite representative of the full sample.
}
\end{figure*}

\begin{figure*}
\begin{center}
\begin{tabular}{c}
\plottwo{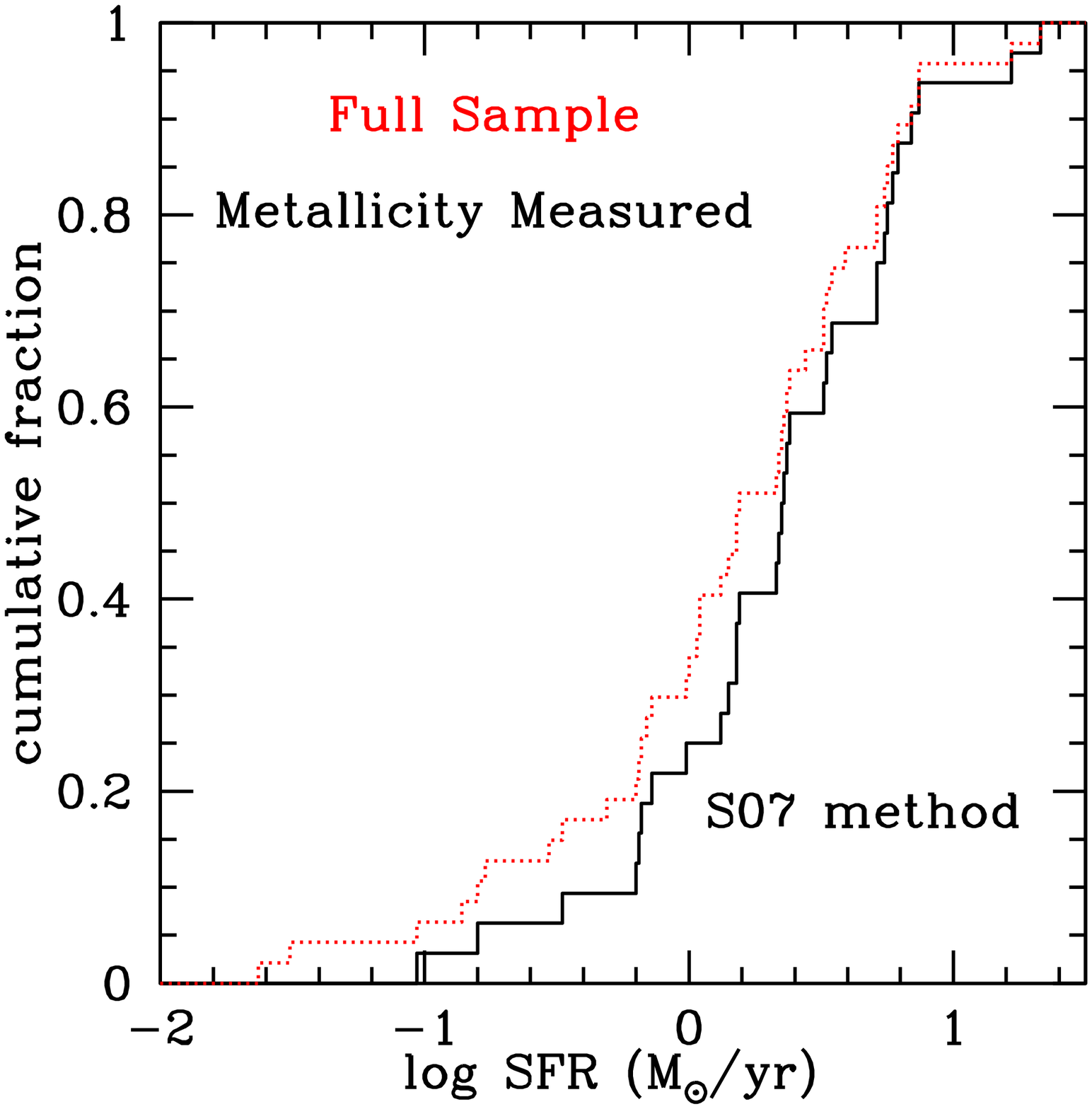}{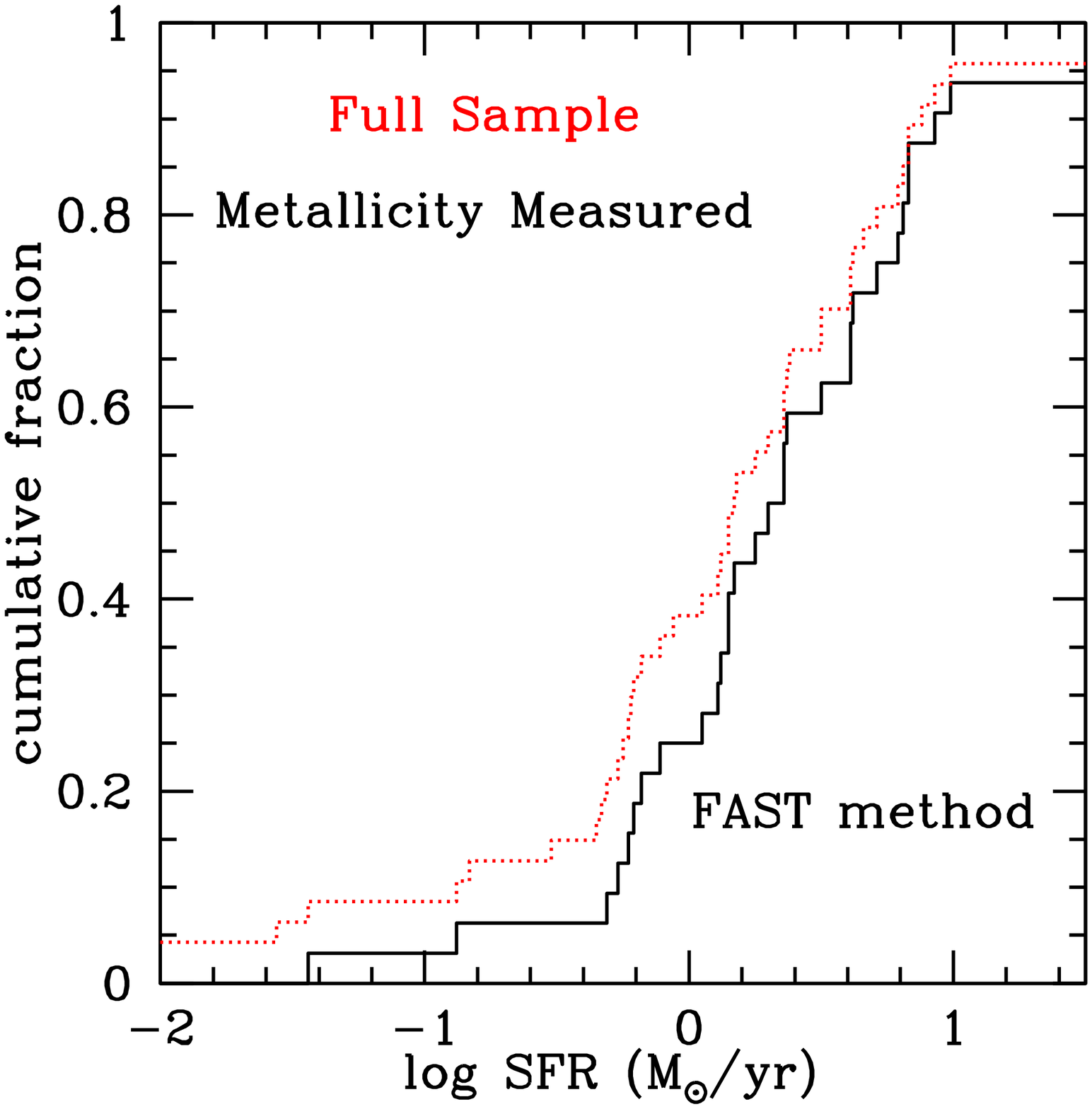}
\end{tabular}
\end{center}
\caption
{\label{fig:repsfr}
%We plot SFR from $u$-band photometry 
The subsample of type~II hosts for which we have measured metallicities 
(solid black) is representative of the full sample of PTF type~II SN hosts
(dotted red) in SFR 
%distribution 
based on $u$-band photometry 
\citep[][left]{salim07} and FAST (right).
}
\end{figure*}

To investigate how representative the subsample for which we have spectra 
and metallicities is of the entire sample of PTF type~II SN hosts 
we compared the distributions of the two samples in galaxy mass, 
characteristic stellar age, and star formation rate, using the 32 (47) 
hosts with (and without) measured metallicities that also lie in the SDSS 
DR8 imaging footprint.  Figures~\ref{fig:representative} and \ref{fig:repsfr} 
show that the two sub-samples have essentially identical distributions in host 
mass (K-S probability 99.9\%), age (57.8\%), and 
SFR (64--74\%, depending on the SFR estimation method).

%----------
%----------

%----------
%----------

\begin{figure*}
\begin{center}
\begin{tabular}{c}
\plottwo{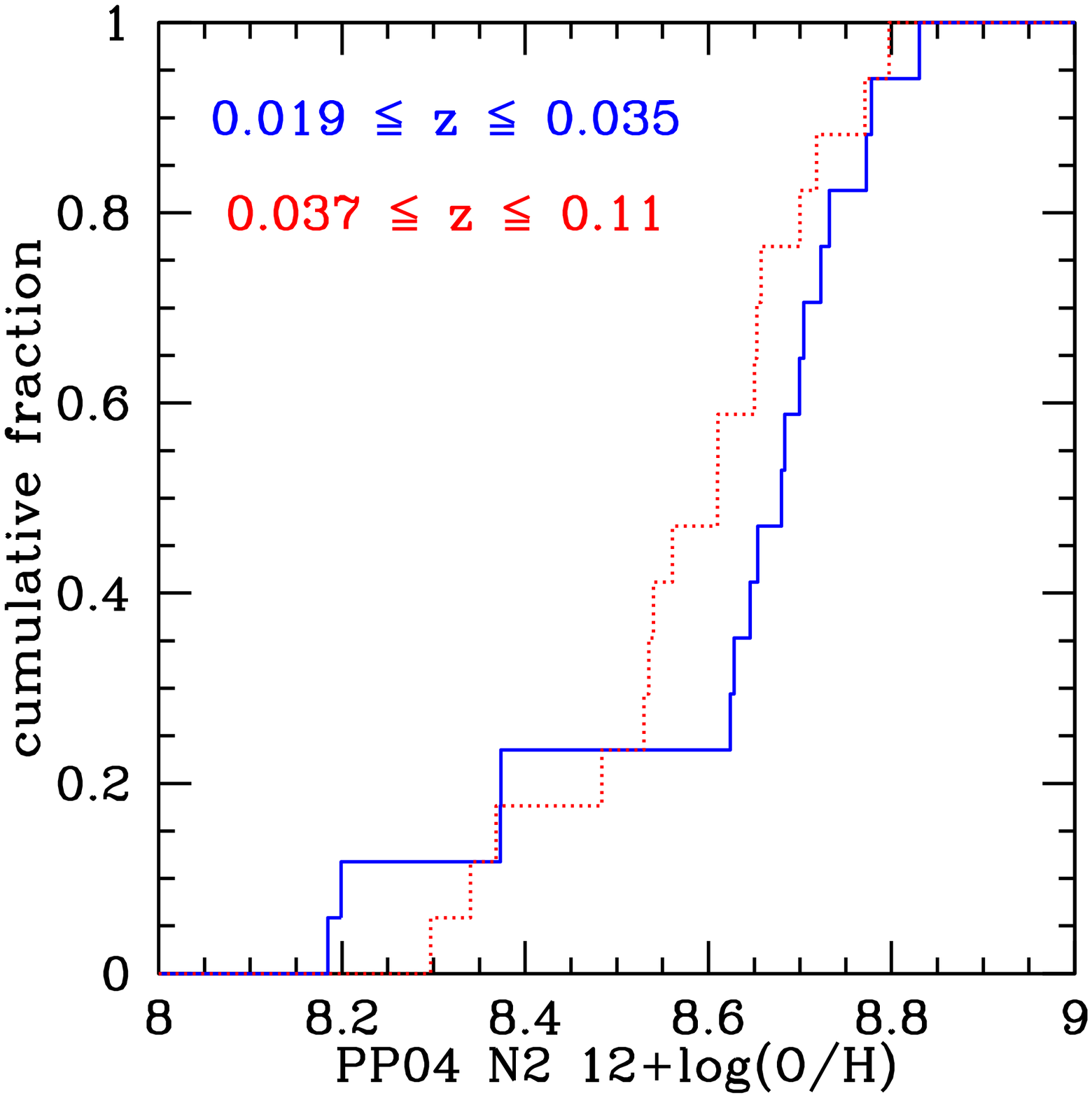}{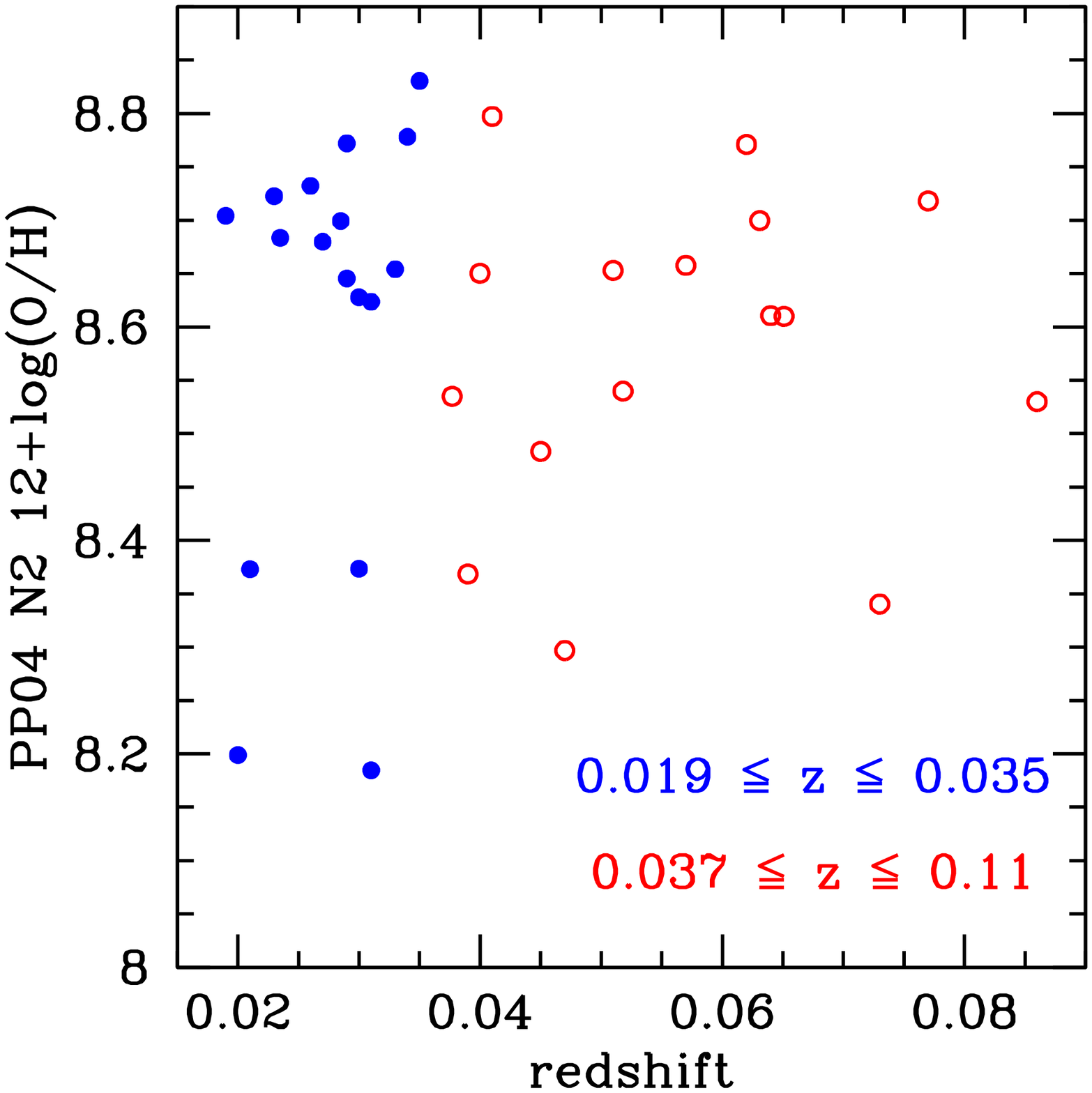}
\end{tabular}
\end{center}
\caption
{\label{fig:PTFzsplit}
The metallicity distribution of the type~II hosts separated into two bins 
in redshift of seventeen hosts each (left).  The green solid (red dotted) line 
is the lower (higher) redshift half of the sample, $0.019 \leq z \leq 0.035$ 
($0.037 \leq z \leq 0.11$).  
%The red dotted 
%line is the higher-redshift half of the sample, $0.037 \leq z \leq 0.11$.
On the right, the distribution is expanded and metallicities are plotted 
against redshift.  Here the lower 
redshift half of the sample is green solid points, and the higher redshift 
is red open points.
A K-S test shows the two are consistent with being drawn from 
the same distribution in metallicity at the 19\% level.  There do not appear 
to be any major selection effects with redshift.
}
\end{figure*}

We also investigated the effects of redshift on the completeness of the 
sample by dividing it into lower and higher redshift subsamples and comparing 
the properties of 
the two.  In Figure~\ref{fig:PTFzsplit} we show the metallicity distribution 
of these two subsamples.  A K-S test indicates that the two have a 19\% 
probability of being drawn from the same metallicity distribution, consistent 
at approximately $1 \sigma$.  The hosts in the two redshift bins have 
essentially identical distributions in host mass (K-S probability 63\%), 
age (91\%), and SFR (91--99\%, depending on the estimation method).

%----------
%----------

\subsection{The type II metallicity sample in context}\label{sec:context}

% 5th paragraph, page 5?
% point 17
%
% Each cut isolated (as far as possible when not derivative):
% Full sample:     927552  
% no SFR flag:     775563 (header)
% SFRavg != -99.:  839426 (header)
% Mmed != -1.:     778804 (header)
% Mmed != INDEF:   927497 (header)
% OHmed != -99.9:  203628 (header)
% Mmed > 2 (and not INDEF):  778695 (header)  (56 between -1 and 2, apparently)
% OHmed != -99.9 and -2.5 < N2 < -0.3:  202399 (header) (just losing 1229.)
% zrange:  > 0.0189995 and < 0.103:  382096
% All cuts/zrange: 130371  (130370, header)
%
% sequential: 
%  zrange: 382096    -58.8%
%  + M 'successful' (Mmed > 2 and not INDEF): 363882   -4.8%
%  + SFR 'successful' (no SFR flag and SFRavg != -99): 362396    -0.40%
%  + Tremonti OH (OHmed != -99.9): 131204   -63.8%
%  + -2.5 < N2 < -0.3: 130371     -0.63%
%
We next place our type~II host sample in context by comparing it to galaxy 
properties in the DR7 SDSS MPA/JHU value-added catalog 
\citep{kauffmann03,tremonti04,brinchmann04,salim07}.  We compare to the 
subset of the DR7 objects which have redshifts within the range of our 
sample, successful estimates of the stellar mass and star formation rate, 
a 12+log(O/H) metallicity estimate, 
and an [\ion{N}{2}]$\lambda 6584$/H$\alpha \lambda 6563$ flux ratio within 
the valid range for the PP04 N2 metallicity diagnostic.  This last 
requirement has almost no effect, reducing the sample by only 0.6\%.  The 
strictest condition by far is the requirement of a valid \citet{tremonti04} 
metallicity estimate, as shown in Table~\ref{table:cuts}.  
Note that the galaxy properties in the MPA/JHU value-added catalog are a 
function of redshift because is not a volume-limited sample.  The selection 
that defines the catalog is reflected in the properties of its constituents.

\begin{figure*}
\begin{center}
\begin{tabular}{c}
\plottwo{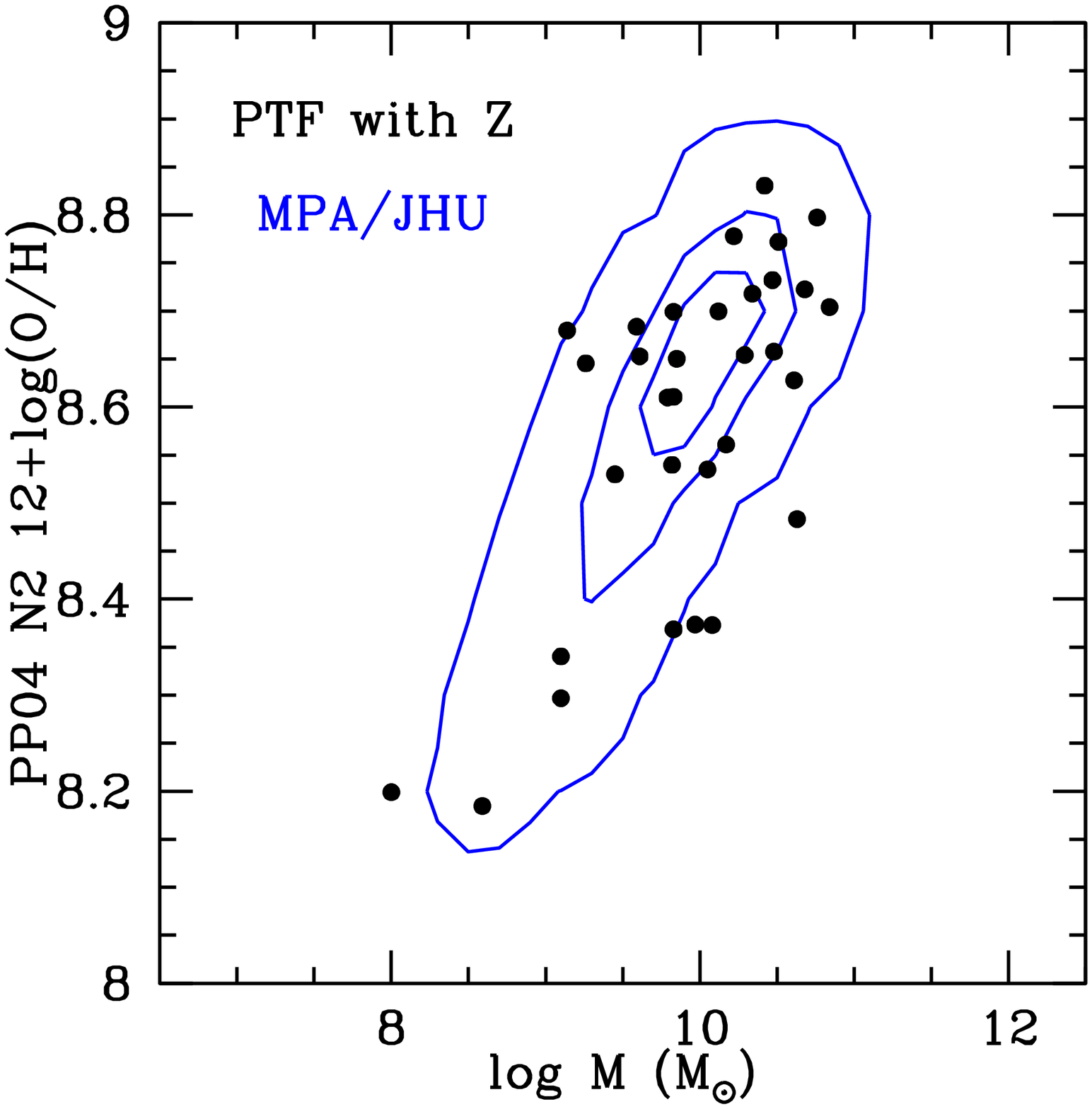}{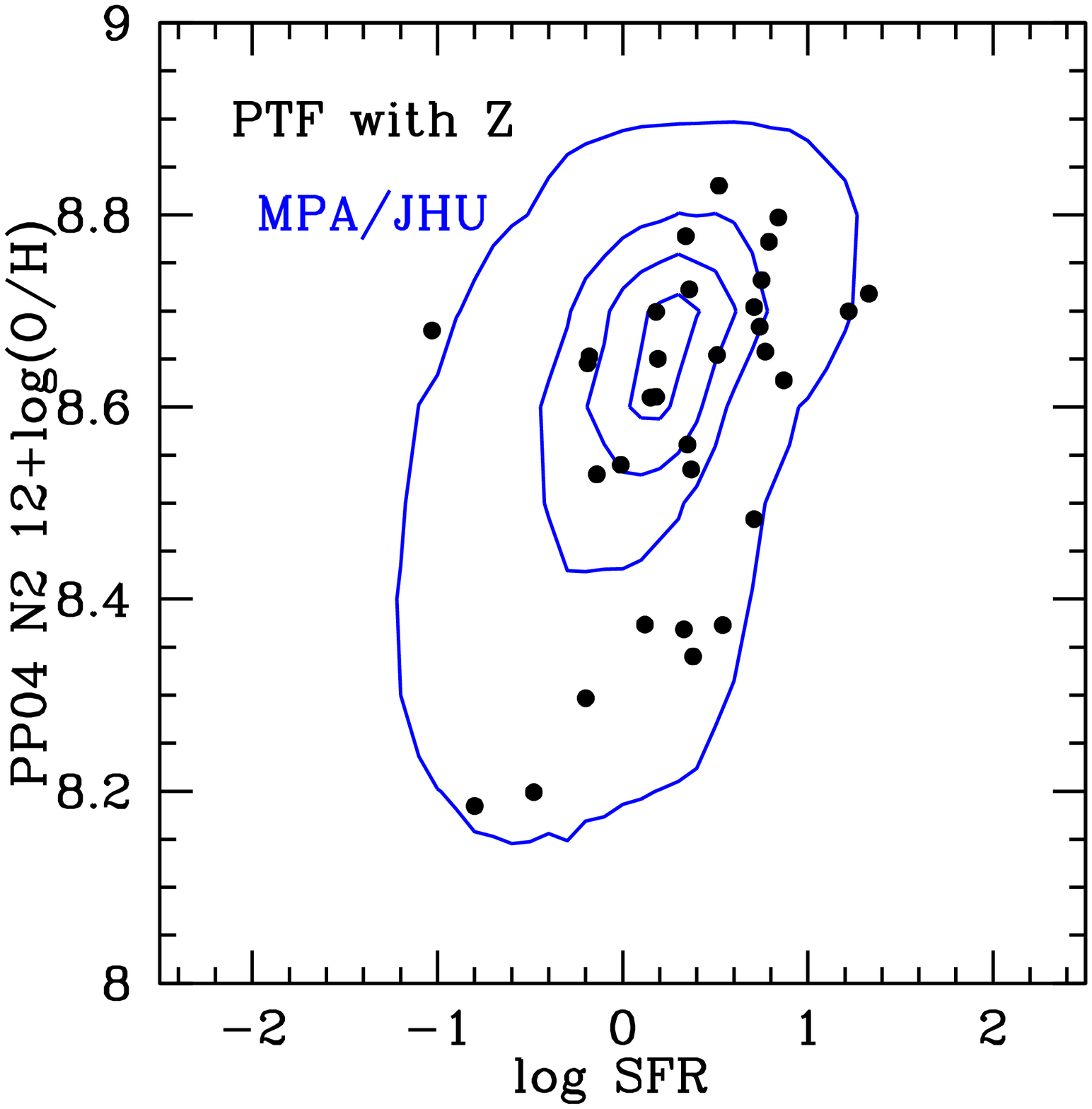}
\end{tabular}
\end{center}
\caption
{\label{fig:contextOHvs}
Measured host metallicities (black points) as a function of 
galaxy mass (left) or SFR (right), overlaid on the distribution in the MPA/JHU 
value-added catalog in the same redshift range (blue contours).  
Our hosts are slightly offset to higher star formation rates, as we would expect 
if type~II SNe trace star formation.
%Our host metallicity vs galaxy mass and SFR (points in black) within the 
%context of a  subset of the MPA/JHU value-added catalog selected to have
%valid SFR, stellar mass, T04 metallicity, and PP04N2 metallicity, and to be
%within the same redshift range as our sample, as described in the text 
%(contours in blue).  Our hosts are slightly offset to higher star formation 
%rates, as we would expect if type~II SNe indeed trace star formation.
}
\end{figure*}

% Apparently 6th paragraph, page 5
% point 18
As shown in Figure~\ref{fig:contextOHvs}, the SN hosts appear to trace the 
MPA/JHU sample well.  Their mass-metallicity relationships are consistent at 
approximately one sigma.  
Strong conclusions should not be drawn from this, as the sample selection 
functions of the PTF type~II host galaxies and the SDSS galaxy 
mass-metallicity sample of the MPA/JHU value-added catalog are each complex 
and based in part on parameters unrelated to the galaxies themselves.  The 
fiber allocation in SDSS prioritized galaxies lower than other target classes, 
including brown dwarfs and quasar candidates \citep[e.g.][]{blanton03}, 
meaning that the selection function of galaxies observed spectroscopically in 
a given field is mediated by the density of other targets in that field rather 
than a simple function of the properties of the galaxies themselves.  To 
zeroth order, SDSS galaxy spectroscopic observations are a function of stellar 
mass, while core-collapse supernovae (and by extension our host sample) are 
instead a function of star formation rate in the recent past.  Inclusion in 
the mass-metallicity sample of the MPA/JHU value-added catalog is not a simple 
function of the galaxy having been spectroscopically observed.  In order to be 
included, the galaxy must have a reasonable fit using the group's Bayesian 
metallicity determination.  The strength of the spectroscopic features that 
enable this determination is greater for galaxies with a higher star formation 
rate, which means that the star formation rate is to a certain extent a hidden 
parameter in this selection function.  
%This function, however, is impossible 
%to forward-model.  Backwards-modeling the selection would be an interesting 
%important study, but is beyond the scope of this paper.  
The PTF followup 
selection function is unpublished, and therefore impossible to model.  
Reconciling the selection functions of these samples sufficiently to draw 
strong conclusions based on comparing their mass-metallicity slopes is not 
possible, nor is it the aim of this study.  Instead, we compare them to place 
our results in context.
%

%----------
%----------

The SDSS galaxy spectra are primarily nuclear spectra but not exclusively so.  
%In contrast, 46\%  (20) of our host spectra are more than 3 arcsec 
%away from the nucleus, as can be seen in Table~\ref{table:AllZ}.  
%Of the 54\% (20) of our sample that are 
In contrast, 50% (17/34) of our type~II host metallicity measurements are 
of locations more than 3 arcsec 
away from the nucleus, as can be seen  in Table~\ref{table:AllZ}.  The SN 
location was within 3 (4,6) arcsec of the galaxy center in all but 
5 (3,1) of the remaining cases.  
In one case (PTF09aux), we were unable to measure [\ion{N}{2}] and H$\alpha$
with sufficient S/N to measure metallicity.  
Type~II SNe have no known metallicity 
dependence, so they are not expected to favor the outer, more metal-poor 
regions of galaxies.

\begin{figure}
\begin{center}
\begin{tabular}{c}
\includegraphics[width=8cm]{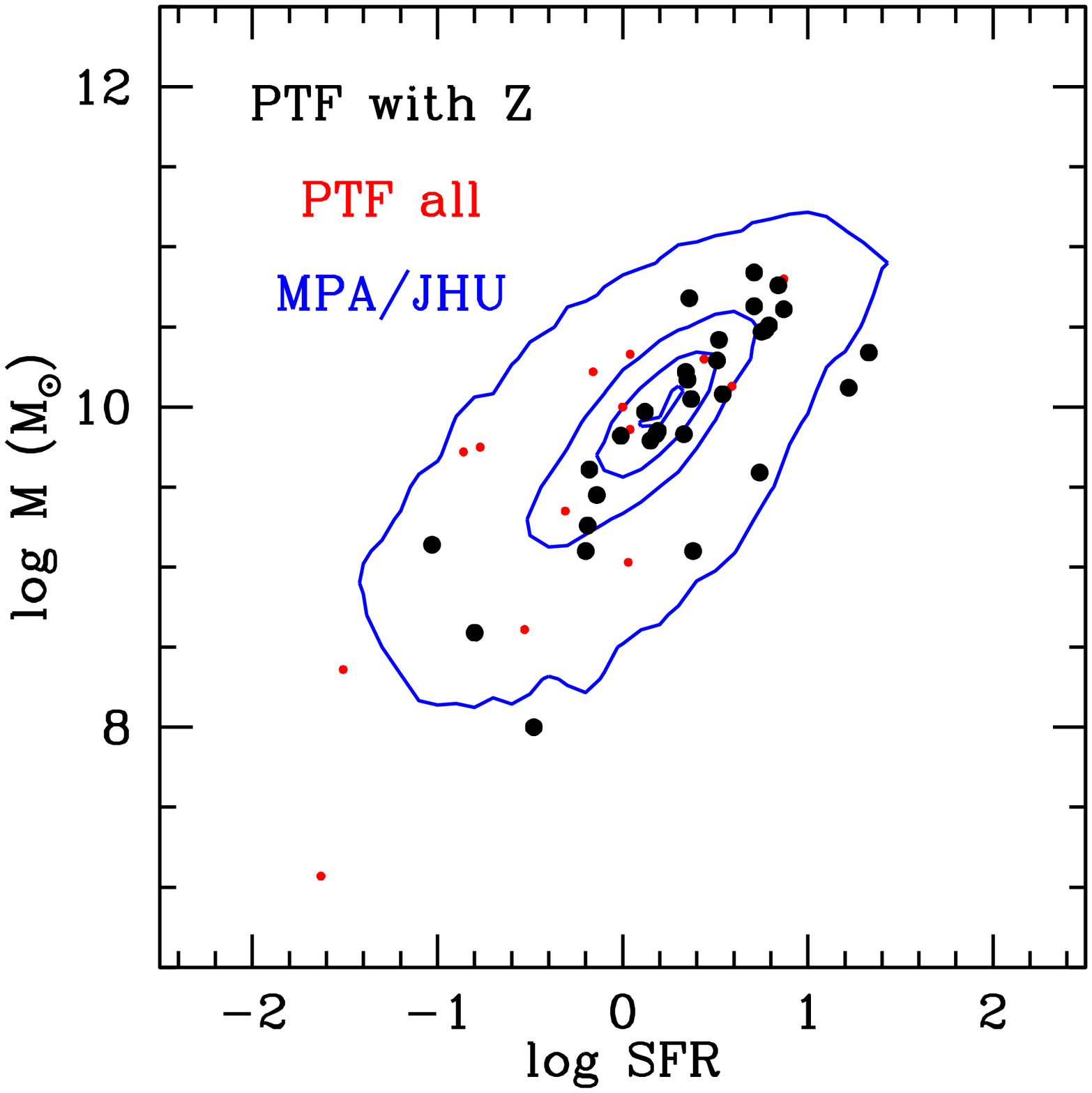}
\end{tabular}
\end{center}
\caption
{\label{fig:contextMvsSFR}
Host mass and SFR for our type~II SN hosts with Z measurements (large black 
points), all first-year PTF type~II SNe hosts in the SDSS photometric footprint
(small red points), overlaid on the distribution in the MPA/JHU catalog 
(blue contours).
%Our host mass vs SFR (large points in black) compared with the full SDSS 
%overlap of the PTF first-year type~II SNe (small points in red) and within 
%the context of the subsample of the MPA/JHU catalog (contours in blue)
}
\end{figure}

Although quantitative conclusions should not be drawn from this similarity 
due to these unavoidable sample selection incongruities, some qualitative 
conclusions can be drawn.  
The type~II SN hosts do not appear to be biased toward lower metallicities at 
a given mass (left panel) as a simplistic interpretation of 
the results of \citet{mannucci10} might suggest.  It has been suggested that 
core-collapse SNe and GRBs may be less frequently observed in higher 
metallicity environments than in lower metallicity environments due to higher 
extinction \citep[e.g.][]{maiolino02,mannucci03,cresci07,campisi11}, but this 
sample does not show evidence for such a bias.  
The SN hosts do, however, appear to be biased toward higher star formation 
rates\footnote{Here we use the SFR calculated from the $u$-band luminosity 
with aperture corrections.}
(right panel) than the MPA/JHU 
galaxy sample, as would be expected if they do indeed trace star formation.
They appear to trace the distribution of the 
MPA/JHU sample well in galaxy mass and star formation rate, as shown in 
Figure~\ref{fig:contextMvsSFR}.

%----------

%----------

\subsection{Comparing with other SNe host samples}\label{sec:comp}

% Is this 1st paragraph, page 6?
% point 19
Work on the metallicity distribution of supernova hosts by 
\citet{prieto08z} and \citet{kelly12} looked at overall galaxy metallicity 
with serendipitous SDSS spectra, without isolating the SN site.  
\citet{kelly12} subdivide their sample into those discovered by 
galaxy-impartial searches and by targeted searches.
Studies by \citet{anderson10} tried to measure abundances at the SN site 
or at a similar galactocentric radius, as we have done in this study.  
These previous studies had uniform spectroscopy 
but the source SN samples were heterogeneous, including SNe discovered in a 
wide variety of ways with very different selection effects.
Our source SN sample is from a single survey with uniform 
selection.  The source survey is areal rather than galaxy-targeted, which 
enables the detection of events in the lowest-mass galaxies, removing or at 
least mitigating a possible bias towards high metallicity environments which 
we expect exists in prior supernova host samples.  We do not attempt to 
define a volume-limited supernova sample here.  Rather, we point out that 
this existing sample represents a substantial step forward from previous 
similar samples for the purpose of providing a comparison for a non-matched 
sample of rare core-collapse events, particularly one thought to depend on 
metallicity such as GRBs or abnormally luminous CCSNe.

% 1st paragraph, page 7?  'turns over'
% point 20
% '2nd paragraph, page 8' but not consistent with 3rd, point 22.
% point 21 

\citet{kelly12} used abundances following \citet{tremonti04} and the 
O3N2 method of \citet{pp04}.  In this paper we use the N2 method of 
\citet{pp04}, so we must convert these to a common scale.  The valid range of 
the conversion defined by \citet{kewley08} from the scale of the O3N2 method 
to the scale of the N2 method does not span the abundances here, so we omit 
three of the 124 hosts in the sample, and convert from O3N2.  
\citet{prieto08z} uses abundances from the method of \citet{tremonti04}, 
we omit the 14 of 152 hosts that have T04 metallicities above or below the 
valid conversion range defined by \citet{kewley08}.

%The conversion defined by \citet{kewley08} from the scale of 
%\citet{tremonti04}
%to the N2 method of \citet{pp04} is non-monotonic at low metallicities, and 
%does not match the reverse conversion.  To ensure monotonicity, we define an 
%ad hoc 
%conversion by fitting to the reverse conversion defined by \citet{kewley08}
%(see Appendix).  This conversion is a fit to a fit rather than a fit to data, 
%but it achieves two purposes:  avoiding a turnover at low metallicity and 
%matching the inverse conversion from PP04N2 to T04.  We use this conversion 
%everywhere in this paper when converting from 
%the scale of the T04 diagnostic to the scale of the PP04N2 method.
%Approximately 9\% of the \citet{prieto08z} sample have T04 metallicities 
%above or below the valid conversion range defined by \citet{kewley08}.  
%Because our conversion is well-behaved at low metallicity, we convert the 
%entire sample instead of excluding that 9\%.  The decision to include these 
%hosts with the lowest and highest metallicity is not highly consequential; 
%the distribution of the full sample with our conversion is consistent (at a 
%K-S probability of 66\%) with the remaining 91\% converted using the method 
%of \citet{kewley08}.

\begin{figure}
\begin{center}
\begin{tabular}{c}
\includegraphics[width=8cm]{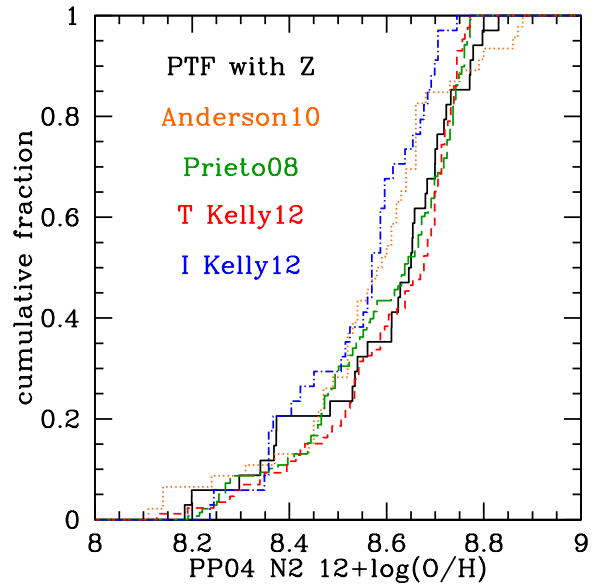}
\end{tabular}
\end{center}
\caption
{\label{fig:SNeCompII}
Distribution of local gas-phase oxygen abundance of our sample of type~II SNe 
(black solid line) compared with existing SN host metallicity samples.   
The dotted orange curve shows the heterogeneous type~II subsample of 
\citet{anderson10}.  The K-S probability of 
this sample being drawn from the same underlying distribution as our PTF 
type~II sample is 11\%. 
Similarly, the green long-dash curve shows the distribution of type~II, IIP, and 
IIn SNe from \citet{prieto08z} (K-S probability 87\%) and the purple dash-dot 
curve shows the type~II sample from \citet{kelly12} (K-S probability 45\%).  
In the latter two cases we converted their T04 scale metallicity to the 
PP04N2 scale using the conversion given in the Appendix.
%In green (long-dash), the \citet{prieto08z} type~II, IIP, and IIn sample with 
%metallicities converted to PP04N2 from T04 using our ad hoc fit to the
%reverse conversion; see the appendix for details (K-S 87\%).
%In purple (dash-dot), the \citet{kelly12} type~II sample with metallicities 
%converted to PP04N2 from T04 using our ad hoc fit (K-S 45\%).
}
\end{figure}

% This must be 2nd paragraph, page 8, 
% but point 21 appears to refer to the above paragraph
We compare our type~II host metallicity distribution with those found by 
\citet{prieto08z}, \citet{kelly12} targeted (T) and galaxy-impartial (I), 
and \citet{anderson10} in Figure~\ref{fig:SNeCompII}.  The K-S test 
probabilities that our type~II sample could be selected from the same 
underlying distribution as those in the earlier studies are 
43\%, 68\%, 4\%, and 21\%, respectively.
The agreement of our SN metallicity distribution with the results of these 
prior studies is striking, given the different sample selection.  The most 
different sample (at a K-S probability of only 4\%) is the one which would 
seem to be the most similar; the heterogeneous galaxy-impartial sample from 
\citet{kelly12}.

%----------
%----------

\subsection{Comparing with SFR metallicity distributions from galaxy population statistics}\label{sec:theodist}

%Gnedin, Niino

% Is this 3rd paragraph page 8?
% point 22
One semi-observational way of determining the global distribution of star 
formation as a function of metallicity is to combine the observed galaxy mass 
function, the observed mean star formation rate as a function of galaxy mass, 
and the observed galaxy mass-metallicity relationship.  
\citet{stanek06} combines the 2MASS and SDSS galaxy stellar mass function 
from \citep{bell03} with the \citet{tremonti04} mass-metallicity relation 
and the \citet{brinchmann04} star formation rate density.
\citet{niino11} calculates the metallicity distribution of star formation
two different ways; first, observationally from the stellar mass
function, the galaxy M-SFR relation, and the galaxy mass-metallicity
relation, nearly identically to \citet{stanek06} except using the 
mass-metallicity relation of \citet{savaglio05}, and second, calculated using 
the relationship between galaxy mass,
metallicity, and SFR defined by \citet{mannucci10}, and assigning 
metallicities based on SFR as well as mass.
Each of the three observed relationships that are inputs to these alternate 
methods of finding the metallicity distribution of star formation has its 
own set of selection effects.  Potential biases or redshift-dependent 
effects in the samples used to define the relation could offset the 
distribution in metallicity.   The width of the distribution may be 
misrepresented by using a mean relationship to translate from one property 
into another, such as the 
mass-metallicity relationship \citep{tremonti04} or the 
three-way relationship between mass, metallicity, and star formation rate 
\citep{laralopez10,mannucci10}, because the scatter around that relationship 
may not be carried through to the final distribution.  

\begin{figure*}
\begin{center}
\begin{tabular}{c}
\plottwo{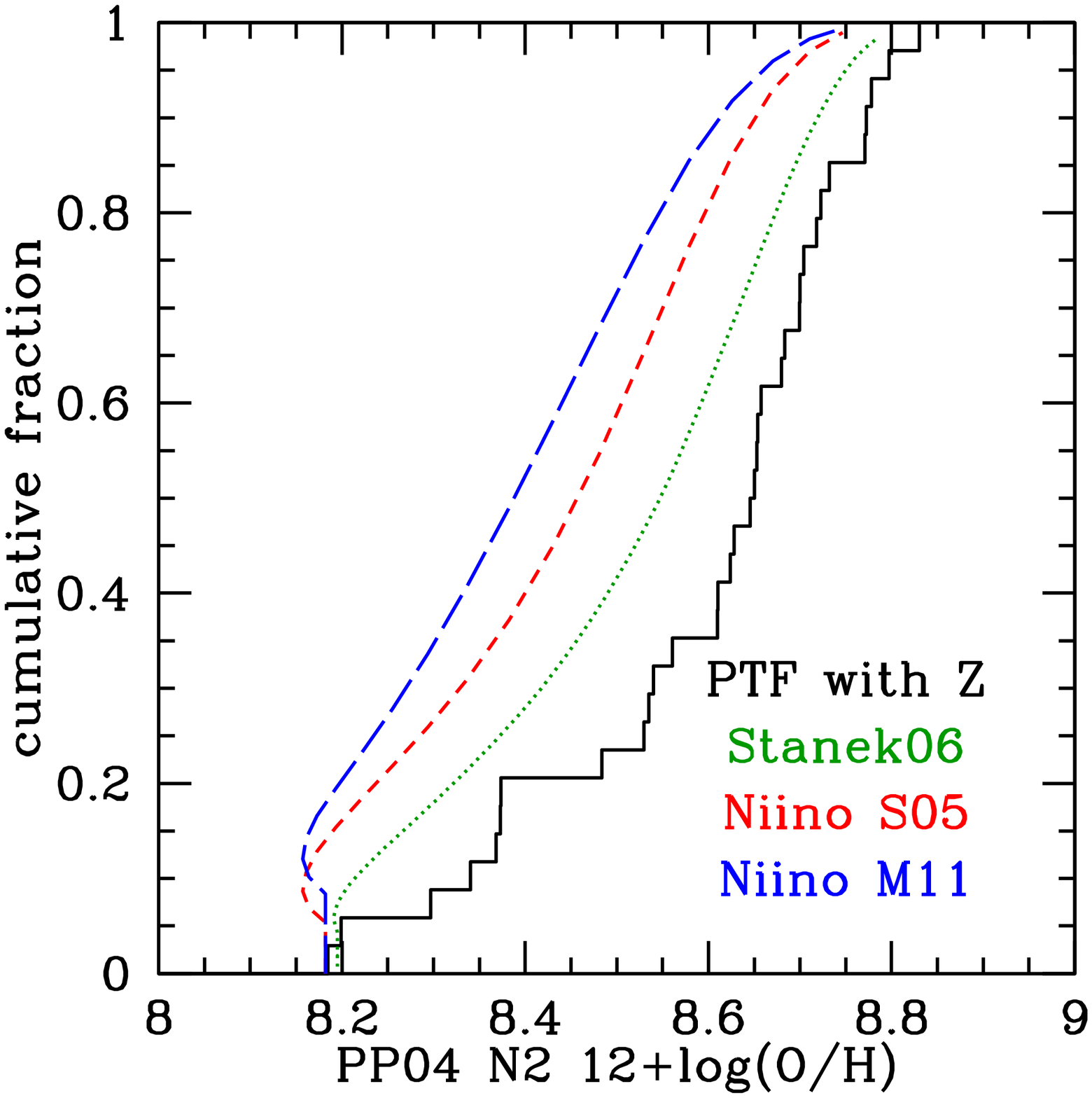}{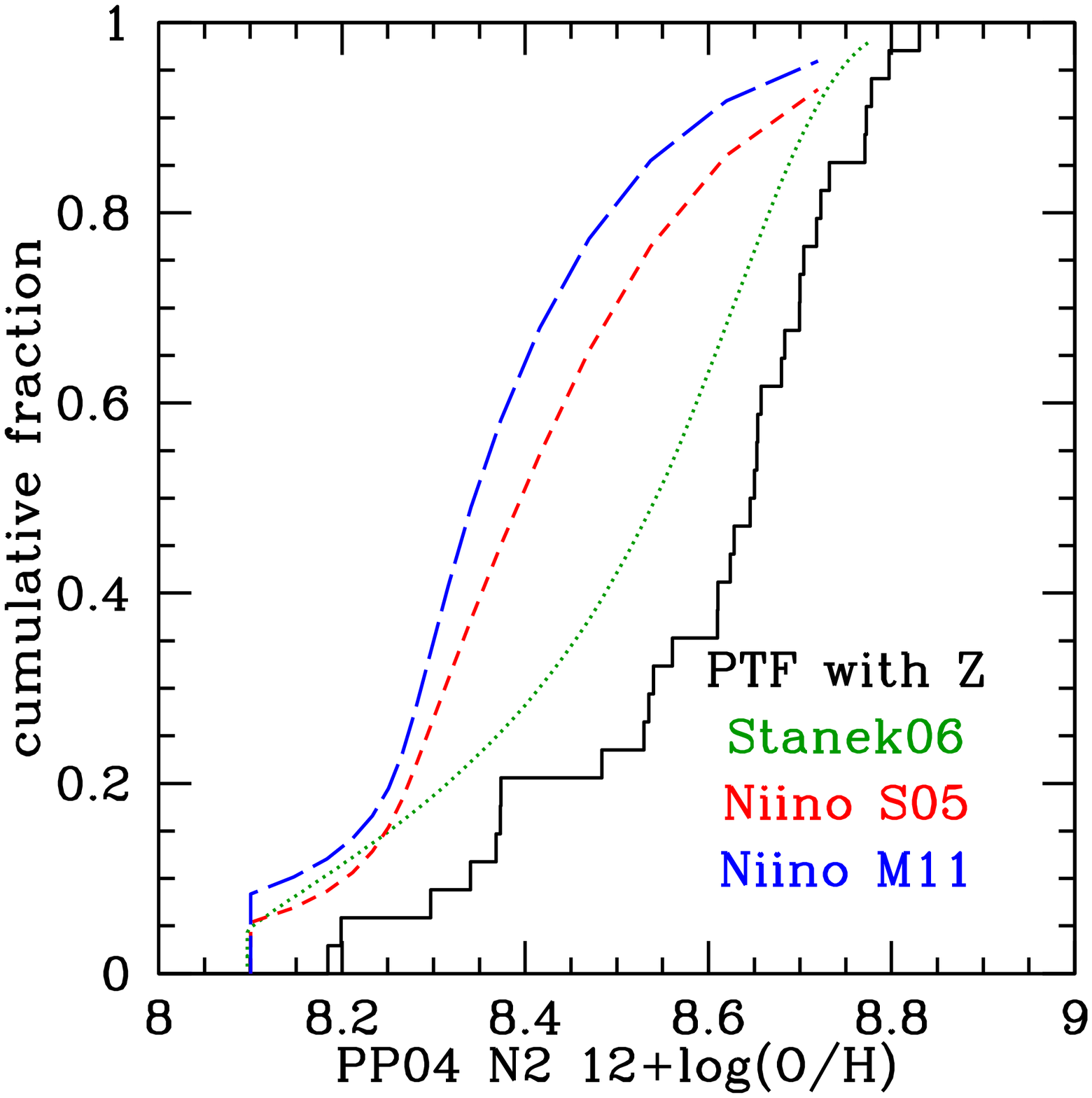}
\end{tabular}
\end{center}
\caption
{\label{fig:GalComp}
%Comparing 
The distribution of star formation as a function of metallicity from 
galaxy population statistics from \citet{stanek06}  (green dotted line) 
and \citet{niino11} with (blue long-dashed line) and without (red short-dashed
line) incorporating the star-formation-rate dependence of the mass-metallicity 
relationship as \citet{mannucci10}, compared to our distribution of site 
metallicities of type~II CCSNe (black solid line).  On the left, the 
metallicity conversions were made with \citet{kewley08}.  The non-monotonicity 
of the conversions from T04 and KK04 to PP04N2 results in the non-physical 
double-values.  On the right, conversions are done with by inverting the 
reverse conversions, as described in Appendix.  This maintains the 
physicality of the distribution function, but may increase inaccuracy at 
higher metallicities.
}
\end{figure*}

In Figure~\ref{fig:GalComp} we plot our distribution against 
those of \citet{stanek06} and \citet{niino11}  
based on galaxy population statistics.  We convert the metallicities using  
\citet{kewley08}.  The non-monotonicity of the conversions from T04 and KK04 
to PP04N2 results in non-physical double-values in the cumulative 
distributions.  On the right, metallicity conversions are done with by 
inverting the reverse conversions, as described in Appendix.  This maintains 
the physicality of the distribution function, but may increase inaccuracy at 
higher metallicities.  Because the forward and reverse conversions do not 
match, multiple conversions between metallicity scales will compound errors.  
This is likely the reason why the distribution from \citet{stanek06} (in green)
does not match the distribution from \citet{niino11} which considers the 
mass-metallicity relation independently of star-formation (in red).  Instead 
of using \citet{tremonti04}, \citet{niino11} uses the mass-metallicity 
relation of \citet{savaglio05} on the KK04 scale, though they 
are otherwise identical.

%----------
%----------

% apparently 4th paragraph page 8
% point 23
Core-collapse SNe, as the deaths of massive, young stars, are a relatively 
good tracer of SFR.  
Using the metallicity distribution of a uniform sample of type~II SN sites 
to approximate the metallicity distribution of star formation, as we have done,
should have almost completely independent selection effects (such as 
extinction from dust) from methods relying on galaxy population statistics. 

\subsection{Selection effects}\label{sec:sel}

One of the primary potential sources of incompleteness in using type~II SNe 
environments to trace the metallicity distribution of global star formation 
will of course be the selection of the sample of type~II SNe.  An ideal sample 
for this purpose would be a complete, volume-limited sample, monitoring a 
fixed region of sky for a fixed period of time, and then eliminating events 
outside the complete sample.  Up until very recently, most supernova surveys 
have monitored large, luminous galaxies rather than regions of the sky, a 
methodology which has the potential to miss any SNe in the very lowest end 
of the galaxy luminosity function.  The Palomar Transient Factory survey 
is areal, which removes the potential bias against extremely low-mass host 
galaxies of targeted surveys.  
Because the survey selection is not yet 
published, however, we are unable to correct for any 
%Malmquist-like 
biases in 
our source sample to get a distribution for a volume-limited sample.  With 
the current rapid growth in depth and breadth of supernova surveys, there is 
great potential for this method of determining the metallicity distribution 
of star formation.

%If there is any correlation between type~II SN peak magnitude and host 
%galaxy metallicity, areal surveys may be subject to Malmquist-like biases 
%due to a metallicity dependence in the SN luminosity.  

% LOSS survey:  (complete) galaxy sample down to mag? and out to 60 Mpc.
% total in-survey light, 1.5 x 10^14 Lsun, total est 4.7x10^14 Lsun (all-sky)
% Galaxy sample not complete at low-luminosity end.  

\begin{figure*}
\begin{center}
\begin{tabular}{c}
\plottwo{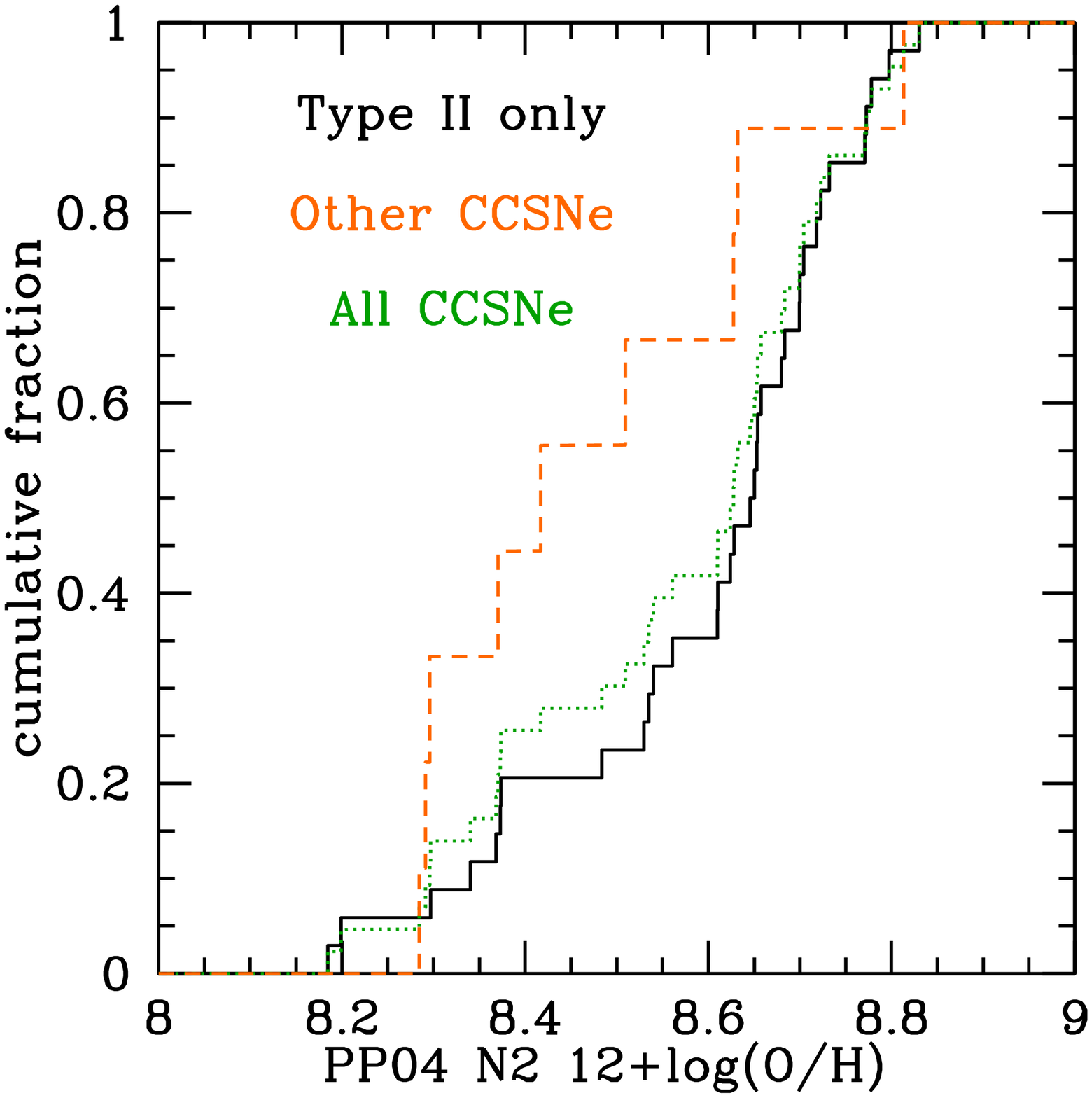}{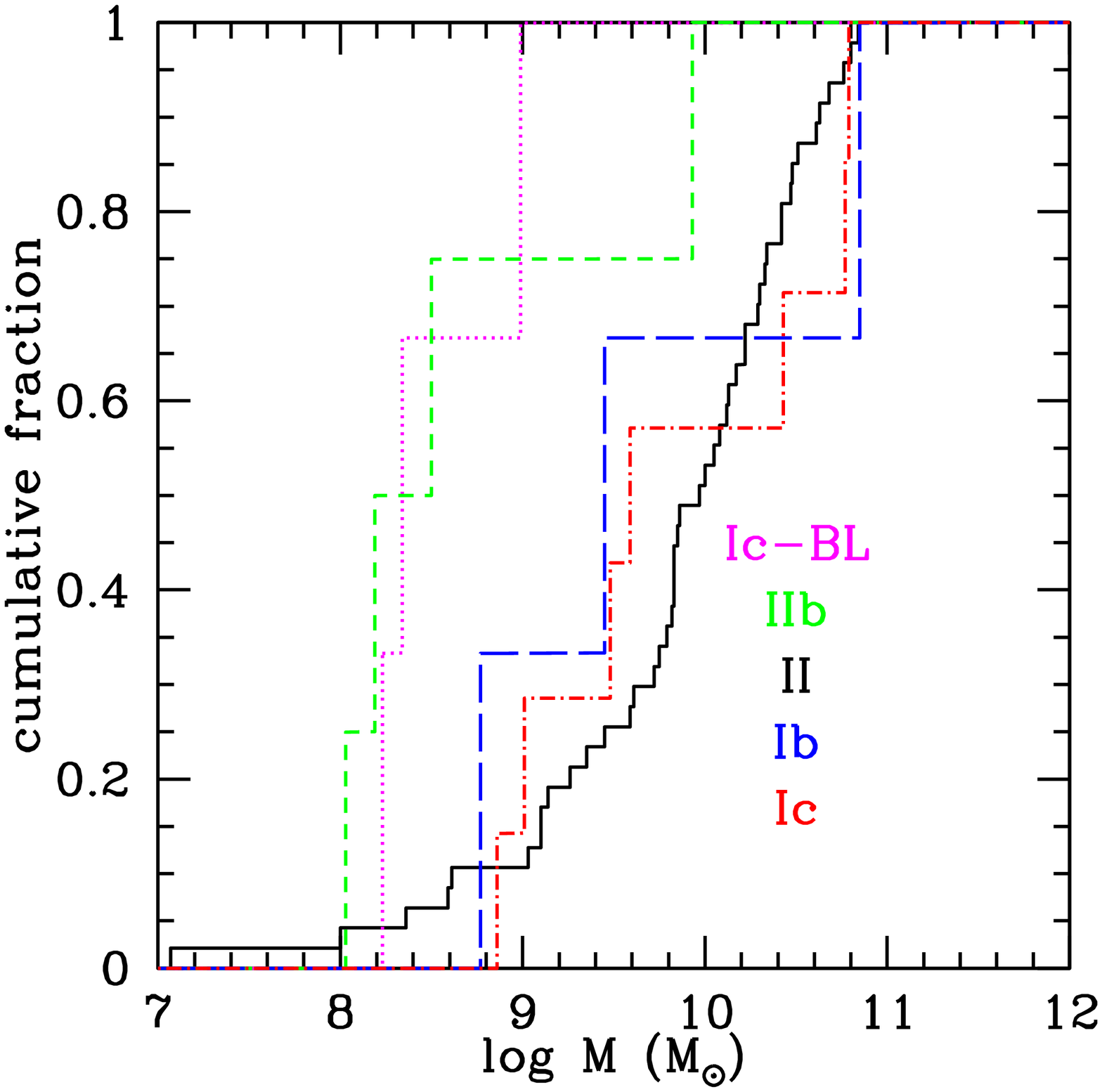}
\end{tabular}
\end{center}
\caption
{\label{fig:subtypes}
The metallicity distribution (left) of type~II SN hosts (solid black), 
all other CCSN hosts (dashed orange) and all CCSN hosts in 
Table~\ref{table:AllZ} (dotted green).  With this small sample of other 
subtypes, the type~II distribution is statistically consistent with the  
other CCSNe (K-S probability 23\%) and with all CCSNe (K-S probability 99\%).
Increasing the sample size, the distribution in photometrically calculated 
host galaxy mass (right) of type~II SN hosts (black solid), 
Ic-BL (pink dotted), IIb (green short-dashed), Ib (blue long-dashed), 
and Ic (red dot-dashed).  These distributions are consistent with the 
overrepresentation of IIb and Ic-BL at low host luminosity found by 
\citet{arcavi10} and with the distributions in host mass found by 
\citet{kelly12}.
}
\end{figure*}

% is this 6th paragraph page 8?
% point 24
% This is definitely 2nd paragraph, page 9
% point 25
% 25.  I think that you mean that 'the relative frequency of type II SN 
% has not yet been found to depend on metallicity, ... '
Another potential source of bias for this method is any dependence of the 
likelihood of a massive star resulting in a type~II SN on metallicity.  
Type~II-P SNe make up around 70\% of all type~II SNe in the 
LOSS survey, which focuses on relatively luminous galaxies \citep{li11}.   
The relative frequency of type~II SNe has not yet been found to depend on 
metallicity, unlike type~Ib and Ic SNe.  
%The relative rate of type~II to type~Ib/c SNe in the PTF first-year CCSN 
%sample \citep{arcavi10} is substantially different from that found by LOSS 
%\citep{li11}.  
We examine only the distribution of type~II hosts, for which 
we have good statistics.  The type~II distribution is consistent with the 
distribution of the nine type Ib/Ic/IIb hosts we have measured with a K-S 
probability of 23\%, and the distribution of the hosts of all CCSNe we have 
measured (including all the type~II hosts) is consistent with the type~II 
distribution with a K-S probability of 99\%, shown in 
Figure~\ref{fig:subtypes} (left).
Completeness corrections to translate between 
type~II hosts and all CCSN hosts may depend slightly on metallicity.

%----------
%----------

% This is apparently ``3rd paragraph, page 9''
% point 26
% dealt with
Several types of CCSNe are known to vary in frequency with metallicity, 
as discussed in \textsection\ref{sec:intro}.
Our very small sample of spectroscopic metallicity measurements of 
hosts of type IIb/Ib/Ic SNe is statistically consistent with these previous 
results \citep{modjaz11,anderson10,kelly12}, but not at very high significance 
level.  Using the host galaxy properties we fit from SDSS photometry gives 
us slightly better statistics, shown in Figure~\ref{fig:subtypes} (right).  
The overrepresentation of types IIb and Ic-BL in low-mass 
hosts is consistent with the results of \citet{arcavi10} of this same sample 
based on host galaxy luminosities, but again, not at very high significance.
The relative distributions of type~II, IIb, Ib, Ic, and Ic-BL in 
photometrically calculated host mass are consistent with the recent results of 
\citet{kelly12}.

\subsection{Strong-line metallicity diagnostics}\label{sec:Zdiagnostics}

There is a substantial literature on the merits and disadvantages of each 
commonly-used method of determining metallicities based on fluxes of strong 
emission lines \citep[e.g.][and references therein]{berg11,kewley08}.  
These methods all rely on simplifying assumptions about the \ion{H}{2} 
regions being examined: uniformity of electron density, cooling dominated 
by oxygen (implying that other cooling species have abundances that vary in 
lockstep with oxygen), and ionization-bounded \ion{H}{2} regions 
\citep[e.g.][]{pagel79}.  
The methods can be classified into rough categories: direct methods, which 
rely on estimates of the electron temperature and require measurements of 
faint auroral lines such as [\ion{O}{3}]$\lambda 4363$\AA{}, 
empirical \citep[e.g.][]{pp04}, theoretical \citep[e.g.][]{kk04}, and a 
combination of empirical and theoretical \citep[e.g.][]{d02}.  All these 
methods are based on high S/N measurements of the line ratios of different 
combinations of emission lines from ions present in the optical region of 
the spectrum ($\simeq 3700-6800$\AA:   
[\ion{O}{2}], [\ion{O}{3}], [\ion{N}{2}], [\ion{S}{2}], 
H$\alpha$, and H$\beta$).

% This is apparently ``5th paragraph, page 9''
% point 27
% dealt with
The slope and intercept of the galaxy mass-metallicity relationship is 
different for each diagnostic \citep{kewley08}, which is a relatively 
straightforward way to demonstrate that they are not all directly measuring 
some platonic ideal of a fundamental oxygen abundance measurement.  
%(If they were, they would all agree.)  
(This problem is independent of the separate question of the exact value of 
the solar oxygen abundance, which also affects how measured gas-phase 
abundances map to stellar abundances.)
Instead, each technique measures a different quantity that correlates well 
with oxygen abundance, but does not directly map to it.  The simplifying 
assumptions that allow us to use each strong-line indicator to estimate 
oxygen abundance are not perfect for all \ion{H}{2} regions.  
Rigorously selecting the best diagnostic for a given 
situation requires better data than are achievable for distant and faint 
targets, and doing a case-by-case selection of strong-line method on 
insufficient data would introduce its own biases.

% 6th paragraph, page 9  adding from above, but point 28 clearly refers 
% to the next paragraph, not this one.
The primary advantage of the strong-line techniques is that they are 
possible with fewer photons, and are therefore feasible to perform on  
large samples for good population statistics.  
Given the scale differences between methods, however, it is crucially 
important to ensure that all metallicities one is comparing are on the 
same scale.  Where possible, we do this by natively determining the 
metallicities in a common scale.  Where impossible, we convert a metallicity 
determined on another scale using the empirical conversions of 
\citet{kewley08} (but see Appendix~\ref{app:conversions}).
We emphasize that any meaningful 
%effect should be rigorous to a change of metallicity measurement.
conclusion should not be affected by a change in the metallicity calibration.

% This paragraph is apparently ``6th paragraph, page 9''.
% point 28
The primary scale we choose for this study is the N2 diagnostic of 
\citet{pp04}, which depends solely on 
[\ion{N}{2}]$\lambda 6584/$H$\alpha \lambda 6563$. 
Three disadvantages of the method are that it shows a larger dispersion 
compared to the direct ($T_e$) method  than most other strong line 
diagnostics, that it has a dependence on the ionization parameter that 
becomes important at low metallicities (PP04N2 12+log(O/H) $< 8.0$)
\citep{lopezsanchez12}, and that 
it loses sensitivity and saturates at high metallicities 
(PP04N2 12+log(O/H) $> 8.86$).
There are a number of key advantages of this diagnostic, however.  
The ratio is very insensitive to reddening due to the close wavelength 
proximity of the lines.  It is monotonic with oxygen abundance.
It depends on lines with relatively high fluxes in star 
forming environments, which means that good metallicity estimates can be 
achieved at relatively low observational expense.  \citet{yin07} find it 
is more consistent with 
$\rm T_e$ methods than the O3N2 diagnostic of \citet{pp04}, and   
\citet{bresolin09} compare a variety of strong-line abundance estimators 
and find that PP04N2 is the closest match in both slope and 
normalization to the oxygen abundance gradient in NGC~300 measured with the 
$\rm T_e$ method and measured with stellar metallicity (blue supergiants).
%the oxygen abundance gradient in NGC~300 measured with 
%stellar metallicity (blue supergiants).
%**************** Not quite true-- closest match to direct Te method.
%**************** Yes true; Te method almost perfectly matched to stellar.

\subsection{Iron abundances}\label{sec:iron}

%Iron is more fundamentally important than oxygen for the late-stage 
Iron is more important than oxygen for the late-stage 
evolution of massive stars, because iron provides much of the opacity for 
radiation-driven stellar winds \citep[][e.g.]{pauldrach86,vink05}.  
Unfortunately, gas-phase iron abundances are difficult to measure, and  
the fraction of iron depleted onto grains is highly variable. 
Even within our own galaxy, measuring iron abundances is challenging 
\citep[e.g.][]{rodriguez02,jensen07,okada08}.  
We measure gas-phase oxygen abundances instead as a proxy for metallicity 
because it is observationally feasible.  

Because iron is the dominant opacity source, we would like to estimate the 
iron abundances implied by the observed oxygen abundances.  To do this we 
make use of three correlations:  first, the PP04N2 diagnostic, which uses a 
function of the flux ratio of the H$\alpha$~$\lambda 6563$ and 
[\ion{N}{2}]$\lambda 6584$ lines that correlates well with direct method 
($\rm T_e$) measures of gas-phase oxygen abundance; second, the tight 
correlation of direct method measures of gas-phase oxygen abundance with 
stellar oxygen abundance; third, the correlation between stellar oxygen 
and stellar iron abundances.

%As mentioned in \textsection~\ref{sec:Zdiagnostics}, 
\citet{bresolin09} find that PP04N2 is the closest match in both slope and 
normalization to the oxygen abundance gradient in NGC~300 measured with 
$\rm T_e$ methods, with an average error of approximately 0.1 dex.  
Although PP04N2 is a good match to direct-method gas-phase oxygen abundance, 
it is not perfect, and a small correction to slope and zero point could be 
made.  For the purpose of this paper, however, we assume that PP04N2 maps 
precisely to direct-method oxygen abundances.
% Apparently '3rd paragraph, page 10'
% point 30
\citeauthor{bresolin09} also find that the gas-phase oxygen abundance gradient 
measured with $\rm T_e$ methods correlates very tightly with with the stellar 
oxygen abundance gradient, with an error of no more than about 0.03 dex.
For the purposes of this paper, we assume that the correspondence between 
$\rm T_e$ gas-phase oxygen abundance and stellar oxygen abundance is 
one-to-one, which is a very good assumption.

\begin{figure*}
\begin{center}
\begin{tabular}{c}
\plottwo{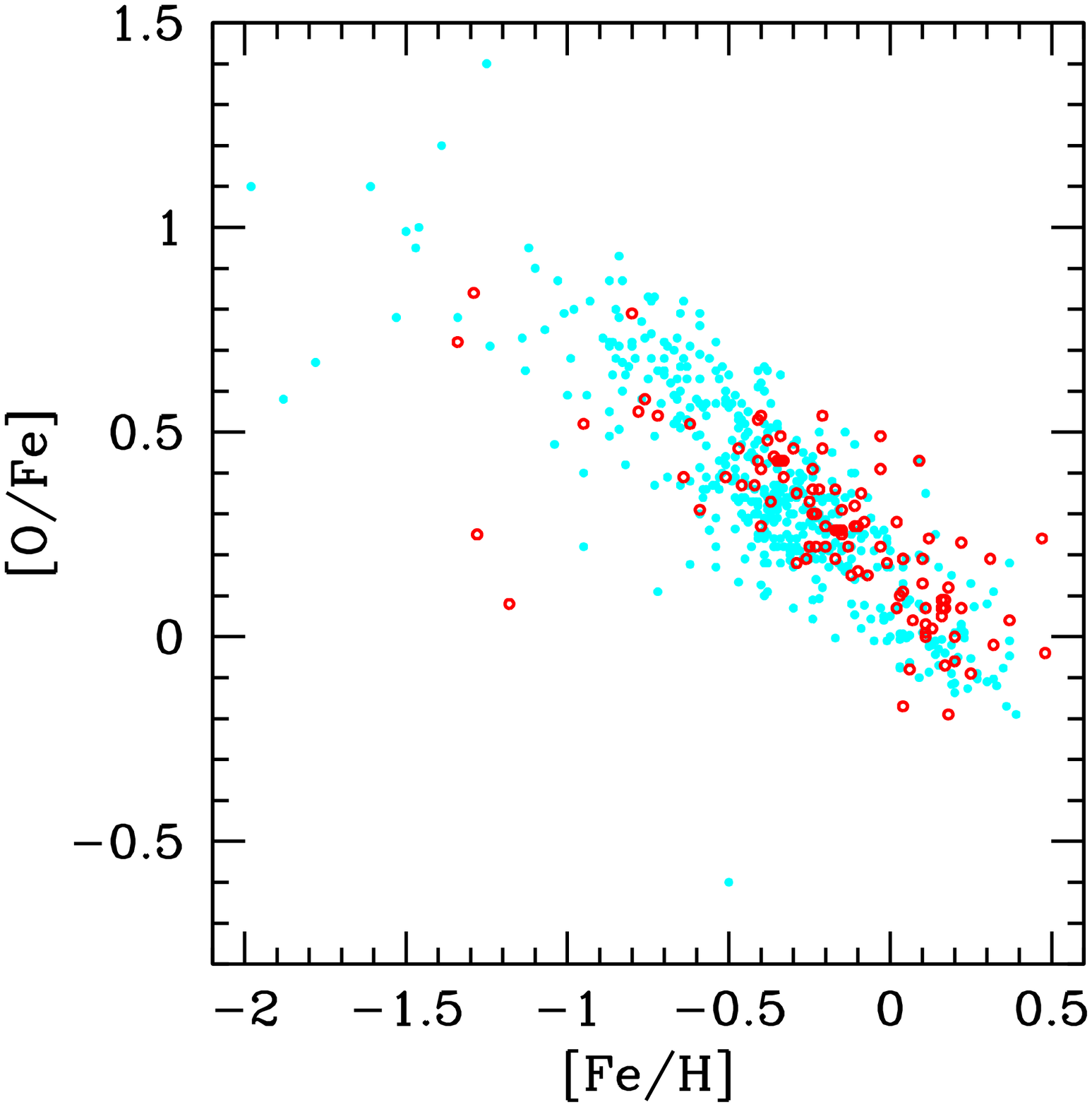}{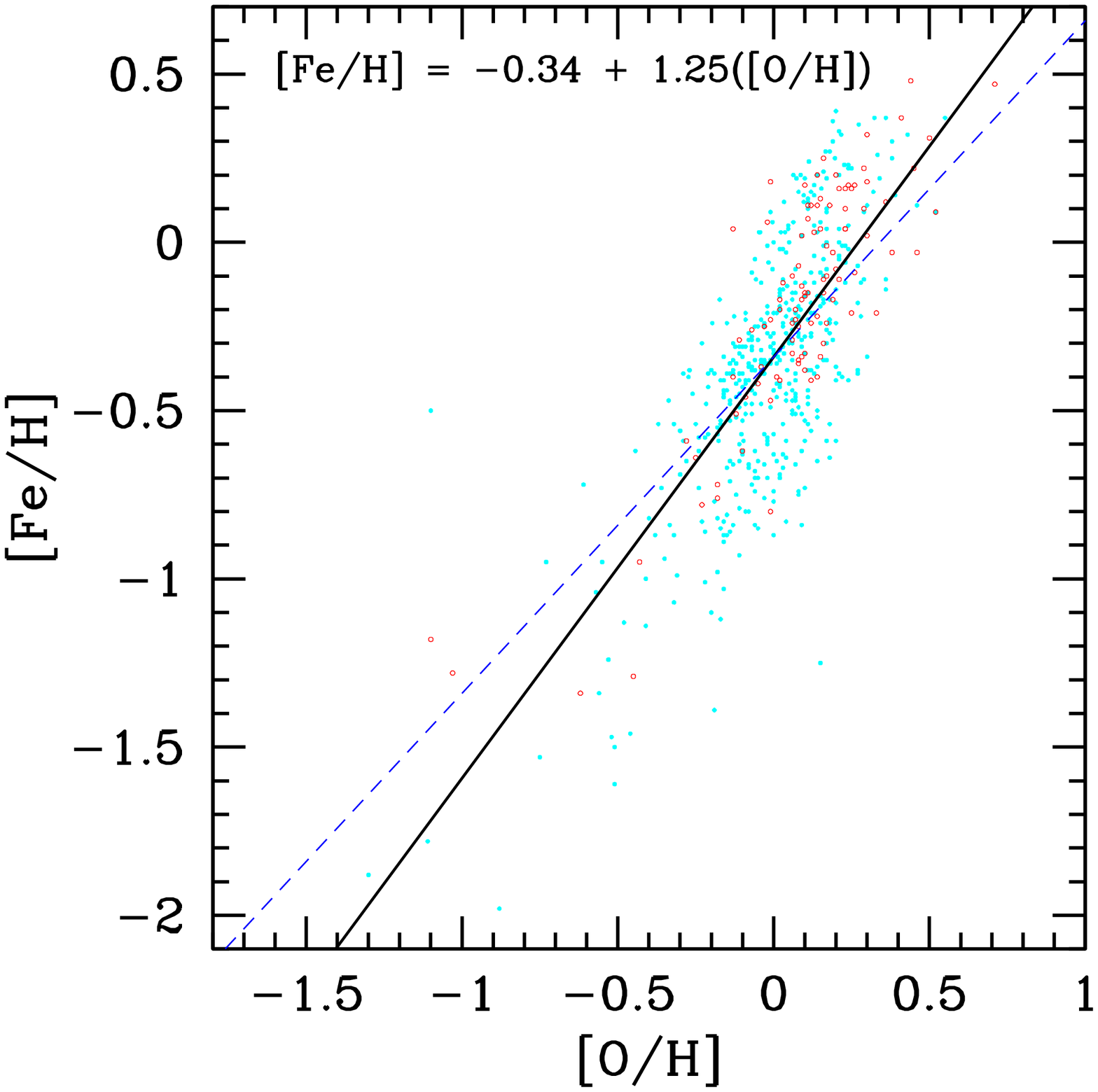}
\end{tabular}
\end{center}
\caption
{\label{fig:ironfit}
To translate oxygen abundances into iron 
abundances, we fit a linear relation (black) to the iron and oxygen 
abundances of Milky Way bulge, disk, and halo stars.  
At low metallicity, $\alpha$-elements like oxygen are enhanced relative to iron
compared to the solar mixture.  
On the left is the well-known relationship between [O/Fe] and [Fe/H].  
The red open points are iron and oxygen abundances of bulge stars 
\citep{fulbright07,lecureur07,rich05,rich07}.
The blue solid points are halo and disk stars 
\citep{bensby04,chen03,reddy03,reddy06}.  On the right we express this in 
terms of [O/H] and [Fe/H] and fit the relation.  
Although the eye is drawn to a steeper trend at higher metallicity than the 
formal fit, this is an illusion based on relatively few points; the fit is 
driven to be flatter by a dense concentration of points with 
$-0.2 < \rm{[O/H]} < 0.2$ and $-0.5 < \rm{[Fe/H]} < 0$.
%The $\alpha$/Fe relationship may flatten out below [Fe/H] = $-1$ 
%\citep[e.g.][]{tinsley79}; excluding all points below that from the fit 
%leads to a steeper relation.  
%We choose here to make the conservative choice and include 
%everything in the fit.  The steeper relation that would result from imposing 
%a low-metallicity cutoff or excluding outliers would have a more dramatic 
%relative effect on the assumed iron abundances, given the measured  
%oxygen abundances.
The blue dashed line has a slope of 1 to guide the eye and intersects with 
the fit at $\rm{[O/H]} = 0$, showing that iron varies more steeply than 
oxygen.  If the slope is steeper than we have fit, this conclusion 
strengthens.  
Equation~\ref{eq:ironconv} can be used to conservatively convert a measured 
gas-phase oxygen abundance to iron abundance for a given solar oxygen 
abundance (modulo the uncertainties of equating gas-phase strong-line 
oxygen abundance indicators to stellar oxygen abundances).  This conversion 
is necessary because at low metallicities, alpha elements such as oxygen are 
enhanced relative to iron, and iron is the main opacity source for line-driven 
winds and thus may drive mass loss for supernova progenitors.
}
\end{figure*}

Stars that have low oxygen abundances have even lower iron abundances.
At low metallicity, $\alpha$-elements like oxygen are enhanced relative to 
iron compared to the solar mixture.  In the galactic disk and halo, at 
[Fe/H]~$ > -1$, [O/Fe] is approximately inversely proportional to [Fe/H], as 
can be seen in the left panel of Figure~\ref{fig:ironfit},
while below [Fe/H] = $-1$, [O/Fe] may flatten out at a constant (and 
lower) relative iron abundance 
\citep[e.g][]{tinsley79,mcwilliam97,johnson07,epstein10} 
\citep[but see e.g.][]{israelian98}.

%----------
%----------

% Apparently 4th paragraph, page 10
% Referee says this paragraph is confusing.
% point 31
To estimate the conversion between stellar oxygen and iron abundance, we 
compared [O/H] to [Fe/H] over a wide range in metallicities using stellar 
abundance measurements from the Milky Way bulge, disk, and halo, from
%\citep{fulbright07,ryde09,rich07,rich05,lecureur07,cunha06,reddy03,reddy06,bensby04,chen03} 
%\citep{fulbright07,rich07,rich05,lecureur07,reddy03,reddy06,bensby04,chen03}. 
\citet{fulbright07,rich07,rich05,lecureur07,reddy03,reddy06,bensby04,chen03}. 
%Noting that [O/H] = [O/Fe] + [Fe/H], 
We fit the relationship between [O/H] 
and [Fe/H] with an unweighted linear fit, as seen in 
Figure~\ref{fig:ironfit}, and find 
\begin{equation}\label{eq:ironfit} \rm [Fe/H] = c_1 + c_2([O/H]),\end{equation} 
where $c_1 = -0.34 \pm 0.01$ and $c_2 = 1.25 \pm 0.05$.  
Although the eye is drawn to a steeper trend at higher metallicity than the 
formal fit shown, this misleading and based on relatively few points; the fit 
is driven to be flatter by a dense concentration of points with 
$-0.2 < \rm{[O/H]} < 0.2$ and $-0.5 < \rm{[Fe/H]} < 0$.
The relationship does not differ substantially between bulge stars (shown in 
red) and halo and disk stars (shown in blue).
If [O/Fe] flattens out below [Fe/H]~$ =-1$, the linear 
relationship we choose to fit may not extend to lower metallicities.  
%We choose to be conservative and not exclude points below [Fe/H]~$ =-1$.  
We would expect to find a slightly steeper relationship were we to exclude 
points below [Fe/H]~$ =-1$, which would mean that type~II SN progenitors in 
low oxygen abundance host regions have even lower iron abundance than we find 
here.  Because there is finite scatter in the measured relationship, however, 
imposing a strict cut at [Fe/H]~$ =-1$ actually drives the fit to be slightly 
flatter by biasing the points with lowest [O/H] to higher [Fe/H].  The exact 
value of the slope is not critically important; the main point is that low 
oxygen abundances imply even lower iron abundances, as can easily be seen in 
the right side of Figure~\ref{fig:ironfit}.  This is equivalent 
to saying that at low metallicities, alpha elements are enhanced relative to 
iron.  

Stellar abundances are measured relative to solar, while gas-phase abundances 
are (nominally) absolute.  Applying the fit to a direct conversion between 
12+log(O/H) and [Fe/H] therefore requires assuming a solar oxygen abundance.
There is currently some dispute over the solar abundance because results 
from atmospheric and interiors methods differ. 
For a given solar oxygen abundance $\rm{O}_\odot$, 
\begin{equation}\label{eq:ironconv} \rm{[Fe/H]} = c_1 - c_2 O_\odot + c_2(12+\log(O/H)). \end{equation} 
Using a solar oxygen abundance of $\rm O_\odot= 8.86$ \citep{delahaye06}, 
the conversion is $\rm{[Fe/H]} = -11.4 + 1.25(12+\log(O/H))$, while 
using $\rm O_\odot= 8.69$ \citep{asplund09}, the conversion is 
$\rm{[Fe/H]} = -11.2 + 1.25(12+\log(O/H))$.

\begin{figure}
\begin{center}
\begin{tabular}{c}
\includegraphics[width=8cm]{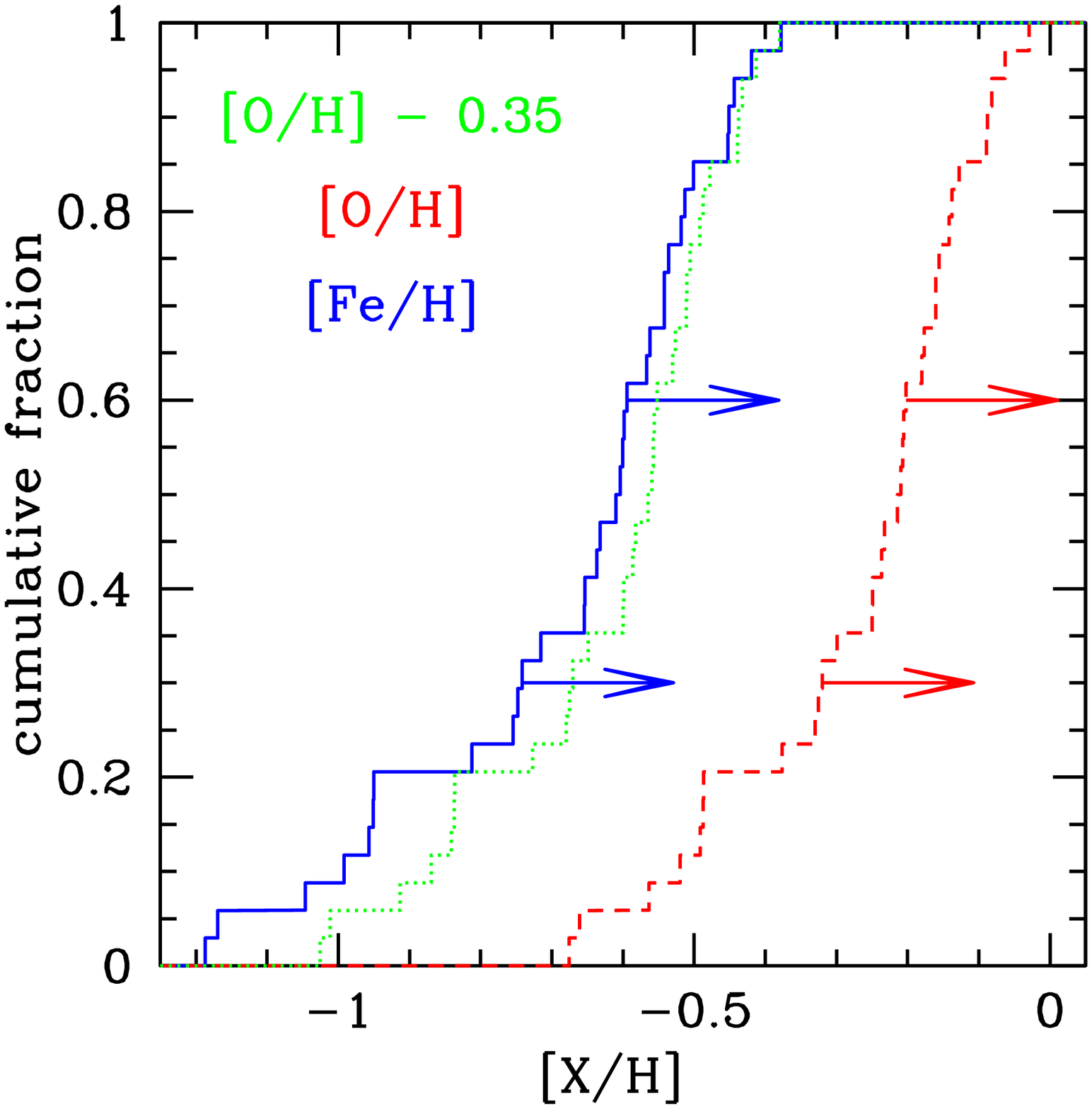}
\end{tabular}
\end{center}
\caption
{\label{fig:irondist}
%We translate the type~II host oxygen abundances to 
The estimated [Fe/H] 
%and plot the 
distribution of the type~II SN sites.  
The red dashed line is the [O/H] distribution, and the blue solid line is the 
[Fe/H] distribution assuming the solar oxygen abundance is 8.86 
\citep{delahaye06}.  The arrows indicate the shift corresponding to assuming 
a solar oxygen abundance of 8.69 \citep{asplund09}.  The green dotted line is 
the oxygen abundance shifted left by 0.35 to line up with the iron at the
high-metallicity end.  Notice that the gap between the iron and the oxygen 
abundance is 50\% wider at the low-metallicity end of the distribution as at 
the high-metallicity end.  Iron varies more steeply than oxygen, and   
a galaxy that has low oxygen abundance has even lower iron abundance.
}
\end{figure}

We apply this fit to transform our oxygen abundance distribution of type~II 
progenitors into an iron abundance distribution, shown in 
Figure~\ref{fig:irondist}.  The median value of [Fe/H] is $-0.60$ using 
the solar value of \citet{delahaye06}.  If another solar oxygen abundance 
$\rm O_\odot$ is assumed, the calculated iron value shifts by 
\begin{equation}\label{eq:ironOshift}c_2(8.86 - \rm{O_\odot}),\end{equation}
so using $\rm O_\odot= 8.69$ \citep{asplund09}, for example, the median 
value of [Fe/H] is $-0.39$.

%----------
%----------

The most important result here is that the gap between the iron and oxygen 
abundances is 50\% wider at the low-metallicity end of the 
distribution than at the high-metallicity end, even though this sample spans 
less than a dex in abundance.  Were we to assume that [O/Fe] flattens out 
below $\rm{[Fe/H]} = -1$ instead of remaining linear, the slope we fit would 
be steeper and the difference would be even greater.
A striking outcome of this translation is that all of the type~II SN 
progenitors in this sample appear to have sub-solar iron abundances.  
Although this is notable, it is not entirely surprising; the sun is more 
enhanced in iron than most Milky Way stars at its oxygen abundance, and all 
of the host galaxies in this sample are smaller than the Milky Way.  It is, 
however, not a secure result.  Given the scatter in the correlations we use 
to define this relation, we caution that conversions of average values are 
uncertain at the level of at least 0.15 dex, and conversions of individual 
values are uncertain at the level of approximately 0.3 dex.

The type~II progenitors all have iron abundances greater than 
$\rm{[Fe/H]} = -1.5$, 
putting them squarely in the regime where winds are primarily driven by iron 
opacity.  For the most metal-poor stars ($Z/Z_\odot < 10^{-3}$), non-iron 
elements such as carbon dominate the radiative driving \citep{vink05}, but 
in the metallicity range of these type~II SN progenitors, iron abundance 
should still be the dominant factor which determines wind strength and mass 
loss.

\section{Conclusions}

% must be paragraph 1, page 11
% point 32
The primary result of this paper is a new progenitor region metallicity 
distribution for a uniform (though not completely unbiased) sample of type~II 
SNe.  Understanding the underlying distribution is important for 
understanding any possible metallicity dependences of different types of 
events associated with massive stars, and it can serve as a probe of the 
metallicity distribution of star formation.

%Although we measured host metallicities for only 34 of 52 type~II SNe in our 
%sample, we find no differences between the hosts with and without metallicity 
%estimates based on their estimated masses, characteristic ages, and star 
%formation rates.

% apparently paragraph 2, page 11
% point 33
The host galaxies of our type~II sample appear to 
%well 
trace galaxies 
%in a redshift-matched sample of 
%also successful parameter fits too many to succinctly list
from the MPA/JHU value-added catalog in mass and metallicity, showing a slight 
bias towards higher star formation rates.  

We find a similarity between the existing host metallicity 
distributions for heterogeneous type~II supernova samples and the 
metallicity distribution we derive.  
Because the existing host metallicity 
distributions are based on supernova samples that are drawn predominantly 
from galaxy-targeted supernova searches, one might naively expect these 
previous distributions might be biased towards higher mass and therefore 
higher metallicity galaxies.
We do not find such a trend.

Comparing to the metallicity distribution of star 
formation rather than to the  metallicity distribution of galaxies as a 
function of mass is the correct way to evaluate a possible metallicity 
dependence of a transient population associated with young stars.
We point out that using CCSNe to trace star formation leads to an almost 
entirely independent way of probing the metallicity distribution of star 
formation from methods involving galaxy population statistics, and we 
compare the metallicity distribution we derive to one of these.  

Finally, we present our host metallicity distribution in terms of iron 
abundance, by converting our oxygen abundance distribution to an iron 
abundance distribution using the $\alpha$/Fe relationship observed in 
Milky Way bulge, disk, and halo stars, noting that iron is more 
important than oxygen for the late-stage evolution of massive stars.  
We show that even though these hosts span less than a dex in oxygen abundance, 
the gap between their iron abundance and oxygen abundance nearly doubles at 
the low-metallicity end compared to the high-metallicity end.
We estimate that $-1.2<$~[Fe/H]~$<0$ for these type~II SN progenitors.  
Though all may have sub-solar iron abundance, none are metal-poor enough that 
elements other than iron will dominate the wind-driving opacity of the 
progenitor star.

%***recast some of these into ...showing that...  

Future improvements to this estimate of the metallicity distribution of 
type~II SNe can be made by performing completeness corrections for any 
selection or followup biases in the source survey.  If the peak luminosity of 
type~II SNe is found to depend on host galaxy metallicity, there may also be 
Malmquist-like biases to correct.

\acknowledgments

We thank B. Andrews, J. Antognini, R. Assef, D. Atlee, C. Epstein, J. Johnson, 
%L. Kewley(?), 
C. Kochanek, 
%Y. Niino(?),
%F. Mannucci(?), 
P. Martini, and L. Watson for discussion, comments, or assistance.
%(We would like to thank C. Epstein and Y. Niino ***?*** for 
%generously sharing data.)
RS is supported by the David G. Price Fellowship in Astronomical 
Instrumentation and the Tuttle Endowment.  
JLP acknowledges support from NASA through Hubble Fellowship grant 
HF-51261.01-A awarded by the STScI, which is operated by AURA, Inc. for NASA, 
under contract NAS 5-26555.  This paper uses data taken with the OSMOS 
spectrograph, built with funding from NSF grant AST-0705170 and from the 
Center for Cosmology and AstroParticle Physics at The Ohio State University.
Funding for SDSS-III has been provided by the Alfred P. Sloan Foundation, the 
Participating Institutions, the National Science Foundation, and the 
U.S.~Department of Energy. The SDSS-III web site is http://www.sdss3.org/.
SDSS-III is managed by the Astrophysical Research Consortium for the Participating Institutions of the SDSS-III Collaboration including the University of Arizona, the Brazilian Participation Group, Brookhaven National Laboratory, University of Cambridge, University of Florida, the French Participation Group, the German Participation Group, the Instituto de Astrofisica de Canarias, the Michigan State/Notre Dame/JINA Participation Group, Johns Hopkins University, Lawrence Berkeley National Laboratory, Max Planck Institute for Astrophysics, New Mexico State University, New York University, Ohio State University, Pennsylvania State University, University of Portsmouth, Princeton University, the Spanish Participation Group, University of Tokyo, University of Utah, Vanderbilt University, University of Virginia, University of Washington, and Yale University.

%Version tracking:  This draft \TeX{}ed on \today.

{\it Facilities:} \facility{Hiltner (OSMOS)}, \facility{Sloan}, \facility{ARC (DIS)}, \facility{Du Pont (WFCCD)}.
% List here:
% http://dopey.mcmaster.ca/cgi-bin/facility_list.cgi

\appendix

\section{Metallicity conversions}\label{app:conversions}

\begin{figure*}
\begin{center}
\begin{tabular}{c}
\plottwo{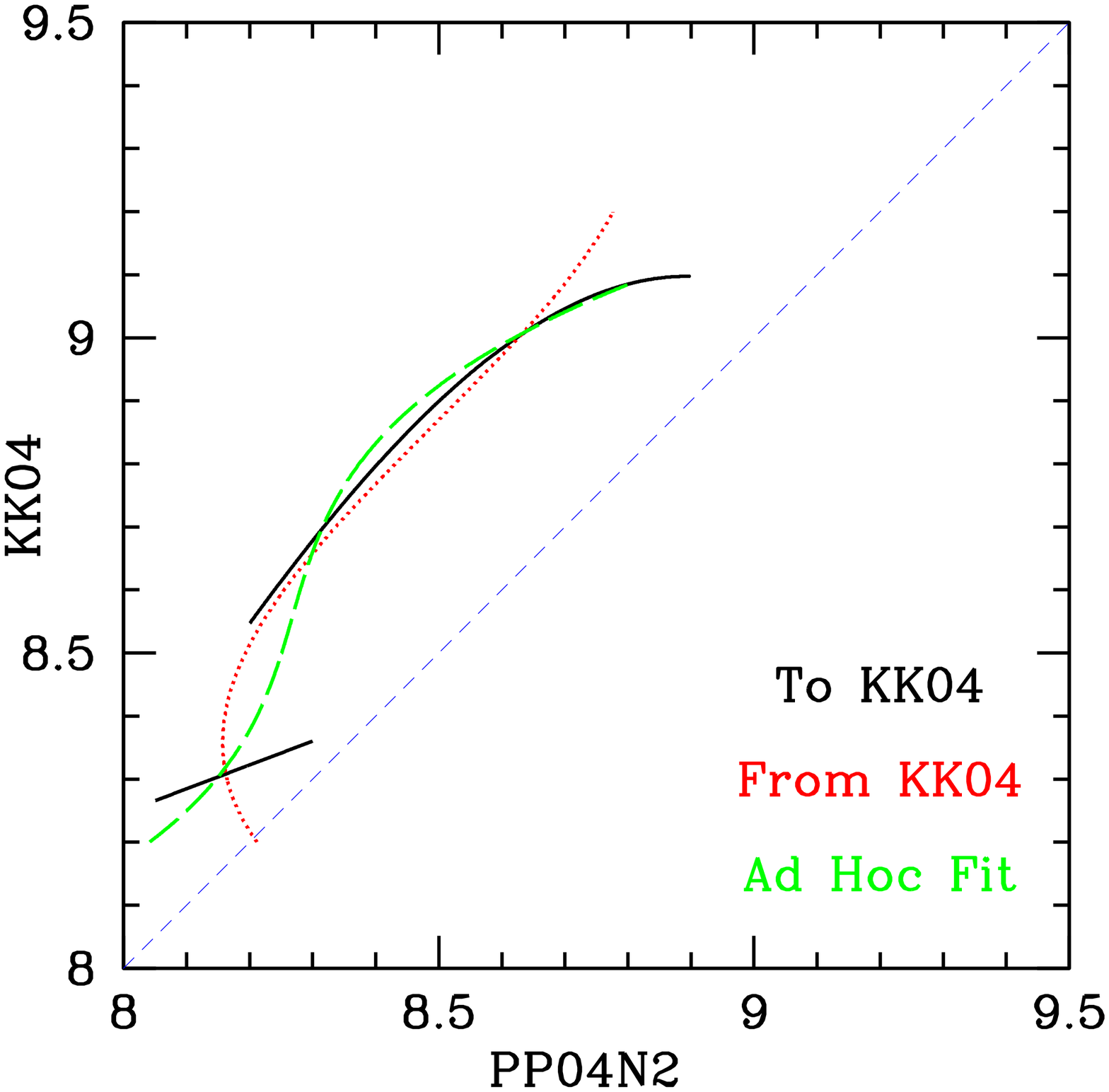}{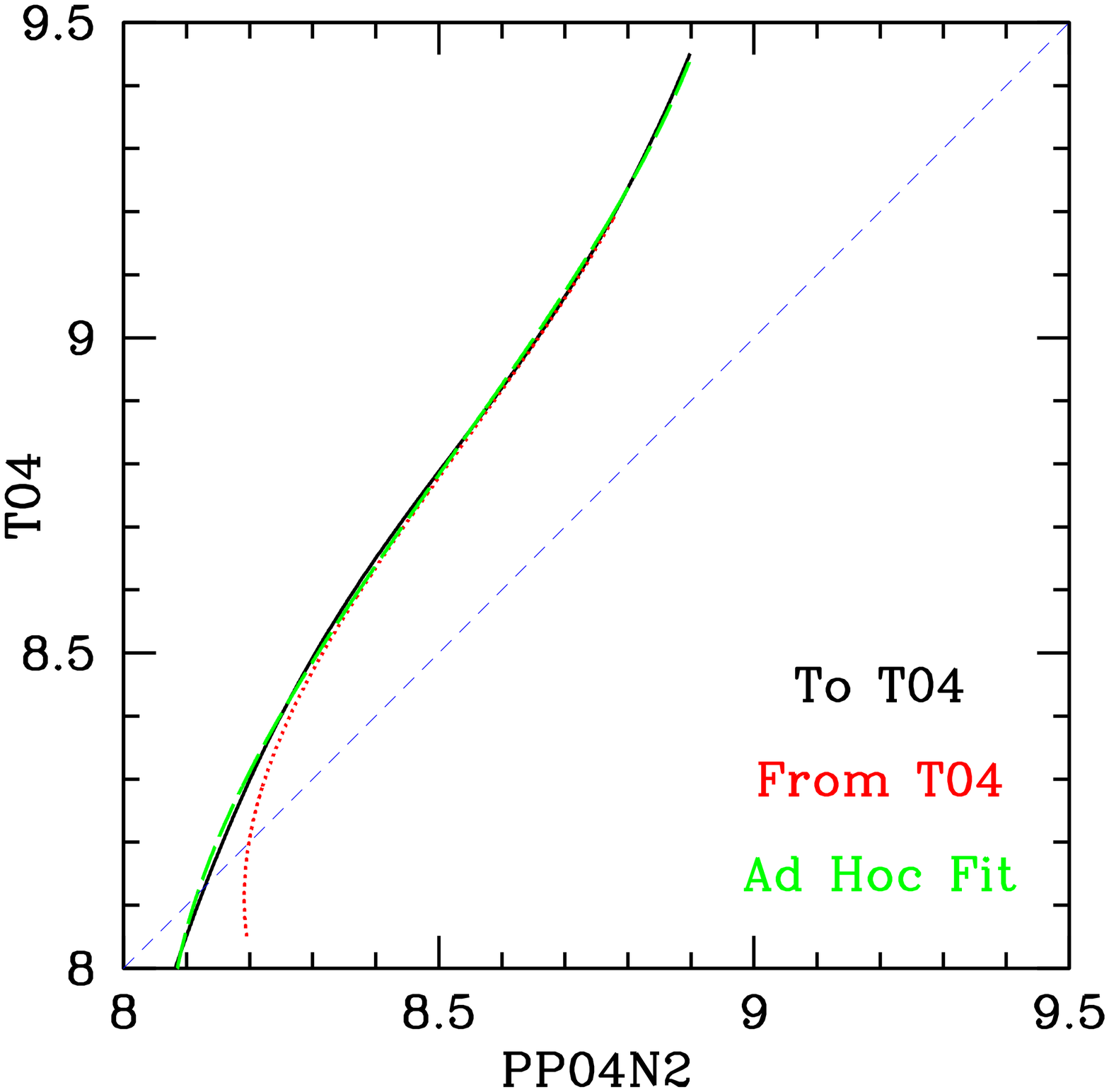}
\end{tabular}
\end{center}
\caption
{\label{fig:convdemo}
  Forward and reverse conversion over the entire valid conversion ranges 
between the metallicity scale of \citet{kk04} and the N2 scale of 
\citet{pp04} (left) and between \citet{tremonti04} and the N2 scale of 
\citet{pp04} (right).  To KK04 and T04 is the black solid line, while from 
KK04 and T04 is the red dotted line.  
The blue dashed line would represent a conversion between two exactly 
equivalent metallicity scales.
The scales are the same as those in Figure 3 in \citet{kewley08}.  
The green long-dashed lines show our fits to
the forward conversions, which we use in Figure~\ref{fig:GalComp} to 
convert from KK04 and T04 to PP04N2 in order to ensure monotonic 
behavior at low metallicities.  
}
\end{figure*}

It is well known that the various strong-line oxygen abundance estimators 
have different scales and zero-points.  
(For excellent pictorial representations of this, see Figure~2 of 
\citet{kewley08} or Figure~12 of \citet{bresolin09}.)
Because of these differences in scale and zero-point, 
it is critically necessary to put 
different estimates on a common scale before comparing abundances.  
\citet{kewley08} determined empirical conversions between many of the 
commonly used scales 
by fitting the trend defined by performing a given two 
diagnostics on a large sample of high S/N galaxy spectra from SDSS.  
While these conversions have been very useful to the community, there can 
be problems using them at very low metallicity.  The forward and reverse 
conversions between two methods 
are not always consistent, or even monotonic, as can be seen by comparing the 
black solid and red dotted lines in Figure~\ref{fig:convdemo}.  
The problem appears to be a consequence of the interaction between the vast 
statistical weight of the abundant high-metallicity galaxies and the 
third-order polynomials (the inversion of which cannot be precisely expressed 
as another third-order polynomial) in which the conversions are expressed.  
The high-metallicity end is tightly pinned, allowing low-metallicity end of 
the third-order polynomial to shift dramatically when the forward and reverse 
conversions are independently fit to the data.

%----------
%----------

Preserving monotonicity in a conversion is important when converting 
a continuous distribution, as can be dramatically seen in 
Figure~\ref{fig:GalComp}.  We are also concerned about the compounding errors 
involved in multiple conversions between different scales, which is perhaps 
the most problematic negative consequence of the difference between the 
forward and reverse conversions, in that even the simplest conversion chain 
(to another scale and back again) results in a different metallicity 
distribution.  For applications involving more than one conversion, 
invertability can be more important than the extra precision from fitting the 
forward and reverse conversions independently to the data.

For the purposes of this paper, we need to avoid the double-value of the 
non-monotonic T04 to PP04N2 conversion at low metallicities in order to 
convert the metallicity distribution of star-formation.  
We do so by defining ad hoc conversions from T04 to PP04N2 and from 
KK04 to PP04N2, as shown in Figure~\ref{fig:convdemo}.  These conversion are 
third-order polynomial fits to the conversion from PP04N2 to T04 and from 
PP04N2 to KK04 given in \citet{kewley08}.
These fits are described by 
\begin{equation}y = a + b x + c x^2 + d x^3, \end{equation}
where $y$ = 12+log(O/H)$_{\rm PP04N2}$.   
For T04, $x$ = 12+log(O/H)$_{\rm T04}$, 
$a = 178.248$, $b = -59.2077$, $c = 6.80078$, and $d = -0.257326$, 
and for KK04, $x$ = 12+log(O/H)$_{\rm KK04}$,
$a = -1657.85$, $b = 583.307$, $c = -68.1052$, and $d = 2.65189$.
Note that these are fits to fits rather than fits to data, and they are 
equivalent to mathematically inverting the conversion from PP04N2 to T04 or 
KK04 and approximating it as a third-order polynomial.  We define them solely 
for the purposes of avoiding a double-valued function at low metallicity and 
matching the inverse conversion from PP04N2 to T04 and from PP04N2 to KK04.

\clearpage
\LongTables
%-------------------------------------------------------
\begin{deluxetable*}{llllrc}
\tablecolumns{6}
\tabletypesize{\scriptsize}
\tablewidth{0pt}
\tablecaption{Observation properties\label{table:obs}}
\tablehead{
   \colhead{Telescope}            &
   \colhead{Instrument}           &
   \colhead{slit width}           &
   \colhead{ruling}               &
   \colhead{$\lambda$ coverage}   &
   \colhead{resolution}           \\
   \colhead{}                     &
   \colhead{}                     &
   \colhead{(arcsec)}             &
   \colhead{(lines/mm)}           &
   \colhead{(\AA)}                &
   \colhead{(\AA)}                  
}
\startdata
APO     & DIS                & 1.5     & B400/R300 (gratings) & 3500--9800  & 7  \\
du Pont & WFCCD              & 1.7     & 400 (blue grism)     & 3700--9200  & 7  \\
Hiltner & OSMOS              & 1.2     & 704 (grism)          & 3960--6870  & 3  \\
SDSS    & fiber spectrograph & 3 (dia) & B640/R440 (grisms)   & 3800--9200  & 3  
\enddata
\end{deluxetable*}
%-------------------------------------------------------

\clearpage
\LongTables
%-------------------------------------------------------
\begin{deluxetable*}{llcccccrrr}
\tablecolumns{11}
\tabletypesize{\scriptsize}
\tablewidth{0pt}
% Would use [\ion{N}{2}]$\lambda 6584$, but screws up at this font size
\tablecaption{Measured host metallicities\label{table:AllZ}}
\tablehead{
   \colhead{SN Name}                                        &
   \colhead{z}                                              &
   \colhead{type}                                           &
   \colhead{H$\alpha$~$\lambda 6563$\tablenotemark{a}}     &
   \colhead{[N~II]$\lambda 6584$\tablenotemark{a}}          &
   \colhead{12+log(O/H)}                                    &
   \colhead{[Fe/H]}                                         &
   \colhead{SN radius\tablenotemark{b}}                     &
   \colhead{SN radius\tablenotemark{c}}                     &
   \colhead{Source}                                         \\
   \colhead{}                                               &
   \colhead{}                                               &
   \colhead{}                                               &
   \colhead{}                                               &
   \colhead{}                                               &
   \colhead{(PP04N2)}                                       &
   \colhead{(8.86)}                                         &
   \colhead{(arcsec)}                                       &
   \colhead{(kpc)}                                          &
   \colhead{}                     
}
\startdata
  PTF09awk &   0.0620 &    Ib &               717$\pm$ 3 &               126$\pm$ 3 & 8.42$\pm$0.03 & $-0.90$ &                    *0.10 &                     0.11 &       SDSS \\ 
  PTF09bce &   0.0230 &    II &               117$\pm$ 3 &                45$\pm$ 4 & 8.72$\pm$0.10 & $-0.51$ &                     5.93 &                     2.76 &      OSMOS \\ 
  PTF09bcl &   0.0620 &    II &               117$\pm$ 9 &                50$\pm$ 8 & 8.77$\pm$0.21 & $-0.45$ & \nodata\tablenotemark{d} & \nodata\tablenotemark{d} &        APO \\ 
  PTF09bgf &   0.0310 &    II &               643$\pm$10 &                38$\pm$11 & 8.18$\pm$0.34 & $-1.19$ &                     1.17 &                     0.72 &     duPont \\ 
  PTF09cjq &   0.0190 &    II &              2982$\pm$25 &              1107$\pm$25 & 8.70$\pm$0.03 & $-0.54$ &                    14.24 &                     5.49 &        APO \\ 
   PTF09cu &   0.0570 &    II &                60$\pm$ 3 &                20$\pm$ 3 & 8.66$\pm$0.20 & $-0.59$ &                     9.14 &                    10.10 &        APO \\ 
  PTF09dah &   0.0238 &   IIb &               207$\pm$ 4 &                31$\pm$ 4 & 8.37$\pm$0.15 & $-0.95$ &                     2.32 &                     1.12 &     duPont \\ 
  PTF09dfk &   0.0160 &    Ib &               601$\pm$ 4 &                65$\pm$ 4 & 8.30$\pm$0.07 & $-1.05$ &                     1.25 &                     0.41 &     duPont \\ 
  PTF09dra &   0.0770 &    II &               145$\pm$ 3 &                55$\pm$ 3 & 8.72$\pm$0.07 & $-0.52$ &                    *3.63 &                     5.29 &       SDSS \\ 
  PTF09due &   0.0290 &    II & \nodata\tablenotemark{e} & \nodata\tablenotemark{e} & 8.77$\pm$0.05 & $-0.45$ &                     9.91 &                     5.76 &    APOgrad \\ 
  PTF09dxv &   0.0330 &   IIb &               277$\pm$ 4 &                88$\pm$ 4 & 8.63$\pm$0.06 & $-0.63$ &                     6.26 &                     4.12 &     duPont \\ 
  PTF09ebq &   0.0235 &    II &              3304$\pm$14 &              1175$\pm$13 & 8.68$\pm$0.01 & $-0.56$ &                     0.69 &                     0.33 &     duPont \\ 
  PTF09ecm &   0.0285 &    II &               149$\pm$ 3 &                55$\pm$ 4 & 8.70$\pm$0.08 & $-0.54$ &                     5.17 &                     2.96 &     duPont \\ 
  PTF09fbf &   0.0210 &    II &               571$\pm$ 3 &                85$\pm$ 3 & 8.37$\pm$0.04 & $-0.95$ &                     4.55 &                     1.93 &     duPont \\ 
  PTF09fma &   0.0310 &    II &               183$\pm$ 5 &                57$\pm$ 6 & 8.62$\pm$0.13 & $-0.64$ & \nodata\tablenotemark{d} & \nodata\tablenotemark{d} &     duPont \\ 
  PTF09fmk &   0.0631 &    II &               451$\pm$ 7 &               166$\pm$ 8 & 8.70$\pm$0.06 & $-0.54$ &                     3.68 &                     4.47 &     duPont \\ 
  PTF09fqa &   0.0300 &    II &                43$\pm$ 1 &                 7$\pm$ 1 & 8.37$\pm$0.22 & $-0.95$ &                    10.76 &                     6.46 &     duPont \\ 
  PTF09fsr &   0.0079 &    Ib &               296$\pm$ 4 &                93$\pm$ 4 & 8.63$\pm$0.06 & $-0.63$ &                    64.61 &                    10.55 &     duPont \\ 
    PTF09g &   0.0400 &    II &               187$\pm$ 3 &                62$\pm$ 3 & 8.65$\pm$0.06 & $-0.60$ &                    *3.83 &                     3.03 &       SDSS \\ 
  PTF09gof &   0.1030 &    II &                62$\pm$ 3 &                17$\pm$ 3 & 8.56$\pm$0.20 & $-0.72$ &                     1.90 &                     3.59 &     duPont \\ 
  PTF09iex &   0.0200 &    II &               278$\pm$ 7 &                18$\pm$ 8 & 8.20$\pm$0.56 & $-1.17$ &                     4.43 &                     1.79 &     duPont \\ 
  PTF09ige &   0.0640 &    II &               322$\pm$ 2 &                97$\pm$ 2 & 8.61$\pm$0.02 & $-0.65$ &                    *5.10 &                     6.28 &       SDSS \\ 
  PTF09igz &   0.0860 &    II &                72$\pm$ 3 &                18$\pm$ 3 & 8.53$\pm$0.19 & $-0.75$ &                     1.69 &                     2.72 &        APO \\ 
  PTF09ism &   0.0290 &    II &               100$\pm$ 2 &                33$\pm$ 2 & 8.65$\pm$0.08 & $-0.61$ &                    *7.47 &                     4.34 &       SDSS \\ 
    PTF09q &   0.0900 &    Ic &               201$\pm$ 4 &                92$\pm$ 4 & 8.81$\pm$0.06 & $-0.40$ &                    *2.88 &                     4.84 &       SDSS \\ 
    PTF09r &   0.0270 &    II &                63$\pm$ 3 &                22$\pm$ 3 & 8.68$\pm$0.18 & $-0.57$ &                     0.79 &                     0.43 &      OSMOS \\ 
   PTF09sh &   0.0377 &    II &               362$\pm$ 9 &                90$\pm$ 9 & 8.53$\pm$0.12 & $-0.75$ &                     9.73 &                     7.27 &        APO \\ 
   PTF09sk &   0.0355 & Ic-BL &               717$\pm$ 3 &                73$\pm$ 2 & 8.28$\pm$0.04 & $-1.06$ &                    *2.77 &                     1.95 &       SDSS \\ 
    PTF09t &   0.0390 &    II &               775$\pm$ 9 &               114$\pm$ 9 & 8.37$\pm$0.10 & $-0.96$ &                     5.84 &                     4.51 &     duPont \\ 
   PTF09tm &   0.0350 &    II &               232$\pm$ 4 &               109$\pm$ 4 & 8.83$\pm$0.04 & $-0.38$ &                    *3.64 &                     2.53 &       SDSS \\ 
   PTF09uj &   0.0651 &    II &                85$\pm$ 2 &                26$\pm$ 2 & 8.61$\pm$0.10 & $-0.65$ &                    *2.72 &                     3.40 &       SDSS \\ 
  PTF10bau &   0.0260 &    II & \nodata\tablenotemark{e} & \nodata\tablenotemark{e} & 8.73$\pm$0.04 & $-0.50$ &                     6.18 &                     3.23 &       grad \\ 
  PTF10bgl &   0.0300 &    II & \nodata\tablenotemark{e} & \nodata\tablenotemark{e} & 8.63$\pm$0.06 & $-0.63$ &                     8.19 &                     4.92 &    OSMgrad \\ 
  PTF10bhu &   0.0360 &    Ic &               152$\pm$ 2 &                36$\pm$ 2 & 8.51$\pm$0.08 & $-0.78$ &                    *1.62 &                     1.16 &       SDSS \\ 
  PTF10bip &   0.0510 &    Ic &               229$\pm$ 6 &                24$\pm$ 5 & 8.29$\pm$0.25 & $-1.05$ &                     1.24 &                     1.23 &     duPont \\ 
  PTF10con &   0.0330 &    II &               132$\pm$ 5 &                44$\pm$ 5 & 8.65$\pm$0.14 & $-0.60$ &                     1.72 &                     1.13 &     duPont \\ 
  PTF10cqh &   0.0410 &    II &               318$\pm$ 7 &               141$\pm$ 7 & 8.80$\pm$0.06 & $-0.42$ &                     9.17 &                     7.43 &     duPont \\ 
  PTF10cwx &   0.0730 &    II &               237$\pm$ 4 &                31$\pm$ 5 & 8.34$\pm$0.19 & $-0.99$ &                     2.56 &                     3.55 &     duPont \\ 
  PTF10cxq &   0.0470 &    II &               286$\pm$ 5 &                31$\pm$ 5 & 8.30$\pm$0.20 & $-1.05$ &                     1.53 &                     1.41 &     duPont \\ 
  PTF10cxx &   0.0340 &    II &               759$\pm$ 4 &               326$\pm$ 4 & 8.78$\pm$0.02 & $-0.44$ &                    *1.84 &                     1.24 &       SDSS \\ 
  PTF10czn &   0.0450 &    II &               611$\pm$13 &               132$\pm$13 & 8.48$\pm$0.12 & $-0.81$ &                    14.98 &                    13.25 &     duPont \\ 
   PTF10hv &   0.0518 &    II &               139$\pm$ 2 &                35$\pm$ 2 & 8.54$\pm$0.06 & $-0.74$ &                    *5.59 &                     5.65 &       SDSS \\ 
    PTF10s &   0.0510 &    II &               271$\pm$ 2 &                90$\pm$ 2 & 8.65$\pm$0.03 & $-0.60$ &                    *1.05 &                     1.04 &       SDSS 
\enddata
\tablenotetext{a}{$10^{-17}$ erg cm$^{-2}$ s$^{-1}$ \AA$^{-1}$}
\tablenotetext{b}{SN radius from the center of the galaxy.  Note that all 
targets are observed at the SN location or equivalent galactocentric radius
except the 12 targets for which we use archival SDSS spectra (*starred).  
For these 12 targets, the distance in this column is the distance between 
the SN location and the fiber center, which is half an arcsecond or less
away from the center of the galaxy in all cases.  Note that for these 
SDSS spectra, most SN locations are within 
the fiber diameter, and only one is more than 2 fiber diameters away.  }
\tablenotetext{c}{Projected physical radius calculated \citep{cosmocalc}
assuming $H_0 = 70$, $\Omega_m = 0.3$, and $\Omega_\Lambda=0.7$.}
\tablenotetext{d}{PTF09bcl and PTF09fma are outside the SDSS photometry,
so we do not determine the coordinates of their host galaxy centers.}
\tablenotetext{e}{For PTF09due, PTF10bau, and PTF10bgl, the metallicity at
the SN location is approximated by fitting a metallicity gradient to points
at other galactocentric radii (see Table~\ref{table:gradlineflux} for
line fluxes and locations) and extrapolating or interpolating the metallicity
at the radius of the SN.}
\end{deluxetable*}
%-------------------------------------------------------

\clearpage
\LongTables
%-------------------------------------------------------
\begin{deluxetable*}{llrrrr}
\tablecolumns{6}
\tabletypesize{\scriptsize}
\tablewidth{0pt}
%\setlength{\tabcolsep}{0.02in}
% Would use [\ion{N}{2}]$\lambda 6584$, but screws up at this font size
\tablecaption{Measured line fluxes and metallicities at non-supernova locations within the galaxy\label{table:gradlineflux}}
\tablehead{
   \colhead{SN Name}                                            &
   \colhead{Source}                                             &
   \colhead{Radius\tablenotemark{a}}                            &
   \colhead{H$\alpha$~$\lambda 6563$\tablenotemark{b}}         &
   \colhead{[N~II]$\lambda 6584$\tablenotemark{b}}              &
   \colhead{12+log(O/H)}                                        \\
   \colhead{}                                                   &
   \colhead{}                                                   &
   \colhead{(arcsec)}                                           &
   \colhead{}                                                   &
   \colhead{}                                                   &
   \colhead{(PP04N2)}                                
}
\startdata
  PTF09due &   APO & 13.0 & 276.0$\pm$1.9 & 105.2$\pm$2.0 & 8.72$\pm$0.02 \\ 
  PTF09due &   APO & 25.0 & 202.2$\pm$1.6 &  46.6$\pm$1.5 & 8.51$\pm$0.04 \\ 
  PTF10bau &  SDSS &  0.0 & 988.5$\pm$6.9 & 379.5$\pm$7.3 & 8.72$\pm$0.02 \\ 
  PTF10bau & OSMOS &  3.1 & 522.3$\pm$1.8 & 209.8$\pm$2.4 & 8.74$\pm$0.01 \\ 
  PTF10bau & OSMOS &  4.8 & 314.7$\pm$2.1 & 124.8$\pm$2.3 & 8.74$\pm$0.02 \\ 
  PTF10bau &  SDSS & 10.5 & 565.2$\pm$1.9 & 215.7$\pm$2.0 & 8.72$\pm$0.01 \\ 
  PTF10bau & OSMOS & 10.6 & 122.8$\pm$1.8 &  57.1$\pm$1.9 & 8.82$\pm$0.04 \\ 
  PTF10bau & OSMOS & 12.9 & 152.7$\pm$1.6 &  60.9$\pm$1.6 & 8.74$\pm$0.03 \\ 
  PTF10bau & OSMOS & 19.8 &  35.9$\pm$1.2 &   6.6$\pm$0.7 & 8.43$\pm$0.14 \\ 
  PTF10bgl & OSMOS &  1.5 & 268.9$\pm$4.4 & 107.7$\pm$4.7 & 8.74$\pm$0.06 \\ 
  PTF10bgl & OSMOS &  5.9 & 164.3$\pm$3.1 &  61.8$\pm$3.4 & 8.71$\pm$0.07 \\ 
  PTF10bgl & OSMOS &  9.1 & 394.4$\pm$2.9 & 116.3$\pm$3.1 & 8.60$\pm$0.03 \\ 
  PTF10bgl & OSMOS & 13.7 & 851.3$\pm$3.0 & 218.5$\pm$3.2 & 8.54$\pm$0.02 \\ 
  PTF10bgl & OSMOS & 21.9 &  68.7$\pm$2.6 &  21.8$\pm$2.5 & 8.63$\pm$0.14 
\enddata
\tablenotetext{a}{Galactocentric radius of the spectrum}
\tablenotetext{b}{$10^{-17}$ erg cm$^{-2}$ s$^{-1}$ \AA$^{-1}$}
\end{deluxetable*}
%-------------------------------------------------------

\clearpage
\LongTables
%-------------------------------------------------------
\begin{deluxetable*}{lrrrrrrr}
\tablecolumns{8}
\tabletypesize{\scriptsize}
\tablewidth{0pt}
%\setlength{\tabcolsep}{0.02in}
%\rotate
\tablecaption{Host properties measured from SDSS photometry\label{table:SDSS}}
\tablehead{
   \colhead{SN}                        &
   \colhead{Gal RA}                    &
   \colhead{Gal Dec}                   &
   \colhead{$f_u$\tablenotemark{a}}       &
   \colhead{$f_g$\tablenotemark{a}}       &
   \colhead{$f_r$\tablenotemark{a}}       &
   \colhead{$f_i$\tablenotemark{a}}       &
   \colhead{$f_z$\tablenotemark{a}}       \\
   \colhead{(PTF)}                     &
   \colhead{(J2000.0)}                 &
   \colhead{(J2000.0)}                 &
   \colhead{(Jy)}                      &
   \colhead{(Jy)}                      &
   \colhead{(Jy)}                      &
   \colhead{(Jy)}                      &
   \colhead{(Jy)}                      
}
\startdata
  09aux &  16:09:15.851 &  +29:17:37.09 & 1.42E-04$\pm$8.5E-06 & 5.87E-04$\pm$1.2E-05 & 1.15E-03$\pm$2.3E-05 & 1.57E-03$\pm$3.2E-05 & 1.99E-03$\pm$6.3E-05\\ 
  09awk &  13:37:56.354 &  +22:55:04.82 & 8.41E-05$\pm$5.4E-06 & 1.96E-04$\pm$4.2E-06 & 2.72E-04$\pm$5.8E-06 & 3.58E-04$\pm$7.9E-06 & 3.89E-04$\pm$1.6E-05\\ 
  09axc &  14:53:13.066 &  +22:14:32.22 & 2.51E-05$\pm$3.0E-06 & 9.85E-05$\pm$2.2E-06 & 1.88E-04$\pm$4.0E-06 & 2.40E-04$\pm$5.3E-06 & 2.78E-04$\pm$1.4E-05\\ 
  09axi &  14:12:40.942 &  +31:04:03.51 & 3.67E-05$\pm$6.1E-06 & 1.09E-04$\pm$2.9E-06 & 1.79E-04$\pm$4.9E-06 & 1.90E-04$\pm$6.4E-06 & 1.74E-04$\pm$1.9E-05\\ 
  09bce &  16:35:18.117 &  +55:38:03.60 & 1.02E-03$\pm$5.2E-05 & 4.05E-03$\pm$8.3E-05 & 8.23E-03$\pm$1.7E-04 & 1.20E-02$\pm$2.4E-04 & 1.55E-02$\pm$4.7E-04\\ 
  09bgf &  14:41:38.351 &  +19:21:43.21 & 5.69E-05$\pm$5.3E-06 & 1.29E-04$\pm$3.1E-06 & 1.65E-04$\pm$4.2E-06 & 1.89E-04$\pm$5.8E-06 & 2.08E-04$\pm$1.6E-05\\ 
   09bw &  15:05:02.037 &  +48:40:03.22 & 1.85E-06$\pm$1.8E-06 & 1.35E-05$\pm$1.1E-06 & 2.80E-05$\pm$1.5E-06 & 3.52E-05$\pm$2.5E-06 & 3.54E-05$\pm$7.6E-06\\ 
  09cjq &  21:16:27.580 &  -00:49:35.16 & 2.97E-03$\pm$1.5E-04 & 1.05E-02$\pm$2.1E-04 & 2.10E-02$\pm$4.2E-04 & 2.96E-02$\pm$6.0E-04 & 3.93E-02$\pm$1.2E-03\\ 
   09ct &  11:42:13.837 &  +10:38:53.86 & 1.03E-05$\pm$2.5E-06 & 2.68E-05$\pm$1.2E-06 & 5.51E-05$\pm$2.2E-06 & 7.46E-05$\pm$3.5E-06 & 9.48E-05$\pm$1.1E-05\\ 
   09cu &  13:15:23.892 &  +46:25:13.40 & 3.85E-04$\pm$2.0E-05 & 1.05E-03$\pm$2.2E-05 & 1.77E-03$\pm$3.6E-05 & 2.38E-03$\pm$4.9E-05 & 2.83E-03$\pm$9.3E-05\\ 
  09cvi &  21:47:09.984 &  +08:18:35.58 & -1.11E-06$\pm$1.2E-06 & 4.06E-06$\pm$4.3E-07 & 4.63E-06$\pm$6.9E-07 & 4.97E-06$\pm$1.1E-06 & 9.74E-06$\pm$3.9E-06\\ 
  09dah &  22:45:17.115 &  +21:49:17.34 & 1.80E-04$\pm$1.3E-05 & 3.79E-04$\pm$8.2E-06 & 5.11E-04$\pm$1.2E-05 & 6.63E-04$\pm$1.6E-05 & 7.64E-04$\pm$4.0E-05\\ 
  09dfk &  23:09:13.471 &  +07:48:16.39 & 1.28E-04$\pm$7.9E-06 & 3.53E-04$\pm$7.5E-06 & 5.40E-04$\pm$1.1E-05 & 6.87E-04$\pm$1.5E-05 & 8.28E-04$\pm$3.2E-05\\ 
  09djl &  16:33:55.969 &  +30:14:16.65 & 4.05E-06$\pm$2.1E-06 & 1.76E-05$\pm$8.0E-07 & 4.63E-05$\pm$1.4E-06 & 5.80E-05$\pm$2.2E-06 & 8.44E-05$\pm$9.8E-06\\ 
  09dra &  15:48:11.297 &  +41:13:31.76 & 2.40E-04$\pm$1.5E-05 & 5.98E-04$\pm$1.3E-05 & 9.95E-04$\pm$2.1E-05 & 1.34E-03$\pm$2.9E-05 & 1.63E-03$\pm$5.9E-05\\ 
  09due &  16:26:53.240 &  +51:33:18.35 & 2.35E-03$\pm$1.2E-04 & 6.12E-03$\pm$1.3E-04 & 9.42E-03$\pm$1.9E-04 & 1.17E-02$\pm$2.4E-04 & 1.41E-02$\pm$4.3E-04\\ 
  09dxv &  23:08:34.828 &  +18:56:19.80 & 3.99E-04$\pm$2.6E-05 & 1.17E-03$\pm$2.4E-05 & 1.94E-03$\pm$4.0E-05 & 2.77E-03$\pm$5.9E-05 & 3.55E-03$\pm$1.3E-04\\ 
  09dzt &  16:03:03.823 &  +21:01:47.28 & 9.55E-05$\pm$9.2E-06 & 2.00E-04$\pm$4.9E-06 & 3.32E-04$\pm$8.7E-06 & 4.26E-04$\pm$1.3E-05 & 5.39E-04$\pm$4.2E-05\\ 
  09ebq &  00:14:01.743 &  +29:25:58.47 & 5.79E-04$\pm$2.9E-05 & 1.44E-03$\pm$2.9E-05 & 2.40E-03$\pm$4.8E-05 & 3.15E-03$\pm$6.5E-05 & 4.01E-03$\pm$1.3E-04\\ 
  09ecm &  01:06:43.123 &  -06:22:46.04 & 3.60E-04$\pm$2.4E-05 & 1.08E-03$\pm$2.3E-05 & 1.83E-03$\pm$3.8E-05 & 2.26E-03$\pm$4.8E-05 & 2.86E-03$\pm$1.0E-04\\ 
  09ejz &  00:55:07.230 &  -06:57:04.82 & 5.63E-05$\pm$8.0E-06 & 1.86E-04$\pm$4.4E-06 & 4.32E-04$\pm$9.5E-06 & 6.28E-04$\pm$1.4E-05 & 8.24E-04$\pm$3.4E-05\\ 
  09fae &  17:26:20.127 &  +72:56:28.74 & 2.23E-05$\pm$3.3E-06 & 3.59E-05$\pm$1.3E-06 & 3.76E-05$\pm$1.9E-06 & 4.74E-05$\pm$2.9E-06 & 8.00E-05$\pm$1.1E-05\\ 
  09fbf &  21:20:38.208 &  +01:02:49.97 & 1.59E-03$\pm$8.1E-05 & 4.36E-03$\pm$9.0E-05 & 6.76E-03$\pm$1.4E-04 & 8.65E-03$\pm$1.8E-04 & 1.06E-02$\pm$3.3E-04\\ 
  09fmk &  23:57:46.435 &  +11:58:44.53 & 1.65E-04$\pm$1.3E-05 & 4.44E-04$\pm$9.6E-06 & 7.74E-04$\pm$1.7E-05 & 1.21E-03$\pm$2.7E-05 & 1.40E-03$\pm$6.1E-05\\ 
  09foy &  23:17:10.983 &  +17:15:04.12 & 1.99E-04$\pm$1.5E-05 & 5.79E-04$\pm$1.2E-05 & 9.82E-04$\pm$2.1E-05 & 1.34E-03$\pm$2.9E-05 & 1.50E-03$\pm$6.9E-05\\ 
  09fqa &  22:25:33.064 &  +18:59:44.12 & 4.48E-04$\pm$2.7E-05 & 1.30E-03$\pm$2.7E-05 & 2.11E-03$\pm$4.4E-05 & 2.71E-03$\pm$5.7E-05 & 3.16E-03$\pm$1.1E-04\\ 
  09fsr &  23:04:56.569 &  +12:19:21.47 & 1.61E-02$\pm$7.8E-04 & 6.23E-02$\pm$1.3E-03 & 1.24E-01$\pm$2.5E-03 & 1.73E-01$\pm$3.5E-03 & 2.13E-01$\pm$6.5E-03\\ 
    09g &  15:16:31.418 &  +54:27:31.04 & 4.12E-04$\pm$2.1E-05 & 9.85E-04$\pm$2.0E-05 & 1.40E-03$\pm$2.9E-05 & 1.66E-03$\pm$3.4E-05 & 1.87E-03$\pm$6.7E-05\\ 
  09gof &  01:22:25.476 &  +03:38:08.80 & 7.57E-05$\pm$8.2E-06 & 1.91E-04$\pm$4.5E-06 & 3.21E-04$\pm$7.5E-06 & 3.74E-04$\pm$1.0E-05 & 4.07E-04$\pm$2.8E-05\\ 
  09gtt &  02:20:37.291 &  +02:24:24.30 & 4.74E-05$\pm$9.5E-06 & 1.39E-04$\pm$4.0E-06 & 1.78E-04$\pm$5.9E-06 & 2.73E-04$\pm$9.8E-06 & 2.85E-04$\pm$4.0E-05\\ 
  09hdo &  00:15:22.815 &  +30:43:16.29 & 4.55E-04$\pm$2.4E-05 & 1.59E-03$\pm$3.3E-05 & 3.17E-03$\pm$6.4E-05 & 4.60E-03$\pm$9.4E-05 & 6.01E-03$\pm$1.9E-04\\ 
  09hzg &  11:50:56.789 &  +21:11:50.06 & 3.64E-04$\pm$2.1E-05 & 1.38E-03$\pm$2.9E-05 & 2.97E-03$\pm$6.0E-05 & 4.45E-03$\pm$9.1E-05 & 6.25E-03$\pm$1.9E-04\\ 
  09iex &  12:02:46.955 &  +02:24:02.61 & 8.43E-05$\pm$9.2E-06 & 1.59E-04$\pm$4.6E-06 & 1.94E-04$\pm$6.7E-06 & 2.28E-04$\pm$1.0E-05 & 3.01E-04$\pm$3.9E-05\\ 
  09ige &  08:55:34.126 &  +32:40:01.34 & 1.39E-04$\pm$8.1E-06 & 3.49E-04$\pm$7.4E-06 & 4.98E-04$\pm$1.1E-05 & 6.26E-04$\pm$1.4E-05 & 6.35E-04$\pm$2.7E-05\\ 
  09igz &  08:53:56.582 &  +33:40:10.68 & 3.04E-05$\pm$4.1E-06 & 7.07E-05$\pm$2.0E-06 & 1.04E-04$\pm$3.4E-06 & 1.29E-04$\pm$5.0E-06 & 1.42E-04$\pm$1.6E-05\\ 
  09ism &  11:44:35.370 &  +10:12:46.55 & 1.98E-04$\pm$1.5E-05 & 4.85E-04$\pm$1.1E-05 & 7.06E-04$\pm$1.6E-05 & 8.73E-04$\pm$2.2E-05 & 9.28E-04$\pm$6.2E-05\\ 
   09ps &  16:14:08.619 &  +55:41:41.78 & 2.18E-05$\pm$3.0E-06 & 4.35E-05$\pm$1.2E-06 & 6.19E-05$\pm$1.8E-06 & 7.92E-05$\pm$2.6E-06 & 6.72E-05$\pm$1.1E-05\\ 
    09q &  12:24:50.022 &  +08:26:01.27 & 1.49E-04$\pm$1.1E-05 & 4.47E-04$\pm$9.6E-06 & 8.69E-04$\pm$1.8E-05 & 1.24E-03$\pm$2.6E-05 & 1.51E-03$\pm$5.2E-05\\ 
    09r &  14:18:58.607 &  +35:23:15.26 & 4.01E-05$\pm$4.0E-06 & 1.37E-04$\pm$3.1E-06 & 2.54E-04$\pm$5.5E-06 & 3.32E-04$\pm$7.6E-06 & 3.98E-04$\pm$1.9E-05\\ 
   09sh &  16:13:58.581 &  +39:31:50.29 & 6.53E-04$\pm$3.6E-05 & 1.46E-03$\pm$3.0E-05 & 2.34E-03$\pm$4.8E-05 & 3.04E-03$\pm$6.4E-05 & 2.83E-03$\pm$1.2E-04\\ 
   09sk &  13:30:51.179 &  +30:20:02.22 & 1.15E-04$\pm$6.7E-06 & 2.52E-04$\pm$5.6E-06 & 3.19E-04$\pm$6.9E-06 & 3.87E-04$\pm$9.1E-06 & 4.14E-04$\pm$1.9E-05\\ 
    09t &  14:15:42.905 &  +16:12:00.93 & 5.39E-04$\pm$2.8E-05 & 1.24E-03$\pm$2.5E-05 & 1.67E-03$\pm$3.4E-05 & 1.99E-03$\pm$4.2E-05 & 2.14E-03$\pm$7.8E-05\\ 
   09tm &  13:46:55.509 &  +61:33:17.33 & 4.21E-04$\pm$2.1E-05 & 1.41E-03$\pm$2.9E-05 & 2.68E-03$\pm$5.4E-05 & 3.76E-03$\pm$7.7E-05 & 4.84E-03$\pm$1.5E-04\\ 
   09uj &  14:20:10.883 &  +53:33:42.11 & 1.14E-04$\pm$8.0E-06 & 2.86E-04$\pm$6.3E-06 & 4.33E-04$\pm$9.5E-06 & 5.28E-04$\pm$1.2E-05 & 5.38E-04$\pm$2.8E-05\\ 
  10bau &  09:16:21.696 &  +17:43:38.08 & 2.47E-03$\pm$1.2E-04 & 6.20E-03$\pm$1.3E-04 & 1.02E-02$\pm$2.0E-04 & 1.28E-02$\pm$2.6E-04 & 1.56E-02$\pm$4.8E-04\\ 
  10bfz &  12:54:41.278 &  +15:24:16.42 & 1.80E-06$\pm$1.1E-06 & 4.07E-06$\pm$4.7E-07 & 6.37E-06$\pm$7.6E-07 & 4.20E-06$\pm$1.4E-06 & 1.20E-06$\pm$5.1E-06\\ 
  10bgl &  10:19:05.166 &  +46:27:16.67 & 3.40E-03$\pm$1.6E-04 & 8.80E-03$\pm$1.8E-04 & 1.27E-02$\pm$2.6E-04 & 1.57E-02$\pm$3.2E-04 & 1.77E-02$\pm$5.4E-04\\ 
  10bhu &  12:55:28.353 &  +53:34:30.63 & 2.17E-04$\pm$1.3E-05 & 5.53E-04$\pm$1.2E-05 & 8.10E-04$\pm$1.7E-05 & 9.83E-04$\pm$2.2E-05 & 1.17E-03$\pm$4.5E-05\\ 
  10bip &  12:34:10.493 &  +08:21:49.67 & 2.75E-05$\pm$3.5E-06 & 9.00E-05$\pm$2.2E-06 & 1.16E-04$\pm$3.0E-06 & 1.46E-04$\pm$4.1E-06 & 1.74E-04$\pm$1.2E-05\\ 
  10bzf &  11:44:02.964 &  +55:41:22.55 & 3.89E-05$\pm$4.7E-06 & 5.33E-05$\pm$2.1E-06 & 8.66E-05$\pm$3.0E-06 & 1.09E-04$\pm$4.5E-06 & 1.37E-04$\pm$1.6E-05\\ 
   10cd &  03:00:33.086 &  +36:15:25.02 & 2.11E-05$\pm$5.1E-06 & 5.21E-05$\pm$2.1E-06 & 9.28E-05$\pm$3.7E-06 & 8.22E-05$\pm$5.4E-06 & 1.02E-04$\pm$1.9E-05\\ 
  10con &  16:11:39.154 &  +00:52:31.87 & 2.36E-04$\pm$2.2E-05 & 6.06E-04$\pm$1.4E-05 & 1.61E-03$\pm$3.5E-05 & 3.58E-03$\pm$7.5E-05 & 2.85E-03$\pm$1.1E-04\\ 
  10cqh &  16:10:36.992 &  -01:43:01.65 & 5.53E-04$\pm$2.9E-05 & 1.98E-03$\pm$4.1E-05 & 3.91E-03$\pm$7.9E-05 & 5.67E-03$\pm$1.2E-04 & 7.45E-03$\pm$2.3E-04\\ 
  10cwx &  12:33:16.405 &  -00:03:12.34 & 4.72E-05$\pm$4.0E-06 & 9.72E-05$\pm$2.6E-06 & 1.23E-04$\pm$3.7E-06 & 1.60E-04$\pm$5.3E-06 & 1.73E-04$\pm$1.7E-05\\ 
  10cxq &  13:48:19.317 &  +13:28:57.27 & 1.03E-04$\pm$6.6E-06 & 2.23E-04$\pm$4.9E-06 & 2.77E-04$\pm$6.4E-06 & 3.22E-04$\pm$8.3E-06 & 3.25E-04$\pm$2.1E-05\\ 
  10cxx &  14:47:27.701 &  +01:55:05.28 & 2.79E-04$\pm$1.5E-05 & 9.62E-04$\pm$2.0E-05 & 1.81E-03$\pm$3.7E-05 & 2.48E-03$\pm$5.1E-05 & 3.31E-03$\pm$1.1E-04\\ 
  10czn &  14:51:17.242 &  +15:26:46.79 & 8.36E-04$\pm$4.4E-05 & 2.63E-03$\pm$5.4E-05 & 4.19E-03$\pm$8.5E-05 & 5.38E-03$\pm$1.1E-04 & 6.08E-03$\pm$2.0E-04\\ 
   10dk &  05:08:21.597 &  +00:12:42.28 & 1.11E-06$\pm$2.0E-06 & 5.16E-06$\pm$6.5E-07 & 9.23E-06$\pm$1.1E-06 & 9.73E-06$\pm$1.7E-06 & 9.92E-06$\pm$7.6E-06\\ 
  10dvb &  17:16:10.672 &  +31:47:32.32 & 1.81E-03$\pm$8.8E-05 & 4.49E-03$\pm$9.2E-05 & 7.00E-03$\pm$1.4E-04 & 8.64E-03$\pm$1.8E-04 & 1.06E-02$\pm$3.2E-04\\ 
   10hv &  14:03:56.535 &  +54:27:27.21 & 1.51E-04$\pm$1.0E-05 & 4.23E-04$\pm$9.2E-06 & 5.92E-04$\pm$1.3E-05 & 7.55E-04$\pm$1.7E-05 & 8.11E-04$\pm$3.7E-05\\ 
   10in &  07:50:00.984 &  +33:06:27.99 & 2.64E-06$\pm$1.8E-06 & 7.66E-06$\pm$7.3E-07 & 9.70E-06$\pm$1.3E-06 & 1.63E-05$\pm$1.8E-06 & 1.16E-05$\pm$6.3E-06\\ 
    10s &  10:37:16.292 &  +38:06:23.57 & 8.95E-05$\pm$6.5E-06 & 2.37E-04$\pm$5.1E-06 & 3.67E-04$\pm$7.9E-06 & 4.55E-04$\pm$1.0E-05 & 5.21E-04$\pm$2.1E-05\\ 
   10ts &  12:33:55.888 &  +13:55:08.31 & 1.70E-04$\pm$1.0E-05 & 4.53E-04$\pm$9.9E-06 & 7.26E-04$\pm$1.5E-05 & 9.44E-04$\pm$2.1E-05 & 1.06E-03$\pm$4.8E-05\\ 
    10u &  10:09:58.780 &  +46:00:33.68 & 6.48E-07$\pm$1.1E-06 & 1.62E-06$\pm$4.8E-07 & 6.87E-06$\pm$7.7E-07 & 9.58E-06$\pm$1.3E-06 & 1.43E-05$\pm$4.4E-06
\enddata
\tablenotetext{a}{3$\sigma$ upper limits for the fluxes of PTF09be and
PTF09gyp are calculated as 3$\sigma_{\rm sky}\sqrt{\pi r^2}$, where $r$ is set to 5 kpc (projected) at the distance of the supernova.}
\end{deluxetable*}
%-------------------------------------------------------

\clearpage
\LongTables
%-------------------------------------------------------
\begin{deluxetable*}{llrrrrrrrrrr}
\tablecolumns{12}
\tabletypesize{\scriptsize}
\tablewidth{0pt}
\tablecaption{Host properties derived from SDSS photometry\label{table:absmags}}
\tablehead{
   \colhead{SN}                        &
   \colhead{type}                      &
   \colhead{SN RA}                     &
   \colhead{SN Dec}                    &
   \colhead{$z$}                       &
   \colhead{$\mu$\tablenotemark{a}}    &
   \colhead{M$_u$\tablenotemark{b}}    &
   \colhead{M$_g$\tablenotemark{b}}    &
   \colhead{M$_r$\tablenotemark{b}}    &
   \colhead{M$_i$\tablenotemark{b}}    &
   \colhead{M$_z$\tablenotemark{b}}    &
   \colhead{M$_{\rm B}$\tablenotemark{b}} \\
   \colhead{(PTF)}                     &
   \colhead{}                          &
   \colhead{(J2000.0)}                 &
   \colhead{(J2000.0)}                 &
   \colhead{}                          &
   \colhead{}                          &
   \colhead{}                          &
   \colhead{}                          &
   \colhead{}                          &
   \colhead{}                          &
   \colhead{}                           
}
\startdata
  09aux &  Ic/Ia & 16:09:15.84 &  +29:17:36.7 & 0.047 & 36.60 & $-18.43$ & $-19.85$ & $-20.49$ & $-20.79$ & $-21.03$ & $-19.48$ \\ 
  09awk &     Ib & 13:37:56.36 &  +22:55:04.8 & 0.062 & 37.22 & $-18.37$ & $-19.18$ & $-19.50$ & $-19.72$ & $-19.84$ & $-18.90$ \\ 
  09axc &     II & 14:53:13.06 &  +22:14:32.2 & 0.115 & 38.64 & $-18.70$ & $-20.07$ & $-20.59$ & $-20.80$ & $-20.94$ & $-19.75$ \\ 
  09axi &     II & 14:12:40.82 &  +31:04:03.3 & 0.064 & 37.29 & $-17.55$ & $-18.60$ & $-19.07$ & $-19.11$ & $-19.02$ & $-18.35$ \\ 
  09bce &     II & 16:35:17.66 &  +55:37:59.1 & 0.023 & 35.01 & $-18.76$ & $-20.22$ & $-20.95$ & $-21.34$ & $-21.62$ & $-19.85$ \\ 
  09bgf &     II & 14:41:38.28 &  +19:21:43.8 & 0.031 & 35.67 & $-16.39$ & $-17.15$ & $-17.38$ & $-17.50$ & $-17.59$ & $-16.91$ \\ 
   09bw &     II & 15:05:02.04 &  +48:40:01.9 & 0.150 & 39.27 & $-16.47$ & $-18.58$ & $-19.11$ & $-19.32$ & $-19.32$ & $-18.25$ \\ 
  09cjq &     II & 21:16:28.48 &  -00:49:39.7 & 0.019 & 34.58 & $-19.76$ & $-21.03$ & $-21.70$ & $-22.01$ & $-22.29$ & $-20.67$ \\ 
   09ct &     II & 11:42:13.80 &  +10:38:53.9 & 0.150 & 39.27 & $-18.35$ & $-19.38$ & $-19.90$ & $-20.17$ & $-20.42$ & $-19.07$ \\ 
   09cu &     II & 13:15:23.14 &  +46:25:08.6 & 0.057 & 37.03 & $-19.80$ & $-20.82$ & $-21.33$ & $-21.61$ & $-21.81$ & $-20.51$ \\ 
  09cvi &     II & 21:47:09.80 &  +08:18:35.6 & 0.030 & 35.59 & $ 99.99$ & $-13.51$ & $-13.56$ & $-13.58$ & $-14.27$ & $-13.18$ \\ 
  09dah &    IIb & 22:45:17.05 &  +21:49:15.2 & 0.024 & 35.08 & $-17.17$ & $-17.85$ & $-18.12$ & $-18.35$ & $-18.48$ & $-17.60$ \\ 
  09dfk &     Ib & 23:09:13.42 &  +07:48:15.4 & 0.016 & 34.21 & $-15.90$ & $-16.89$ & $-17.29$ & $-17.51$ & $-17.68$ & $-16.60$ \\ 
  09djl &     II & 16:33:55.94 &  +30:14:16.3 & 0.184 & 39.75 & $-17.92$ & $-19.52$ & $-20.20$ & $-20.39$ & $-20.79$ & $-19.20$ \\ 
  09dra &     II & 15:48:11.47 &  +41:13:28.2 & 0.077 & 37.71 & $-19.98$ & $-20.93$ & $-21.41$ & $-21.68$ & $-21.91$ & $-20.63$ \\ 
  09due &     II & 16:26:52.36 &  +51:33:23.9 & 0.029 & 35.52 & $-20.23$ & $-21.18$ & $-21.62$ & $-21.83$ & $-22.03$ & $-20.90$ \\ 
  09dxv &    IIb & 23:08:34.73 &  +18:56:13.7 & 0.033 & 35.81 & $-19.38$ & $-20.24$ & $-20.61$ & $-20.86$ & $-21.04$ & $-19.97$ \\ 
  09dzt &     Ic & 16:03:04.20 &  +21:01:47.2 & 0.087 & 38.00 & $-19.62$ & $-20.30$ & $-20.72$ & $-20.85$ & $-21.09$ & $-20.05$ \\ 
  09ebq &     II & 00:14:01.69 &  +29:25:58.5 & 0.024 & 35.05 & $-18.32$ & $-19.23$ & $-19.74$ & $-19.99$ & $-20.23$ & $-18.93$ \\ 
  09ecm &     II & 01:06:43.16 &  -06:22:40.9 & 0.029 & 35.48 & $-18.82$ & $-19.74$ & $-20.15$ & $-20.27$ & $-20.45$ & $-19.47$ \\ 
  09ejz &  Ic/Ia & 00:55:07.29 &  -06:57:05.4 & 0.110 & 38.54 & $-19.72$ & $-20.92$ & $-21.59$ & $-21.88$ & $-22.13$ & $-20.57$ \\ 
  09fae &    IIb & 17:26:20.33 &  +72:56:30.6 & 0.067 & 37.40 & $-17.07$ & $-17.52$ & $-17.59$ & $-17.68$ & $-18.29$ & $-17.29$ \\ 
  09fbf &     II & 21:20:38.44 &  +01:02:52.9 & 0.021 & 34.80 & $-19.36$ & $-20.30$ & $-20.69$ & $-20.90$ & $-21.08$ & $-20.02$ \\ 
  09fmk &     II & 23:57:46.19 &  +11:58:45.3 & 0.063 & 37.26 & $-19.49$ & $-20.40$ & $-20.95$ & $-21.24$ & $-21.41$ & $-20.08$ \\ 
  09foy &     II & 23:17:10.58 &  +17:15:03.2 & 0.060 & 37.15 & $-19.39$ & $-20.41$ & $-20.89$ & $-21.16$ & $-21.28$ & $-20.11$ \\ 
  09fqa &     II & 22:25:32.33 &  +18:59:41.4 & 0.030 & 35.59 & $-18.66$ & $-19.69$ & $-20.15$ & $-20.38$ & $-20.52$ & $-19.39$ \\ 
  09fsr &     Ib & 23:04:52.98 &  +12:19:59.0 & 0.008 & 32.67 & $-19.89$ & $-21.19$ & $-21.81$ & $-22.09$ & $-22.25$ & $-20.84$ \\ 
    09g &     II & 15:16:31.48 &  +54:27:34.7 & 0.040 & 36.23 & $-19.07$ & $-19.90$ & $-20.25$ & $-20.41$ & $-20.54$ & $-19.65$ \\ 
  09gof &     II & 01:22:25.60 &  +03:38:08.4 & 0.103 & 38.38 & $-19.56$ & $-20.43$ & $-20.85$ & $-20.97$ & $-21.06$ & $-20.18$ \\ 
  09gtt &     II & 02:20:37.70 &  +02:24:13.2 & 0.041 & 36.29 & $-16.92$ & $-17.92$ & $-18.14$ & $-18.56$ & $-18.59$ & $-17.62$ \\ 
  09hdo &     II & 00:15:23.20 &  +30:43:19.3 & 0.047 & 36.60 & $-19.83$ & $-21.06$ & $-21.69$ & $-22.02$ & $-22.29$ & $-20.71$ \\ 
  09hzg &     II & 11:50:57.74 &  +21:11:49.4 & 0.028 & 35.44 & $-18.21$ & $-19.62$ & $-20.37$ & $-20.77$ & $-21.12$ & $-19.25$ \\ 
  09iex &     II & 12:02:46.86 &  +02:24:06.8 & 0.020 & 34.70 & $-15.79$ & $-16.40$ & $-16.60$ & $-16.74$ & $-17.03$ & $-16.16$ \\ 
  09ige &     II & 08:55:34.24 &  +32:39:57.0 & 0.064 & 37.29 & $-19.07$ & $-19.91$ & $-20.22$ & $-20.43$ & $-20.44$ & $-19.66$ \\ 
  09igz &     II & 08:53:56.70 &  +33:40:11.5 & 0.086 & 37.97 & $-18.09$ & $-18.89$ & $-19.21$ & $-19.38$ & $-19.49$ & $-18.63$ \\ 
  09ism &     II & 11:44:35.87 &  +10:12:43.7 & 0.029 & 35.52 & $-17.82$ & $-18.62$ & $-18.95$ & $-19.12$ & $-19.15$ & $-18.37$ \\ 
   09ps &     Ic & 16:14:08.62 &  +55:41:41.4 & 0.106 & 38.46 & $-18.09$ & $-18.79$ & $-19.11$ & $-19.25$ & $-19.14$ & $-18.54$ \\ 
    09q &     Ic & 12:24:50.11 &  +08:25:58.8 & 0.090 & 38.07 & $-19.92$ & $-21.07$ & $-21.67$ & $-21.98$ & $-22.21$ & $-20.73$ \\ 
    09r &     II & 14:18:58.63 &  +35:23:16.0 & 0.027 & 35.36 & $-15.63$ & $-16.90$ & $-17.53$ & $-17.80$ & $-18.00$ & $-16.57$ \\ 
   09sh &     II & 16:13:58.08 &  +39:31:58.1 & 0.038 & 36.10 & $-19.36$ & $-20.18$ & $-20.68$ & $-20.92$ & $-20.86$ & $-19.92$ \\ 
   09sk &  Ic-BL & 13:30:51.15 &  +30:20:04.9 & 0.035 & 35.97 & $-17.38$ & $-18.13$ & $-18.36$ & $-18.55$ & $-18.62$ & $-17.88$ \\ 
    09t &     II & 14:15:43.29 &  +16:11:59.1 & 0.039 & 36.18 & $-19.30$ & $-20.08$ & $-20.38$ & $-20.54$ & $-20.63$ & $-19.84$ \\ 
   09tm &     II & 13:46:55.94 &  +61:33:15.6 & 0.035 & 35.94 & $-18.78$ & $-20.05$ & $-20.69$ & $-21.03$ & $-21.30$ & $-19.69$ \\ 
   09uj &     II & 14:20:11.15 &  +53:33:41.0 & 0.065 & 37.33 & $-18.80$ & $-19.68$ & $-20.07$ & $-20.25$ & $-20.28$ & $-19.42$ \\ 
  10bau &     II & 09:16:21.29 &  +17:43:40.2 & 0.026 & 35.28 & $-20.08$ & $-21.00$ & $-21.50$ & $-21.71$ & $-21.92$ & $-20.71$ \\ 
  10bfz &  Ic-BL & 12:54:41.27 &  +15:24:17.0 & 0.150 & 39.27 & $-16.29$ & $-17.09$ & $-17.34$ & $-16.87$ & $-15.52$ & $-16.82$ \\ 
  10bgl &     II & 10:19:04.70 &  +46:27:23.3 & 0.030 & 35.59 & $-20.67$ & $-21.62$ & $-22.00$ & $-22.20$ & $-22.33$ & $-21.34$ \\ 
  10bhu &     Ic & 12:55:28.44 &  +53:34:28.7 & 0.036 & 36.00 & $-18.12$ & $-19.03$ & $-19.42$ & $-19.60$ & $-19.79$ & $-18.75$ \\ 
  10bip &     Ic & 12:34:10.52 &  +08:21:48.5 & 0.051 & 36.78 & $-16.72$ & $-17.87$ & $-18.10$ & $-18.32$ & $-18.51$ & $-17.56$ \\ 
  10bzf &  Ic-BL & 11:44:02.99 &  +55:41:27.6 & 0.050 & 36.73 & $-16.81$ & $-17.21$ & $-17.77$ & $-17.94$ & $-18.22$ & $-16.97$ \\ 
   10cd &     II & 03:00:32.93 &  +36:15:25.4 & 0.045 & 36.52 & $-17.11$ & $-17.71$ & $-18.13$ & $-17.81$ & $-17.94$ & $-17.51$ \\ 
  10con &     II & 16:11:39.09 &  +00:52:33.3 & 0.033 & 35.81 & $-18.55$ & $-19.41$ & $-20.35$ & $-21.07$ & $-20.77$ & $-19.08$ \\ 
  10cqh &     II & 16:10:37.60 &  -01:43:00.7 & 0.041 & 36.29 & $-20.13$ & $-21.25$ & $-21.81$ & $-22.09$ & $-22.31$ & $-20.93$ \\ 
  10cwx &     II & 12:33:16.53 &  -00:03:10.6 & 0.073 & 37.59 & $-18.12$ & $-18.79$ & $-19.05$ & $-19.18$ & $-19.32$ & $-18.53$ \\ 
  10cxq &     II & 13:48:19.32 &  +13:28:58.8 & 0.047 & 36.60 & $-17.96$ & $-18.67$ & $-18.86$ & $-18.98$ & $-18.99$ & $-18.43$ \\ 
  10cxx &     II & 14:47:27.78 &  +01:55:03.8 & 0.034 & 35.87 & $-18.40$ & $-19.65$ & $-20.26$ & $-20.56$ & $-20.86$ & $-19.30$ \\ 
  10czn &     II & 14:51:16.23 &  +15:26:43.6 & 0.045 & 36.50 & $-20.23$ & $-21.33$ & $-21.78$ & $-22.00$ & $-22.13$ & $-21.02$ \\ 
   10dk &     II & 05:08:21.54 &  +00:12:42.9 & 0.074 & 37.62 & $-14.43$ & $-15.89$ & $-16.37$ & $-16.35$ & $-16.34$ & $-15.63$ \\ 
  10dvb &     II & 17:16:12.25 &  +31:47:36.0 & 0.023 & 35.00 & $-19.50$ & $-20.40$ & $-20.83$ & $-21.02$ & $-21.22$ & $-20.11$ \\ 
   10hv &     II & 14:03:56.18 &  +54:27:31.1 & 0.052 & 36.81 & $-18.56$ & $-19.56$ & $-19.88$ & $-20.12$ & $-20.21$ & $-19.27$ \\ 
   10in &    IIb & 07:50:01.24 &  +33:06:23.8 & 0.070 & 37.50 & $-15.08$ & $-16.08$ & $-16.27$ & $-16.73$ & $-16.36$ & $-15.80$ \\ 
    10s &     II & 10:37:16.30 &  +38:06:23.2 & 0.051 & 36.78 & $-18.00$ & $-18.94$ & $-19.36$ & $-19.56$ & $-19.71$ & $-18.66$ \\ 
   10ts &     II & 12:33:56.40 &  +13:55:08.3 & 0.046 & 36.55 & $-18.52$ & $-19.47$ & $-19.92$ & $-20.16$ & $-20.28$ & $-19.18$ \\ 
    10u &     II & 10:09:58.42 &  +46:00:35.2 & 0.150 & 39.27 & $-15.69$ & $-16.48$ & $-17.68$ & $-17.96$ & $-18.38$ & $-16.38$ 
\enddata
\tablenotetext{a}{Assuming $H_0 = 70$, $\Omega_m = 0.3$, and $\Omega_\Lambda=0.7$}
\tablenotetext{b}{Magnitudes are corrected for galactic extinction but not
intrinsic extinction.}
\end{deluxetable*}
%-------------------------------------------------------

\clearpage
\LongTables
%-------------------------------------------------------
\begin{deluxetable*}{llrrrrrr}
\tablecolumns{8}
\tabletypesize{\scriptsize}
\tablewidth{0pt}
\tablecaption{Host galaxy properties fit from SDSS photometry\label{table:FAST}}
\tablehead{
   \colhead{SN Name}              &
   \colhead{type}                 &
   \colhead{M}                    &
   \colhead{SFR (FAST)}           &
   \colhead{SFR (u)}              &
   \colhead{SSFR}                 &
   \colhead{Age}                  \\
   \colhead{}                     &
   \colhead{}                     &
   \colhead{$\log$ M$_\odot$}     &
   \colhead{$\log$ M$_\odot /$yr} &
   \colhead{$\log$ M$_\odot /$yr} &
   \colhead{$\log$ yr$^{-1}$}     &
   \colhead{$\log$ yr}
}
\startdata
  PTF09aux &  Ic/Ia & $10.43_{-0.07}^{+0.03}$ & $ -0.59_{-0.00}^{+0.43}$ & $-0.06$ & $-11.03_{-0.00}^{+0.45}$ & $  9.70_{-0.16}^{+0.02}$ \\ 
  PTF09awk &     Ib & $ 9.45_{-0.14}^{+0.18}$ & $  0.61_{-0.75}^{+0.35}$ & $ 0.44$ & $ -8.85_{-0.83}^{+0.47}$ & $  8.80_{-0.42}^{+0.56}$ \\ 
  PTF09axc &     II & $10.33_{-0.09}^{+0.01}$ & $ -0.25_{-0.00}^{+0.33}$ & $ 0.04$ & $-10.58_{-0.00}^{+0.36}$ & $  9.60_{-0.14}^{+0.01}$ \\ 
  PTF09axi &     II & $ 9.35_{-0.04}^{+0.09}$ & $ -0.33_{-0.16}^{+0.03}$ & $-0.31$ & $ -9.68_{-0.24}^{+0.00}$ & $  9.30_{-0.06}^{+0.11}$ \\ 
  PTF09bce &     II & $10.68_{-0.09}^{+0.08}$ & $  0.11_{-0.95}^{+0.83}$ & $ 0.36$ & $-10.58_{-1.01}^{+0.90}$ & $  9.60_{-0.37}^{+0.20}$ \\ 
  PTF09bgf &     II & $ 8.59_{-0.19}^{+0.06}$ & $ -0.88_{-0.00}^{+0.74}$ & $-0.80$ & $ -9.47_{-0.00}^{+0.87}$ & $  9.20_{-0.68}^{+0.06}$ \\ 
   PTF09bw &     II & $ 9.75_{-0.15}^{+0.02}$ & $ -0.83_{-0.00}^{+0.34}$ & $-0.77$ & $-10.58_{-0.00}^{+0.36}$ & $  9.60_{-0.19}^{+0.02}$ \\ 
  PTF09cjq &     II & $10.84_{-0.14}^{+0.05}$ & $  0.62_{-0.37}^{+0.98}$ & $ 0.71$ & $-10.22_{-0.36}^{+1.09}$ & $  9.50_{-0.54}^{+0.15}$ \\ 
   PTF09ct &     II & $10.00_{-0.19}^{+0.10}$ & $ -0.22_{-0.27}^{+1.46}$ & $-0.00$ & $-10.22_{-0.36}^{+1.62}$ & $  9.50_{-0.95}^{+0.10}$ \\ 
   PTF09cu &     II & $10.48_{-0.19}^{+0.09}$ & $  0.81_{-0.51}^{+0.89}$ & $ 0.77$ & $ -9.68_{-0.54}^{+1.08}$ & $  9.30_{-0.71}^{+0.22}$ \\ 
  PTF09cvi &     II & $ 7.07_{-0.15}^{+0.17}$ & $<-2.06_{-0.42}^{+0.38}$ & $-1.63$ & $ -9.13_{-0.55}^{+0.41}$ & $  9.00_{-0.34}^{+0.34}$ \\ 
  PTF09dah &    IIb & $ 8.50_{-0.13}^{+0.38}$ & $  1.70_{-1.72}^{+0.02}$ & $ 0.83$ & $ -6.80_{-2.05}^{+0.00}$ & $  6.80_{-0.08}^{+2.10}$ \\ 
  PTF09dfk &     Ib & $ 8.77_{-0.14}^{+0.07}$ & $ -0.91_{-0.24}^{+0.72}$ & $-0.80$ & $ -9.68_{-0.24}^{+0.83}$ & $  9.30_{-0.52}^{+0.13}$ \\ 
  PTF09djl &     II & $10.22_{-0.06}^{+0.09}$ & $ -0.35_{-0.36}^{+0.06}$ & $-0.16$ & $-10.58_{-0.45}^{+0.00}$ & $  9.60_{-0.09}^{+0.10}$ \\ 
  PTF09dra &     II & $10.34_{-0.14}^{+0.22}$ & $  1.62_{-1.28}^{+0.32}$ & $ 1.33$ & $ -8.72_{-1.50}^{+0.45}$ & $  8.70_{-0.41}^{+0.80}$ \\ 
  PTF09due &     II & $10.51_{-0.18}^{+0.07}$ & $  0.83_{-0.25}^{+0.82}$ & $ 0.79$ & $ -9.68_{-0.24}^{+0.96}$ & $  9.30_{-0.64}^{+0.14}$ \\ 
  PTF09dxv &    IIb & $ 9.93_{-0.11}^{+0.19}$ & $  1.20_{-1.01}^{+0.26}$ & $ 0.95$ & $ -8.72_{-1.20}^{+0.34}$ & $  8.70_{-0.31}^{+0.70}$ \\ 
  PTF09dzt &     Ic & $ 9.48_{-0.10}^{+0.36}$ & $  2.58_{-1.24}^{+0.13}$ & $ 1.70$ & $ -6.89_{-1.60}^{+0.09}$ & $  6.90_{-0.12}^{+1.61}$ \\ 
  PTF09ebq &     II & $ 9.59_{-0.15}^{+0.24}$ & $  0.99_{-1.11}^{+0.30}$ & $ 0.74$ & $ -8.60_{-1.32}^{+0.43}$ & $  8.60_{-0.41}^{+0.81}$ \\ 
  PTF09ecm &     II & $ 9.83_{-0.09}^{+0.06}$ & $  0.15_{-0.19}^{+0.50}$ & $ 0.18$ & $ -9.68_{-0.24}^{+0.55}$ & $  9.30_{-0.36}^{+0.11}$ \\ 
  PTF09ejz &  Ic/Ia & $10.79_{-0.08}^{+0.05}$ & $  0.22_{-0.06}^{+0.85}$ & $ 0.50$ & $-10.58_{-0.00}^{+0.90}$ & $  9.60_{-0.33}^{+0.05}$ \\ 
  PTF09fae &    IIb & $ 8.03_{-0.13}^{+0.23}$ & $  0.56_{-0.41}^{+0.33}$ & $ 0.06$ & $ -7.47_{-0.59}^{+0.38}$ & $  7.50_{-0.42}^{+0.63}$ \\ 
  PTF09fbf &     II & $10.08_{-0.13}^{+0.10}$ & $  0.61_{-0.40}^{+0.63}$ & $ 0.54$ & $ -9.47_{-0.45}^{+0.75}$ & $  9.20_{-0.51}^{+0.22}$ \\ 
  PTF09fmk &     II & $10.12_{-0.16}^{+0.23}$ & $  1.52_{-1.14}^{+0.31}$ & $ 1.22$ & $ -8.60_{-1.32}^{+0.43}$ & $  8.60_{-0.44}^{+0.84}$ \\ 
  PTF09foy &     II & $10.30_{-0.11}^{+0.07}$ & $  0.38_{-0.26}^{+0.70}$ & $ 0.44$ & $ -9.92_{-0.30}^{+0.79}$ & $  9.40_{-0.40}^{+0.11}$ \\ 
  PTF09fqa &     II & $ 9.97_{-0.10}^{+0.07}$ & $  0.05_{-0.23}^{+0.56}$ & $ 0.12$ & $ -9.92_{-0.30}^{+0.63}$ & $  9.40_{-0.36}^{+0.10}$ \\ 
  PTF09fsr &     Ib & $10.85_{-0.02}^{+0.03}$ & $  0.27_{-0.02}^{+0.03}$ & $ 0.50$ & $-10.58_{-0.00}^{+0.00}$ & $  9.60_{-0.08}^{+0.01}$ \\ 
    PTF09g &     II & $ 9.85_{-0.16}^{+0.05}$ & $  0.17_{-0.02}^{+0.82}$ & $ 0.19$ & $ -9.68_{-0.00}^{+0.96}$ & $  9.30_{-0.61}^{+0.04}$ \\ 
  PTF09gof &     II & $10.17_{-0.10}^{+0.03}$ & $  0.25_{-0.00}^{+0.40}$ & $ 0.35$ & $ -9.92_{-0.00}^{+0.45}$ & $  9.40_{-0.25}^{+0.01}$ \\ 
  PTF09gtt &     II & $ 9.03_{-0.20}^{+0.13}$ & $  0.18_{-0.73}^{+0.43}$ & $ 0.03$ & $ -8.85_{-0.83}^{+0.58}$ & $  8.80_{-0.55}^{+0.51}$ \\ 
  PTF09hdo &     II & $10.80_{-0.14}^{+0.07}$ & $  0.88_{-0.65}^{+0.83}$ & $ 0.87$ & $ -9.92_{-0.66}^{+0.94}$ & $  9.40_{-0.58}^{+0.22}$ \\ 
  PTF09hzg &     II & $10.42_{-0.19}^{+0.09}$ & $  0.50_{-0.65}^{+1.02}$ & $ 0.51$ & $ -9.92_{-0.66}^{+1.20}$ & $  9.40_{-0.70}^{+0.28}$ \\ 
  PTF09iex &     II & $ 8.00_{-0.41}^{+0.21}$ & $ -0.27_{-0.53}^{+1.67}$ & $-0.48$ & $ -8.27_{-0.71}^{+1.67}$ & $  8.30_{-1.71}^{+0.61}$ \\ 
  PTF09ige &     II & $ 9.83_{-0.09}^{+0.04}$ & $  0.15_{-0.01}^{+0.50}$ & $ 0.18$ & $ -9.68_{-0.00}^{+0.55}$ & $  9.30_{-0.32}^{+0.03}$ \\ 
  PTF09igz &     II & $ 9.45_{-0.58}^{+0.10}$ & $ -0.23_{-0.15}^{+2.35}$ & $-0.14$ & $ -9.68_{-0.24}^{+2.88}$ & $  9.30_{-2.51}^{+0.10}$ \\ 
  PTF09ism &     II & $ 9.26_{-0.14}^{+0.09}$ & $ -0.21_{-0.15}^{+0.65}$ & $-0.19$ & $ -9.47_{-0.21}^{+0.75}$ & $  9.20_{-0.54}^{+0.13}$ \\ 
   PTF09ps &     Ic & $ 8.86_{-0.11}^{+0.54}$ & $  2.06_{-2.37}^{+0.15}$ & $ 1.16$ & $ -6.80_{-2.88}^{+0.10}$ & $  6.80_{-0.12}^{+2.52}$ \\ 
    PTF09q &     Ic & $10.77_{-0.14}^{+0.05}$ & $  0.55_{-0.35}^{+1.00}$ & $ 0.67$ & $-10.22_{-0.36}^{+1.09}$ & $  9.50_{-0.55}^{+0.11}$ \\ 
    PTF09r &     II & $ 9.14_{-0.09}^{+0.05}$ & $ -1.44_{-0.03}^{+0.82}$ & $-1.03$ & $-10.58_{-0.00}^{+0.90}$ & $  9.60_{-0.30}^{+0.05}$ \\ 
   PTF09sh &     II & $10.05_{-0.07}^{+0.08}$ & $  0.37_{-0.18}^{+0.35}$ & $ 0.37$ & $ -9.68_{-0.24}^{+0.39}$ & $  9.30_{-0.24}^{+0.11}$ \\ 
   PTF09sk &  Ic-BL & $ 8.99_{-0.42}^{+0.08}$ & $ -0.30_{-0.17}^{+2.10}$ & $-0.31$ & $ -9.29_{-0.18}^{+2.49}$ & $  9.10_{-2.31}^{+0.17}$ \\ 
    PTF09t &     II & $ 9.83_{-0.14}^{+0.08}$ & $  0.36_{-0.15}^{+0.64}$ & $ 0.33$ & $ -9.47_{-0.21}^{+0.75}$ & $  9.20_{-0.54}^{+0.11}$ \\ 
   PTF09tm &     II & $10.42_{-0.15}^{+0.08}$ & $  0.50_{-0.66}^{+0.83}$ & $ 0.52$ & $ -9.92_{-0.66}^{+0.94}$ & $  9.40_{-0.57}^{+0.25}$ \\ 
   PTF09uj &     II & $ 9.79_{-0.05}^{+0.08}$ & $  0.12_{-0.17}^{+0.21}$ & $ 0.15$ & $ -9.68_{-0.24}^{+0.21}$ & $  9.30_{-0.19}^{+0.10}$ \\ 
  PTF10bau &     II & $10.47_{-0.24}^{+0.07}$ & $  0.79_{-0.25}^{+0.99}$ & $ 0.75$ & $ -9.68_{-0.24}^{+1.19}$ & $  9.30_{-0.82}^{+0.16}$ \\ 
  PTF10bfz &  Ic-BL & $ 8.34_{-0.53}^{+0.20}$ & $ -0.79_{-0.14}^{+3.08}$ & $-0.82$ & $ -9.13_{-0.34}^{+3.03}$ & $  9.00_{-3.00}^{+0.21}$ \\ 
  PTF10bgl &     II & $10.61_{-0.11}^{+0.07}$ & $  0.93_{-0.20}^{+0.62}$ & $ 0.87$ & $ -9.68_{-0.24}^{+0.70}$ & $  9.30_{-0.45}^{+0.11}$ \\ 
  PTF10bhu &     Ic & $ 9.59_{-0.18}^{+0.07}$ & $ -0.09_{-0.21}^{+0.82}$ & $-0.04$ & $ -9.68_{-0.24}^{+0.96}$ & $  9.30_{-0.64}^{+0.12}$ \\ 
  PTF10bip &     Ic & $ 9.01_{-0.08}^{+0.13}$ & $ -0.12_{-0.69}^{+0.24}$ & $-0.23$ & $ -9.13_{-0.79}^{+0.28}$ & $  9.00_{-0.23}^{+0.41}$ \\ 
  PTF10bzf &  Ic-BL & $ 8.23_{-0.07}^{+0.25}$ & $  1.14_{-0.58}^{+0.29}$ & $ 0.52$ & $ -7.09_{-0.77}^{+0.20}$ & $  7.10_{-0.21}^{+0.89}$ \\ 
   PTF10cd &     II & $ 8.61_{-0.32}^{+0.10}$ & $ -0.52_{-0.08}^{+2.17}$ & $-0.53$ & $ -9.13_{-0.16}^{+2.43}$ & $  9.00_{-2.32}^{+0.12}$ \\ 
  PTF10con &     II & $10.29_{-0.49}^{+0.07}$ & $  0.36_{-0.62}^{+1.42}$ & $ 0.51$ & $ -9.92_{-0.66}^{+1.86}$ & $  9.40_{-1.34}^{+0.21}$ \\ 
  PTF10cqh &     II & $10.76_{-0.15}^{+0.06}$ & $  0.83_{-0.31}^{+0.83}$ & $ 0.84$ & $ -9.92_{-0.30}^{+0.94}$ & $  9.40_{-0.53}^{+0.16}$ \\ 
  PTF10cwx &     II & $ 9.10_{-0.41}^{+0.17}$ & $  0.61_{-0.83}^{+1.59}$ & $ 0.38$ & $ -8.49_{-0.98}^{+1.79}$ & $  8.50_{-1.83}^{+0.70}$ \\ 
  PTF10cxq &     II & $ 9.10_{-0.12}^{+0.09}$ & $ -0.18_{-0.11}^{+0.56}$ & $-0.20$ & $ -9.29_{-0.18}^{+0.69}$ & $  9.10_{-0.50}^{+0.11}$ \\ 
  PTF10cxx &     II & $10.22_{-0.14}^{+0.08}$ & $  0.30_{-0.66}^{+0.82}$ & $ 0.34$ & $ -9.92_{-0.66}^{+0.94}$ & $  9.40_{-0.53}^{+0.22}$ \\ 
  PTF10czn &     II & $10.63_{-0.07}^{+0.07}$ & $  0.71_{-0.24}^{+0.40}$ & $ 0.71$ & $ -9.92_{-0.30}^{+0.45}$ & $  9.40_{-0.23}^{+0.11}$ \\ 
   PTF10dk &     II & $ 8.36_{-0.62}^{+0.17}$ & $ -1.56_{-0.15}^{+2.50}$ & $-1.51$ & $ -9.92_{-0.30}^{+3.12}$ & $  9.40_{-2.60}^{+0.12}$ \\ 
  PTF10dvb &     II & $10.13_{-0.20}^{+0.11}$ & $  0.66_{-0.39}^{+0.80}$ & $ 0.59$ & $ -9.47_{-0.45}^{+0.98}$ & $  9.20_{-0.70}^{+0.22}$ \\ 
   PTF10hv &     II & $ 9.82_{-0.13}^{+0.04}$ & $ -0.11_{-0.00}^{+0.72}$ & $-0.01$ & $ -9.92_{-0.00}^{+0.79}$ & $  9.40_{-0.48}^{+0.01}$ \\ 
   PTF10in &    IIb & $ 8.19_{-0.60}^{+0.29}$ & $ -0.42_{-1.09}^{+1.42}$ & $-0.59$ & $ -8.60_{-1.32}^{+1.80}$ & $  8.60_{-1.88}^{+0.83}$ \\ 
    PTF10s &     II & $ 9.61_{-0.13}^{+0.05}$ & $ -0.31_{-0.02}^{+0.83}$ & $-0.18$ & $ -9.92_{-0.00}^{+0.94}$ & $  9.40_{-0.52}^{+0.03}$ \\ 
   PTF10ts &     II & $ 9.86_{-0.12}^{+0.04}$ & $ -0.06_{-0.05}^{+0.81}$ & $ 0.04$ & $ -9.92_{-0.00}^{+0.94}$ & $  9.40_{-0.50}^{+0.04}$ \\ 
    PTF10u &     II & $ 9.72_{-0.25}^{+0.21}$ & $ -3.47_{-0.13}^{+2.58}$ & $-0.86$ & $-13.18_{-0.00}^{+2.60}$ & $ 10.00_{-0.46}^{+0.00}$ 
\enddata
\end{deluxetable*}
%-------------------------------------------------------

%\clearpage
%\LongTables
%-------------------------------------------------------
\begin{deluxetable*}{lrc}
\tablecolumns{3}
\tabletypesize{\scriptsize}
\tablewidth{0pt}
\tablecaption{Cuts to MPA/JHU comparison sample \label{table:cuts}}
\tablehead{
   \colhead{Criterion}            &
   \colhead{Details}              &
   \colhead{Sample\tablenotemark{a}}
}
\startdata
Within redshift range of sample  & 0.0189995 $<$ z $<$ 0.103       & 382095 \\
Successful stellar mass estimate & Mmed $>$ 2 and not INDEF        & 363881 \\
Successful SFR estimate          & SFRavg != $-99$ and no SFR flag & 362395 \\
12+log(O/H) metallicity estimate & OHmed != $-99.9$                & 131203 \\
N2 flux ratio in valid range     & -2.5 $<$ N2 $<$ -0.3            & \bf{130370}
\enddata
\tablenotetext{a}{Remaining sample shown.  Cuts are sequential.
The sample size of the full MPA/JHU value-added catalog is 927552.}
\end{deluxetable*}
%-------------------------------------------------------

%************************Figures


\begin{thebibliography}{100}
\expandafter\ifx\csname natexlab\endcsname\relax\def\natexlab#1{#1}\fi

\bibitem[{{Abazajian} {et~al.}(2009){Abazajian}, {Adelman-McCarthy},
  {Ag{\"u}eros}, {Allam}, {Allende Prieto}, {An}, {Anderson}, {Anderson},
  {Annis}, {Bahcall}, \& et~al.}]{dr7}
{Abazajian}, K.~N., {et~al.} 2009, \apjs, 182, 543

\bibitem[{{Aihara} {et~al.}(2011){Aihara}, {Allende Prieto}, {An}, {Anderson},
  {Aubourg}, {Balbinot}, {Beers}, {Berlind}, {Bickerton}, {Bizyaev}, {Blanton},
  {Bochanski}, {Bolton}, {Bovy}, {Brandt}, {Brinkmann}, {Brown}, {Brownstein},
  {Busca}, {Campbell}, {Carr}, {Chen}, {Chiappini}, {Comparat}, {Connolly},
  {Cortes}, {Croft}, {Cuesta}, {da Costa}, {Davenport}, {Dawson}, {Dhital},
  {Ealet}, {Ebelke}, {Edmondson}, {Eisenstein}, {Escoffier}, {Esposito},
  {Evans}, {Fan}, {Femen{\'{\i}}a Castell{\'a}}, {Font-Ribera}, {Frinchaboy},
  {Ge}, {Gillespie}, {Gilmore}, {Gonz{\'a}lez Hern{\'a}ndez}, {Gott}, {Gould},
  {Grebel}, {Gunn}, {Hamilton}, {Harding}, {Harris}, {Hawley}, {Hearty}, {Ho},
  {Hogg}, {Holtzman}, {Honscheid}, {Inada}, {Ivans}, {Jiang}, {Johnson},
  {Jordan}, {Jordan}, {Kazin}, {Kirkby}, {Klaene}, {Knapp}, {Kneib},
  {Kochanek}, {Koesterke}, {Kollmeier}, {Kron}, {Lampeitl}, {Lang}, {Le Goff},
  {Lee}, {Lin}, {Long}, {Loomis}, {Lucatello}, {Lundgren}, {Lupton}, {Ma},
  {MacDonald}, {Mahadevan}, {Maia}, {Makler}, {Malanushenko}, {Malanushenko},
  {Mandelbaum}, {Maraston}, {Margala}, {Masters}, {McBride}, {McGehee},
  {McGreer}, {M{\'e}nard}, {Miralda-Escud{\'e}}, {Morrison}, {Mullally},
  {Muna}, {Munn}, {Murayama}, {Myers}, {Naugle}, {Fausti Neto}, {Cuong Nguyen},
  {Nichol}, {O'Connell}, {Ogando}, {Olmstead}, {Oravetz}, {Padmanabhan},
  {Palanque-Delabrouille}, {Pan}, {Pandey}, {P{\^a}ris}, {Percival},
  {Petitjean}, {Pfaffenberger}, {Pforr}, {Phleps}, {Pichon}, {Pieri}, {Prada},
  {Price-Whelan}, {Raddick}, {Ramos}, {Reyl{\'e}}, {Rich}, {Richards}, {Rix},
  {Robin}, {Rocha-Pinto}, {Rockosi}, {Roe}, {Rollinde}, {Ross}, {Ross},
  {Rossetto}, {S{\'a}nchez}, {Sayres}, {Schlegel}, {Schlesinger}, {Schmidt},
  {Schneider}, {Sheldon}, {Shu}, {Simmerer}, {Simmons}, {Sivarani}, {Snedden},
  {Sobeck}, {Steinmetz}, {Strauss}, {Szalay}, {Tanaka}, {Thakar}, {Thomas},
  {Tinker}, {Tofflemire}, {Tojeiro}, {Tremonti}, {Vandenberg}, {Vargas
  Maga{\~n}a}, {Verde}, {Vogt}, {Wake}, {Wang}, {Weaver}, {Weinberg}, {White},
  {White}, {Yanny}, {Yasuda}, {Yeche}, \& {Zehavi}}]{dr8}
{Aihara}, H., {et~al.} 2011, \apjs, 193, 29

\bibitem[{{Anderson} {et~al.}(2010){Anderson}, {Covarrubias}, {James}, {Hamuy},
  \& {Habergham}}]{anderson10}
{Anderson}, J.~P., {Covarrubias}, R.~A., {James}, P.~A., {Hamuy}, M., \&
  {Habergham}, S.~M. 2010, \mnras, 407, 2660

\bibitem[{{Anderson} {et~al.}(2012){Anderson}, {Habergham}, {James}, \&
  {Hamuy}}]{anderson12}
{Anderson}, J.~P., {Habergham}, S.~M., {James}, P.~A., \& {Hamuy}, M. 2012,
  \mnras, 424, 1372

\bibitem[{{Anderson} \& {James}(2008)}]{anderson08}
{Anderson}, J.~P., \& {James}, P.~A. 2008, \mnras, 390, 1527

\bibitem[{{Arcavi} {et~al.}(2010){Arcavi}, {Gal-Yam}, {Kasliwal}, {Quimby},
  {Ofek}, {Kulkarni}, {Nugent}, {Cenko}, {Bloom}, {Sullivan}, {Howell},
  {Poznanski}, {Filippenko}, {Law}, {Hook}, {J{\"o}nsson}, {Blake}, {Cooke},
  {Dekany}, {Rahmer}, {Hale}, {Smith}, {Zolkower}, {Velur}, {Walters},
  {Henning}, {Bui}, {McKenna}, \& {Jacobsen}}]{arcavi10}
{Arcavi}, I., {et~al.} 2010, \apj, 721, 777

\bibitem[{{Asplund} {et~al.}(2009){Asplund}, {Grevesse}, {Sauval}, \&
  {Scott}}]{asplund09}
{Asplund}, M., {Grevesse}, N., {Sauval}, A.~J., \& {Scott}, P. 2009, \araa, 47,
  481

\bibitem[{{Assef} {et~al.}(2008){Assef}, {Kochanek}, {Brodwin}, {Brown},
  {Caldwell}, {Cool}, {Eisenhardt}, {Eisenstein}, {Gonzalez}, {Jannuzi},
  {Jones}, {McKenzie}, {Murray}, \& {Stern}}]{assef08}
{Assef}, R.~J., {et~al.} 2008, \apj, 676, 286

\bibitem[{{Badenes} {et~al.}(2009){Badenes}, {Harris}, {Zaritsky}, \&
  {Prieto}}]{badenes09}
{Badenes}, C., {Harris}, J., {Zaritsky}, D., \& {Prieto}, J.~L. 2009, \apj,
  700, 727

\bibitem[{{Bell} {et~al.}(2003){Bell}, {McIntosh}, {Katz}, \&
  {Weinberg}}]{bell03}
{Bell}, E.~F., {McIntosh}, D.~H., {Katz}, N., \& {Weinberg}, M.~D. 2003, \apjs,
  149, 289

\bibitem[{{Bensby} {et~al.}(2004){Bensby}, {Feltzing}, \&
  {Lundstr{\"o}m}}]{bensby04}
{Bensby}, T., {Feltzing}, S., \& {Lundstr{\"o}m}, I. 2004, \aap, 415, 155

\bibitem[{{Berg} {et~al.}(2011){Berg}, {Skillman}, \& {Marble}}]{berg11}
{Berg}, D.~A., {Skillman}, E.~D., \& {Marble}, A.~R. 2011, \apj, 738, 2

\bibitem[{{Bertin} \& {Arnouts}(1996)}]{sextractor}
{Bertin}, E., \& {Arnouts}, S. 1996, \aaps, 117, 393

\bibitem[{{Blanton} {et~al.}(2003){Blanton}, {Lin}, {Lupton}, {Maley}, {Young},
  {Zehavi}, \& {Loveday}}]{blanton03}
{Blanton}, M.~R., {Lin}, H., {Lupton}, R.~H., {Maley}, F.~M., {Young}, N.,
  {Zehavi}, I., \& {Loveday}, J. 2003, \aj, 125, 2276

\bibitem[{{Bresolin} {et~al.}(2009){Bresolin}, {Gieren}, {Kudritzki},
  {Pietrzy{\'n}ski}, {Urbaneja}, \& {Carraro}}]{bresolin09}
{Bresolin}, F., {Gieren}, W., {Kudritzki}, R., {Pietrzy{\'n}ski}, G.,
  {Urbaneja}, M.~A., \& {Carraro}, G. 2009, \apj, 700, 309

\bibitem[{{Brinchmann} {et~al.}(2004){Brinchmann}, {Charlot}, {White},
  {Tremonti}, {Kauffmann}, {Heckman}, \& {Brinkmann}}]{brinchmann04}
{Brinchmann}, J., {Charlot}, S., {White}, S.~D.~M., {Tremonti}, C.,
  {Kauffmann}, G., {Heckman}, T., \& {Brinkmann}, J. 2004, \mnras, 351, 1151

\bibitem[{{Bruzual} \& {Charlot}(2003)}]{bruzual03}
{Bruzual}, G., \& {Charlot}, S. 2003, \mnras, 344, 1000

\bibitem[{{Campisi} {et~al.}(2011){Campisi}, {Tapparello}, {Salvaterra},
  {Mannucci}, \& {Colpi}}]{campisi11}
{Campisi}, M.~A., {Tapparello}, C., {Salvaterra}, R., {Mannucci}, F., \&
  {Colpi}, M. 2011, \mnras, 417, 1013

\bibitem[{{Chatzopoulos} {et~al.}(2011){Chatzopoulos}, {Wheeler}, {Vinko},
  {Quimby}, {Robinson}, {Miller}, {Foley}, {Perley}, {Yuan}, {Akerlof}, \&
  {Bloom}}]{chatzopoulos11}
{Chatzopoulos}, E., {et~al.} 2011, \apj, 729, 143

\bibitem[{{Chen} {et~al.}(2003){Chen}, {Zhao}, {Nissen}, {Bai}, \&
  {Qiu}}]{chen03}
{Chen}, Y.~Q., {Zhao}, G., {Nissen}, P.~E., {Bai}, G.~S., \& {Qiu}, H.~M. 2003,
  \apj, 591, 925

\bibitem[{{Chomiuk} {et~al.}(2011){Chomiuk}, {Chornock}, {Soderberg}, {Berger},
  {Chevalier}, {Foley}, {Huber}, {Narayan}, {Rest}, {Gezari}, {Kirshner},
  {Riess}, {Rodney}, {Smartt}, {Stubbs}, {Tonry}, {Wood-Vasey}, {Burgett},
  {Chambers}, {Czekala}, {Flewelling}, {Forster}, {Kaiser}, {Kudritzki},
  {Magnier}, {Martin}, {Morgan}, {Neill}, {Price}, {Roth}, {Sanders}, \&
  {Wainscoat}}]{chomiuk11}
{Chomiuk}, L., {et~al.} 2011, \apj, 743, 114

\bibitem[{{Cresci} {et~al.}(2007){Cresci}, {Mannucci}, {Della Valle}, \&
  {Maiolino}}]{cresci07}
{Cresci}, G., {Mannucci}, F., {Della Valle}, M., \& {Maiolino}, R. 2007, \aap,
  462, 927

\bibitem[{{Delahaye} \& {Pinsonneault}(2006)}]{delahaye06}
{Delahaye}, F., \& {Pinsonneault}, M.~H. 2006, \apj, 649, 529

\bibitem[{{Denicol{\'o}} {et~al.}(2002){Denicol{\'o}}, {Terlevich}, \&
  {Terlevich}}]{d02}
{Denicol{\'o}}, G., {Terlevich}, R., \& {Terlevich}, E. 2002, \mnras, 330, 69

\bibitem[{{Eldridge} {et~al.}(2008){Eldridge}, {Izzard}, \&
  {Tout}}]{eldridge08}
{Eldridge}, J.~J., {Izzard}, R.~G., \& {Tout}, C.~A. 2008, \mnras, 384, 1109

\bibitem[{{Epstein} {et~al.}(2010){Epstein}, {Johnson}, {Dong}, {Udalski},
  {Gould}, \& {Becker}}]{epstein10}
{Epstein}, C.~R., {Johnson}, J.~A., {Dong}, S., {Udalski}, A., {Gould}, A., \&
  {Becker}, G. 2010, \apj, 709, 447

\bibitem[{{Fulbright} {et~al.}(2007){Fulbright}, {McWilliam}, \&
  {Rich}}]{fulbright07}
{Fulbright}, J.~P., {McWilliam}, A., \& {Rich}, R.~M. 2007, \apj, 661, 1152

\bibitem[{{Georgy} {et~al.}(2012){Georgy}, {Ekstr{\"o}m}, {Meynet}, {Massey},
  {Levesque}, {Hirschi}, {Eggenberger}, \& {Maeder}}]{georgy12}
{Georgy}, C., {Ekstr{\"o}m}, S., {Meynet}, G., {Massey}, P., {Levesque}, E.~M.,
  {Hirschi}, R., {Eggenberger}, P., \& {Maeder}, A. 2012, arXiv:1203.5243

\bibitem[{{Georgy} {et~al.}(2009){Georgy}, {Meynet}, {Walder}, {Folini}, \&
  {Maeder}}]{georgy09}
{Georgy}, C., {Meynet}, G., {Walder}, R., {Folini}, D., \& {Maeder}, A. 2009,
  \aap, 502, 611

\bibitem[{{Gunn} {et~al.}(2006){Gunn}, {Siegmund}, {Mannery}, {Owen}, {Hull},
  {Leger}, {Carey}, {Knapp}, {York}, {Boroski}, {Kent}, {Lupton}, {Rockosi},
  {Evans}, {Waddell}, {Anderson}, {Annis}, {Barentine}, {Bartoszek}, {Bastian},
  {Bracker}, {Brewington}, {Briegel}, {Brinkmann}, {Brown}, {Carr},
  {Czarapata}, {Drennan}, {Dombeck}, {Federwitz}, {Gillespie}, {Gonzales},
  {Hansen}, {Harvanek}, {Hayes}, {Jordan}, {Kinney}, {Klaene}, {Kleinman},
  {Kron}, {Kresinski}, {Lee}, {Limmongkol}, {Lindenmeyer}, {Long}, {Loomis},
  {McGehee}, {Mantsch}, {Neilsen}, {Neswold}, {Newman}, {Nitta}, {Peoples},
  {Pier}, {Prieto}, {Prosapio}, {Rivetta}, {Schneider}, {Snedden}, \&
  {Wang}}]{gunn06}
{Gunn}, J.~E., {et~al.} 2006, \aj, 131, 2332

\bibitem[{{Han} {et~al.}(2010){Han}, {Hammer}, {Liang}, {Flores}, {Rodrigues},
  {Hou}, \& {Wei}}]{han10}
{Han}, X.~H., {Hammer}, F., {Liang}, Y.~C., {Flores}, H., {Rodrigues}, M.,
  {Hou}, J.~L., \& {Wei}, J.~Y. 2010, \aap, 514, A24

\bibitem[{{Heger} \& {Woosley}(2002)}]{heger02}
{Heger}, A., \& {Woosley}, S.~E. 2002, \apj, 567, 532

\bibitem[{{Israelian} {et~al.}(1998){Israelian}, {Garc{\'{\i}}a L{\'o}pez}, \&
  {Rebolo}}]{israelian98}
{Israelian}, G., {Garc{\'{\i}}a L{\'o}pez}, R.~J., \& {Rebolo}, R. 1998, \apj,
  507, 805

\bibitem[{{Jensen} \& {Snow}(2007)}]{jensen07}
{Jensen}, A.~G., \& {Snow}, T.~P. 2007, \apj, 669, 378

\bibitem[{{Johnson} {et~al.}(2007){Johnson}, {Gal-Yam}, {Leonard}, {Simon},
  {Udalski}, \& {Gould}}]{johnson07}
{Johnson}, J.~A., {Gal-Yam}, A., {Leonard}, D.~C., {Simon}, J.~D., {Udalski},
  A., \& {Gould}, A. 2007, \apjl, 655, L33

\bibitem[{{Kasen} \& {Woosley}(2009)}]{kasen09}
{Kasen}, D., \& {Woosley}, S.~E. 2009, \apj, 703, 2205

\bibitem[{{Kauffmann} {et~al.}(2003){Kauffmann}, {Heckman}, {White}, {Charlot},
  {Tremonti}, {Brinchmann}, {Bruzual}, {Peng}, {Seibert}, {Bernardi},
  {Blanton}, {Brinkmann}, {Castander}, {Cs{\'a}bai}, {Fukugita}, {Ivezic},
  {Munn}, {Nichol}, {Padmanabhan}, {Thakar}, {Weinberg}, \&
  {York}}]{kauffmann03}
{Kauffmann}, G., {et~al.} 2003, \mnras, 341, 33

\bibitem[{{Kelly} \& {Kirshner}(2012)}]{kelly12}
{Kelly}, P.~L., \& {Kirshner}, R.~P. 2012, \apj, 759, 107

\bibitem[{{Kelly} {et~al.}(2008){Kelly}, {Kirshner}, \& {Pahre}}]{kelly08}
{Kelly}, P.~L., {Kirshner}, R.~P., \& {Pahre}, M. 2008, \apj, 687, 1201

\bibitem[{{Kessler} {et~al.}(2009){Kessler}, {Becker}, {Cinabro}, {Vanderplas},
  {Frieman}, {Marriner}, {Davis}, {Dilday}, {Holtzman}, {Jha}, {Lampeitl},
  {Sako}, {Smith}, {Zheng}, {Nichol}, {Bassett}, {Bender}, {Depoy}, {Doi},
  {Elson}, {Filippenko}, {Foley}, {Garnavich}, {Hopp}, {Ihara}, {Ketzeback},
  {Kollatschny}, {Konishi}, {Marshall}, {McMillan}, {Miknaitis}, {Morokuma},
  {M{\"o}rtsell}, {Pan}, {Prieto}, {Richmond}, {Riess}, {Romani}, {Schneider},
  {Sollerman}, {Takanashi}, {Tokita}, {van der Heyden}, {Wheeler}, {Yasuda}, \&
  {York}}]{kessler09}
{Kessler}, R., {et~al.} 2009, \apjs, 185, 32

\bibitem[{{Kewley} \& {Dopita}(2002)}]{kd02}
{Kewley}, L.~J., \& {Dopita}, M.~A. 2002, \apjs, 142, 35

\bibitem[{{Kewley} \& {Ellison}(2008)}]{kewley08}
{Kewley}, L.~J., \& {Ellison}, S.~L. 2008, \apj, 681, 1183

\bibitem[{{Kobulnicky} \& {Kewley}(2004)}]{kk04}
{Kobulnicky}, H.~A., \& {Kewley}, L.~J. 2004, \apj, 617, 240

\bibitem[{{Kocevski} \& {West}(2011)}]{kocevski11}
{Kocevski}, D., \& {West}, A.~A. 2011, \apjl, 735, L8

\bibitem[{{Koz{\l}owski} {et~al.}(2010){Koz{\l}owski}, {Kochanek}, {Stern},
  {Prieto}, {Stanek}, {Thompson}, {Assef}, {Drake}, {Szczygie{\l}},
  {Wo{\'z}niak}, {Nugent}, {Ashby}, {Beshore}, {Brown}, {Dey}, {Griffith},
  {Harrison}, {Jannuzi}, {Larson}, {Madsen}, {Pilecki}, {Pojma{\'n}ski},
  {Skowron}, {Vestrand}, \& {Wren}}]{kozlowski10}
{Koz{\l}owski}, S., {et~al.} 2010, \apj, 722, 1624

\bibitem[{{Kriek} {et~al.}(2009){Kriek}, {van Dokkum}, {Labb{\'e}}, {Franx},
  {Illingworth}, {Marchesini}, \& {Quadri}}]{kriek09}
{Kriek}, M., {van Dokkum}, P.~G., {Labb{\'e}}, I., {Franx}, M., {Illingworth},
  G.~D., {Marchesini}, D., \& {Quadri}, R.~F. 2009, \apj, 700, 221

\bibitem[{{Kudritzki} \& {Puls}(2000)}]{kudritzki00}
{Kudritzki}, R., \& {Puls}, J. 2000, \araa, 38, 613

\bibitem[{{Langer} {et~al.}(2007){Langer}, {Norman}, {de Koter}, {Vink},
  {Cantiello}, \& {Yoon}}]{langer07}
{Langer}, N., {Norman}, C.~A., {de Koter}, A., {Vink}, J.~S., {Cantiello}, M.,
  \& {Yoon}, S. 2007, \aap, 475, L19

\bibitem[{{Lara-L{\'o}pez} {et~al.}(2010){Lara-L{\'o}pez}, {Cepa},
  {Bongiovanni}, {P{\'e}rez Garc{\'{\i}}a}, {Ederoclite}, {Casta{\~n}eda},
  {Fern{\'a}ndez Lorenzo}, {Povi{\'c}}, \& {S{\'a}nchez-Portal}}]{laralopez10}
{Lara-L{\'o}pez}, M.~A., {et~al.} 2010, \aap, 521, L53+

\bibitem[{{Lecureur} {et~al.}(2007){Lecureur}, {Hill}, {Zoccali}, {Barbuy},
  {G{\'o}mez}, {Minniti}, {Ortolani}, \& {Renzini}}]{lecureur07}
{Lecureur}, A., {Hill}, V., {Zoccali}, M., {Barbuy}, B., {G{\'o}mez}, A.,
  {Minniti}, D., {Ortolani}, S., \& {Renzini}, A. 2007, \aap, 465, 799

\bibitem[{{Leloudas} {et~al.}(2011){Leloudas}, {Gallazzi}, {Sollerman},
  {Stritzinger}, {Fynbo}, {Hjorth}, {Malesani}, {Micha{\l}owski},
  {Milvang-Jensen}, \& {Smith}}]{leloudas11}
{Leloudas}, G., {et~al.} 2011, \aap, 530, A95

\bibitem[{{Levesque} {et~al.}(2010){Levesque}, {Kewley}, {Berger}, \& {Jabran
  Zahid}}]{levesque10b}
{Levesque}, E.~M., {Kewley}, L.~J., {Berger}, E., \& {Jabran Zahid}, H. 2010,
  \aj, 140, 1557

\bibitem[{{Li} {et~al.}(2011){Li}, {Leaman}, {Chornock}, {Filippenko},
  {Poznanski}, {Ganeshalingam}, {Wang}, {Modjaz}, {Jha}, {Foley}, \&
  {Smith}}]{li11}
{Li}, W., {et~al.} 2011, \mnras, 412, 1441

\bibitem[{{L{\'o}pez-S{\'a}nchez} {et~al.}(2012){L{\'o}pez-S{\'a}nchez},
  {Dopita}, {Kewley}, {Zahid}, {Nicholls}, \&
  {Scharw{\"a}chter}}]{lopezsanchez12}
{L{\'o}pez-S{\'a}nchez}, {\'A}.~R., {Dopita}, M.~A., {Kewley}, L.~J., {Zahid},
  H.~J., {Nicholls}, D.~C., \& {Scharw{\"a}chter}, J. 2012, \mnras, 426, 2630

\bibitem[{{Maiolino} {et~al.}(2002){Maiolino}, {Vanzi}, {Mannucci}, {Cresci},
  {Ghinassi}, \& {Della Valle}}]{maiolino02}
{Maiolino}, R., {Vanzi}, L., {Mannucci}, F., {Cresci}, G., {Ghinassi}, F., \&
  {Della Valle}, M. 2002, \aap, 389, 84

\bibitem[{{Mannucci} {et~al.}(2010){Mannucci}, {Cresci}, {Maiolino}, {Marconi},
  \& {Gnerucci}}]{mannucci10}
{Mannucci}, F., {Cresci}, G., {Maiolino}, R., {Marconi}, A., \& {Gnerucci}, A.
  2010, \mnras, 408, 2115

\bibitem[{{Mannucci} {et~al.}(2011){Mannucci}, {Salvaterra}, \&
  {Campisi}}]{mannucci11}
{Mannucci}, F., {Salvaterra}, R., \& {Campisi}, M.~A. 2011, \mnras, 414, 1263

\bibitem[{{Mannucci} {et~al.}(2003){Mannucci}, {Maiolino}, {Cresci}, {Della
  Valle}, {Vanzi}, {Ghinassi}, {Ivanov}, {Nagar}, \&
  {Alonso-Herrero}}]{mannucci03}
{Mannucci}, F., {et~al.} 2003, \aap, 401, 519

\bibitem[{{Martini} {et~al.}(2011){Martini}, {Stoll}, {Derwent}, {Zhelem},
  {Atwood}, {Gonzalez}, {Mason}, {O'Brien}, {Pappalardo}, {Pogge}, {Ward}, \&
  {Wong}}]{martini11}
{Martini}, P., {et~al.} 2011, \pasp, 123, 187

\bibitem[{{McGaugh}(1991)}]{m91}
{McGaugh}, S.~S. 1991, \apj, 380, 140

\bibitem[{{McWilliam}(1997)}]{mcwilliam97}
{McWilliam}, A. 1997, \araa, 35, 503

\bibitem[{{Modjaz} {et~al.}(2010){Modjaz}, {Filippenko}, {Silverman},
  {Kleiser}, \& {Morton}}]{2010ayZ}
{Modjaz}, M., {Filippenko}, A.~V., {Silverman}, J.~M., {Kleiser}, I.~K.~W., \&
  {Morton}, A.~J.~L. 2010, The Astronomer's Telegram, 2503, 1

\bibitem[{{Modjaz} {et~al.}(2011){Modjaz}, {Kewley}, {Bloom}, {Filippenko},
  {Perley}, \& {Silverman}}]{modjaz11}
{Modjaz}, M., {Kewley}, L., {Bloom}, J.~S., {Filippenko}, A.~V., {Perley}, D.,
  \& {Silverman}, J.~M. 2011, \apjl, 731, L4+

\bibitem[{{Modjaz} {et~al.}(2008){Modjaz}, {Kewley}, {Kirshner}, {Stanek},
  {Challis}, {Garnavich}, {Greene}, {Kelly}, \& {Prieto}}]{modjaz08}
{Modjaz}, M., {et~al.} 2008, \aj, 135, 1136

\bibitem[{{Murphy} {et~al.}(2011){Murphy}, {Jennings}, {Williams}, {Dalcanton},
  \& {Dolphin}}]{murphy11}
{Murphy}, J.~W., {Jennings}, Z.~G., {Williams}, B., {Dalcanton}, J.~J., \&
  {Dolphin}, A.~E. 2011, \apjl, 742, L4

\bibitem[{{Neill} {et~al.}(2009){Neill}, {Sullivan}, {Howell}, {Conley},
  {Seibert}, {Martin}, {Barlow}, {Foster}, {Friedman}, {Morrissey}, {Neff},
  {Schiminovich}, {Wyder}, {Bianchi}, {Donas}, {Heckman}, {Lee}, {Madore},
  {Milliard}, {Rich}, \& {Szalay}}]{neill09}
{Neill}, J.~D., {et~al.} 2009, \apj, 707, 1449

\bibitem[{{Neill} {et~al.}(2011){Neill}, {Sullivan}, {Gal-Yam}, {Quimby},
  {Ofek}, {Wyder}, {Howell}, {Nugent}, {Seibert}, {Martin}, {Overzier},
  {Barlow}, {Foster}, {Friedman}, {Morrissey}, {Neff}, {Schiminovich},
  {Bianchi}, {Donas}, {Heckman}, {Lee}, {Madore}, {Milliard}, {Rich}, \&
  {Szalay}}]{neill11}
---. 2011, \apj, 727, 15

\bibitem[{{Niino}(2011)}]{niino11}
{Niino}, Y. 2011, \mnras, 417, 567

\bibitem[{{Ober} {et~al.}(1983){Ober}, {El Eid}, \& {Fricke}}]{ober83}
{Ober}, W.~W., {El Eid}, M.~F., \& {Fricke}, K.~J. 1983, \aap, 119, 61

\bibitem[{{Okada} {et~al.}(2008){Okada}, {Onaka}, {Miyata}, {Okamoto}, {Sakon},
  {Shibai}, \& {Takahashi}}]{okada08}
{Okada}, Y., {Onaka}, T., {Miyata}, T., {Okamoto}, Y.~K., {Sakon}, I.,
  {Shibai}, H., \& {Takahashi}, H. 2008, \apj, 682, 416

\bibitem[{{Pagel} {et~al.}(1979){Pagel}, {Edmunds}, {Blackwell}, {Chun}, \&
  {Smith}}]{pagel79}
{Pagel}, B.~E.~J., {Edmunds}, M.~G., {Blackwell}, D.~E., {Chun}, M.~S., \&
  {Smith}, G. 1979, \mnras, 189, 95

\bibitem[{{Pauldrach} {et~al.}(1986){Pauldrach}, {Puls}, \&
  {Kudritzki}}]{pauldrach86}
{Pauldrach}, A., {Puls}, J., \& {Kudritzki}, R.~P. 1986, \aap, 164, 86

\bibitem[{{Pettini} \& {Pagel}(2004)}]{pp04}
{Pettini}, M., \& {Pagel}, B.~E.~J. 2004, \mnras, 348, L59

\bibitem[{{Prantzos} \& {Boissier}(2003)}]{prantzos03}
{Prantzos}, N., \& {Boissier}, S. 2003, \aap, 406, 259

\bibitem[{{Prieto} \& {Filippenko}(2010)}]{2010aydisc}
{Prieto}, J., \& {Filippenko}, A.~V. 2010, Central Bureau Electronic Telegrams,
  2224, 3

\bibitem[{{Prieto} {et~al.}(2008){Prieto}, {Stanek}, \& {Beacom}}]{prieto08z}
{Prieto}, J.~L., {Stanek}, K.~Z., \& {Beacom}, J.~F. 2008, \apj, 673, 999

\bibitem[{{Rau} {et~al.}(2009){Rau}, {Kulkarni}, {Law}, {Bloom}, {Ciardi},
  {Djorgovski}, {Fox}, {Gal-Yam}, {Grillmair}, {Kasliwal}, {Nugent}, {Ofek},
  {Quimby}, {Reach}, {Shara}, {Bildsten}, {Cenko}, {Drake}, {Filippenko},
  {Helfand}, {Helou}, {Howell}, {Poznanski}, \& {Sullivan}}]{rau09}
{Rau}, A., {et~al.} 2009, \pasp, 121, 1334

\bibitem[{{Reddy} {et~al.}(2006){Reddy}, {Lambert}, \& {Allende
  Prieto}}]{reddy06}
{Reddy}, B.~E., {Lambert}, D.~L., \& {Allende Prieto}, C. 2006, \mnras, 367,
  1329

\bibitem[{{Reddy} {et~al.}(2003){Reddy}, {Tomkin}, {Lambert}, \& {Allende
  Prieto}}]{reddy03}
{Reddy}, B.~E., {Tomkin}, J., {Lambert}, D.~L., \& {Allende Prieto}, C. 2003,
  \mnras, 340, 304

\bibitem[{{Rich} \& {Origlia}(2005)}]{rich05}
{Rich}, R.~M., \& {Origlia}, L. 2005, \apj, 634, 1293

\bibitem[{{Rich} {et~al.}(2007){Rich}, {Origlia}, \& {Valenti}}]{rich07}
{Rich}, R.~M., {Origlia}, L., \& {Valenti}, E. 2007, \apjl, 665, L119

\bibitem[{{Rodr{\'{\i}}guez}(2002)}]{rodriguez02}
{Rodr{\'{\i}}guez}, M. 2002, \aap, 389, 556

\bibitem[{{Salim} {et~al.}(2007){Salim}, {Rich}, {Charlot}, {Brinchmann},
  {Johnson}, {Schiminovich}, {Seibert}, {Mallery}, {Heckman}, {Forster},
  {Friedman}, {Martin}, {Morrissey}, {Neff}, {Small}, {Wyder}, {Bianchi},
  {Donas}, {Lee}, {Madore}, {Milliard}, {Szalay}, {Welsh}, \& {Yi}}]{salim07}
{Salim}, S., {et~al.} 2007, \apjs, 173, 267

\bibitem[{{Sanders} {et~al.}(2012){Sanders}, {Soderberg}, {Valenti}, {Foley},
  {Chornock}, {Chomiuk}, {Berger}, {Smartt}, {Hurley}, {Barthelmy}, {Levesque},
  {Narayan}, {Botticella}, {Briggs}, {Connaughton}, {Terada}, {Gehrels},
  {Golenetskii}, {Mazets}, {Cline}, {von Kienlin}, {Boynton}, {Chambers},
  {Grav}, {Heasley}, {Hodapp}, {Jedicke}, {Kaiser}, {Kirshner}, {Kudritzki},
  {Luppino}, {Lupton}, {Magnier}, {Monet}, {Morgan}, {Onaka}, {Price},
  {Stubbs}, {Tonry}, {Wainscoat}, \& {Waterson}}]{sanders12}
{Sanders}, N.~E., {et~al.} 2012, \apj, 756, 184

\bibitem[{{Savaglio} {et~al.}(2005){Savaglio}, {Glazebrook}, {Le Borgne},
  {Juneau}, {Abraham}, {Chen}, {Crampton}, {McCarthy}, {Carlberg}, {Marzke},
  {Roth}, {J{\o}rgensen}, \& {Murowinski}}]{savaglio05}
{Savaglio}, S., {et~al.} 2005, \apj, 635, 260

\bibitem[{{Smartt}(2009)}]{smartt09review}
{Smartt}, S.~J. 2009, \araa, 47, 63

\bibitem[{{Stanek} {et~al.}(2006){Stanek}, {Gnedin}, {Beacom}, {Gould},
  {Johnson}, {Kollmeier}, {Modjaz}, {Pinsonneault}, {Pogge}, \&
  {Weinberg}}]{stanek06}
{Stanek}, K.~Z., {et~al.} 2006, \actaa, 56, 333

\bibitem[{{Stoll} {et~al.}(2011){Stoll}, {Prieto}, {Stanek}, {Pogge},
  {Szczygie{\l}}, {Pojma{\'n}ski}, {Antognini}, \& {Yan}}]{stoll11}
{Stoll}, R., {Prieto}, J.~L., {Stanek}, K.~Z., {Pogge}, R.~W., {Szczygie{\l}},
  D.~M., {Pojma{\'n}ski}, G., {Antognini}, J., \& {Yan}, H. 2011, \apj, 730, 34

\bibitem[{{Stoll} {et~al.}(2010){Stoll}, {Martini}, {Derwent}, {Gonzalez},
  {O'Brien}, {Pappalardo}, {Pogge}, {Wong}, \& {Zhelem}}]{stoll10}
{Stoll}, R., {et~al.} 2010, in \procspie, Vol. 7735, 154

\bibitem[{{Tinsley}(1979)}]{tinsley79}
{Tinsley}, B.~M. 1979, \apj, 229, 1046

\bibitem[{{Tremonti} {et~al.}(2004){Tremonti}, {Heckman}, {Kauffmann},
  {Brinchmann}, {Charlot}, {White}, {Seibert}, {Peng}, {Schlegel}, {Uomoto},
  {Fukugita}, \& {Brinkmann}}]{tremonti04}
{Tremonti}, C.~A., {et~al.} 2004, \apj, 613, 898

\bibitem[{{Uomoto} {et~al.}(1999){Uomoto}, {Smee}, {Rockosi}, {Burles}, {Pope},
  {Friedman}, {Brinkmann}, {Gunn}, {Nichol}, \& {SDSS
  Collaboration}}]{uomoto99}
{Uomoto}, A., {et~al.} 1999, in Bulletin of the American Astronomical Society,
  Vol.~31, American Astronomical Society Meeting Abstracts, 1501

\bibitem[{{Vergani} {et~al.}(2011){Vergani}, {Flores}, {Covino}, {Fugazza},
  {Gorosabel}, {Levan}, {Puech}, {Salvaterra}, {Tello}, {de Ugarte Postigo},
  {D'Avanzo}, {D'Elia}, {Fern{\'a}ndez}, {Fynbo}, {Ghirlanda},
  {Jel{\'{\i}}nek}, {Lundgren}, {Malesani}, {Palazzi}, {Piranomonte},
  {Rodrigues}, {S{\'a}nchez-Ram{\'{\i}}rez}, {Terr{\'o}n}, {Th{\"o}ne},
  {Antonelli}, {Campana}, {Castro-Tirado}, {Goldoni}, {Hammer}, {Hjorth},
  {Jakobsson}, {Kaper}, {Melandri}, {Milvang-Jensen}, {Sollerman},
  {Tagliaferri}, {Tanvir}, {Wiersema}, \& {Wijers}}]{vergani11}
{Vergani}, S.~D., {et~al.} 2011, \aap, 535, A127

\bibitem[{{Vink} \& {de Koter}(2005)}]{vink05}
{Vink}, J.~S., \& {de Koter}, A. 2005, \aap, 442, 587

\bibitem[{{Wright}(2006)}]{cosmocalc}
{Wright}, E.~L. 2006, \pasp, 118, 1711

\bibitem[{{Yates} {et~al.}(2012){Yates}, {Kauffmann}, \& {Guo}}]{yates12}
{Yates}, R.~M., {Kauffmann}, G., \& {Guo}, Q. 2012, \mnras, 422, 215

\bibitem[{{Yin} {et~al.}(2007){Yin}, {Liang}, {Hammer}, {Brinchmann}, {Zhang},
  {Deng}, \& {Flores}}]{yin07}
{Yin}, S.~Y., {Liang}, Y.~C., {Hammer}, F., {Brinchmann}, J., {Zhang}, B.,
  {Deng}, L.~C., \& {Flores}, H. 2007, \aap, 462, 535

\bibitem[{{York} {et~al.}(2000){York}, {Adelman}, {Anderson}, {Anderson},
  {Annis}, {Bahcall}, {Bakken}, {Barkhouser}, {Bastian}, {Berman}, {Boroski},
  {Bracker}, {Briegel}, {Briggs}, {Brinkmann}, {Brunner}, {Burles}, {Carey},
  {Carr}, {Castander}, {Chen}, {Colestock}, {Connolly}, {Crocker}, {Csabai},
  {Czarapata}, {Davis}, {Doi}, {Dombeck}, {Eisenstein}, {Ellman}, {Elms},
  {Evans}, {Fan}, {Federwitz}, {Fiscelli}, {Friedman}, {Frieman}, {Fukugita},
  {Gillespie}, {Gunn}, {Gurbani}, {de Haas}, {Haldeman}, {Harris}, {Hayes},
  {Heckman}, {Hennessy}, {Hindsley}, {Holm}, {Holmgren}, {Huang}, {Hull},
  {Husby}, {Ichikawa}, {Ichikawa}, {Ivezi{\'c}}, {Kent}, {Kim}, {Kinney},
  {Klaene}, {Kleinman}, {Kleinman}, {Knapp}, {Korienek}, {Kron}, {Kunszt},
  {Lamb}, {Lee}, {Leger}, {Limmongkol}, {Lindenmeyer}, {Long}, {Loomis},
  {Loveday}, {Lucinio}, {Lupton}, {MacKinnon}, {Mannery}, {Mantsch}, {Margon},
  {McGehee}, {McKay}, {Meiksin}, {Merelli}, {Monet}, {Munn}, {Narayanan},
  {Nash}, {Neilsen}, {Neswold}, {Newberg}, {Nichol}, {Nicinski}, {Nonino},
  {Okada}, {Okamura}, {Ostriker}, {Owen}, {Pauls}, {Peoples}, {Peterson},
  {Petravick}, {Pier}, {Pope}, {Pordes}, {Prosapio}, {Rechenmacher}, {Quinn},
  {Richards}, {Richmond}, {Rivetta}, {Rockosi}, {Ruthmansdorfer}, {Sandford},
  {Schlegel}, {Schneider}, {Sekiguchi}, {Sergey}, {Shimasaku}, {Siegmund},
  {Smee}, {Smith}, {Snedden}, {Stone}, {Stoughton}, {Strauss}, {Stubbs},
  {SubbaRao}, {Szalay}, {Szapudi}, {Szokoly}, {Thakar}, {Tremonti}, {Tucker},
  {Uomoto}, {Vanden Berk}, {Vogeley}, {Waddell}, {Wang}, {Watanabe},
  {Weinberg}, {Yanny}, {Yasuda}, \& {SDSS Collaboration}}]{york00}
{York}, D.~G., {et~al.} 2000, \aj, 120, 1579

\bibitem[{{Young} {et~al.}(2010){Young}, {Smartt}, {Valenti}, {Pastorello},
  {Benetti}, {Benn}, {Bersier}, {Botticella}, {Corradi}, {Harutyunyan},
  {Hrudkova}, {Hunter}, {Mattila}, {de Mooij}, {Navasardyan}, {Snellen},
  {Tanvir}, \& {Zampieri}}]{young10}
{Young}, D.~R., {et~al.} 2010, \aap, 512, A70

\bibitem[{{Zaritsky} {et~al.}(1994){Zaritsky}, {Kennicutt}, \& {Huchra}}]{z94}
{Zaritsky}, D., {Kennicutt}, Jr., R.~C., \& {Huchra}, J.~P. 1994, \apj, 420, 87

\end{thebibliography}
\end{document}